\documentclass[useAMS,usenatbib]{mn2e}

\usepackage{graphicx}
\usepackage{longtable}
\usepackage{amssymb,amsmath}
\newcommand{\zlbg}{$z{\sim}5.7$}
\newcommand{\zid}{$z{\sim} 6$}
\newcommand{\zlae}{$z{\sim}5.7$}

\newcommand{\Lya}{\mbox{Ly$\alpha$}}

\newcommand{\msun}{\mbox{$M_{\sun}$}}

\newcommand{\kms}{\mbox{km s$^{-1}$}}
\newcommand{\cm}{cm$^{-2}$}
\newcommand{\civ}{\hbox{C\,{\sc iv}}}
\newcommand{\siiv}{\hbox{Si\,{\sc iv}}}

\newcommand{\oi}{\hbox{O\,{\sc i}}}

\newcommand{\nciv}{\mbox{$N_{\text \civ}$}}

\def\ltsima{$\,\buildrel{<}\over {\sim}\,$}
\def\simlt{\lower.5ex\hbox{\ltsima}}
\def\gtsima{$\,\buildrel{>}\over {\sim}\,$}
\def\simgt{\lower.5ex\hbox{\gtsima}}

\title[Environment of $z{\sim}5.7$ \civ\ absorption systems]{Large-scale environment of 
$z{\sim}5.7$ \civ\ absorption systems I: projected distribution of galaxies\thanks{
Based on data collected at Subaru Telescope, which is operated by the National Astronomical Observatory of Japan}}
\author[D\'{i}az et. al.]{
C. Gonzalo D\'{i}az$^{1}$\thanks{E-mail:gdiaz@swin.edu.au}, 
Yusei Koyama$^{2}$, Emma V. Ryan-Weber$^{1}$, Jeff Cooke$^{1}$,
\newauthor \, Masami Ouchi$^{3,4}$, Kazuhiro Shimasaku$^{5,6}$, and Fumiaki Nakata $^{7}$ \\
\\
$^{1}$Centre for Astrophysics and Supercomputing, Swinburne University of Technology, Hawthorn, VIC 3122, Australia\\
$^{2}$National Astronomical Observatory of Japan, Mitaka, Tokyo 181-8588, Japan\\
$^{3}$Institute for Cosmic Ray Research, The University of Tokyo, Kashiwa, Chiba 277-8582, Japan\\
$^{4}$Kavli Institute for the Physics and Mathematics of the Universe (WPI), The University of Tokyo, Kashiwa, Chiba 277-8583, Japan\\
$^{5}$Research Center for the Early Universe (WPI), University of Tokyo, Bunkyo, Tokyo 113-0033, Japan\\
$^{6}$Department of Astronomy, Graduate School of Science, The University of Tokyo, Bunkyo-ku, Tokyo 113-0033, Japan\\
$^{7}$Subaru Telescope, National Astronomical Observatory of Japan, Hilo, HI 96720, USA}

\begin{document}

\date{Accepted. Received}

\pagerange{\pageref{firstpage}--\pageref{lastpage}} \pubyear{2010}

\maketitle

\label{firstpage}

\begin{abstract}

Metal absorption systems are products of star formation.
They are believed to be associated with massive star forming galaxies, 
which have significantly enriched their surroundings.
To test this idea with high column density \civ\ absorption systems at $z{\sim}5.7$,
we study the projected distribution of galaxies and characterise the environment 
of \civ\ systems in two independent quasar lines-of-sight:
J103027.01+052455.0 and J113717.73+354956.9. 
Using wide field photometry (${\sim} 80 {\times}60 {\it h}^{-1}$ comoving Mpc), 
we select bright (M$_{\rm UV}(1350$\AA$)\simlt{-}21.0$ mag.) 
Lyman break galaxies (LBGs) at $z{\sim}5.7$ in a redshift 
slice $\Delta z {\sim}0.2$ and we compare their 
projected distribution with \zlbg\ narrow-band 
selected Lyman alpha emitters (LAEs, $\Delta z {\sim}0.08$).

We find that the \civ\ systems are located more than 
10${\it h}^{-1}$ projected comoving Mpc
from the main concentrations of LBGs
and no candidate is closer than 
${\sim}5{\it h}^{-1}$ projected comoving Mpc.
In contrast, an excess of LAEs
--lower mass galaxies-- is found
on scales of ${\sim}10{\it h}^{-1}$ comoving Mpc, 
suggesting that LAEs are the primary 
candidates for the source of the \civ\ systems.
Furthermore, the closest object 
to the system in the field J1030+0524
is a faint LAE at a projected distance of
212${\it h}^{-1}$ physical kpc. 
However, this work cannot rule out undiscovered 
lower mass galaxies as the origin of these 
absorption systems.

We conclude that, in contrast with lower
redshift examples ($z{\simlt}3.5$), 
strong \civ\ absorption systems at \zlbg\ 
trace low-to-intermediate density environments 
dominated by low-mass galaxies.
Moreover, the excess of LAEs associated with
high levels of ionizing flux agrees with
the idea that faint galaxies dominate the ionizing
photon budget at this redshift.

\end{abstract}


\section{Introduction}

There is significant observational evidence 
that cosmic reionization of hydrogen
is largely completed by $z{\sim}5.7$
\citep[e.g.][]{ouchi2010, kashikawa2011,larson2011,komatsu2011,caruana2012,finkelstein2012b,zahn2012,jensen2013}.
\citet{zahn2012} combined WMAP7 and 
South Pole Telescope data to model the duration of
the epoch of reionization (EoR) by including
the kinetic Sunyaev-Zel'dovich effect in their analysis.
They report for the most conservative case that the EoR
begins at $z_{EoR}{<}13.1$  and
is over by $z_{EoR}{>}5.8$ at $95\%$ confidence level.
Moreover, studies of narrow-band selected \Lya\ emitters
(LAEs) have found that the normalisation of the 
\Lya\ luminosity function of LAEs
is higher at $z{\sim}5.7$ than $z{\sim}6.5$ 
\citep[e.g.][]{ouchi2010,hu2010,clement2012}.
Since \Lya\ is a resonant line, 
a small amount of neutral hydrogen in the intergalactic medium (IGM) 
is able to reduce the number of 
\Lya\ photons that are transmitted.
Therefore, a higher transmission of \Lya\
photons at $z{=}5.7$ would result from a lower fraction 
of neutral-to-ionized hydrogen in the IGM than at $z\sim6.5$,
suggesting that the main reionization process took 
place at $z_{EoR}{\simgt}6$.
\citet[][2011]{stark2010} reported a decrease in the fraction
of Lyman break galaxies \citep[LBGs,][]{steidel1996} with \Lya\ in emission
from $z{=}6$ to $z{=}3$, at fixed luminosity, in line 
with the observational and theoretical expectation 
of the evolution of star-forming galaxies over this
period.
However, several spectroscopic campaigns 
for the identification of $z{\simgt} 7$ LBGs 
have shown very low numbers of 
\Lya\ detections \citep{fontana2010, pentericci2011, schenker2012a, ono2012a, caruana2012, bunker2013, treu2013}.
This suppression of the \Lya\ emission
could indicate
a more neutral IGM at $z{>}6$
\citep[although see][]{bolton2013}.
Finally, a classic piece of evidence that
cosmic reionization was probably 
complete by $z{\sim}6$
comes from the \Lya\ forest in the spectra 
of high redshift QSOs. 
They show several examples of complete Gunn--Peterson 
troughs \citep{gunnpeterson1965} at $z{>}6$ and an optical depth 
decreasing with cosmic time \citep[e.g.][]{becker2001, fan2006a, goto2011, mortlock2011}.

\citet{becker2007} show that the distribution 
of optical depths in the \Lya\ forest is better reproduced 
by models that account for an inhomogeneous 
ionizing background and non-uniform IGM temperature,
which implies that reionization was certainly not homogeneous 
nor instantaneous \citep[e.g.][]{schroeder2013}.
Moreover, a heterogeneous reionization is 
also predicted by theoretical studies 
\citep[e.g.][]{trac2008,mesinger2009, finlator2009b, choudhury2009, griffen2013}. 
Hydrodynamical simulations combined with high redshift \Lya\
forest data suggest an extended process of reionization
that occurs in a ``photon-starved'' regime \citep{bolton2007}. 
Semi-numerical simulations have explored the implications of 
this result on the topology of the reionization process.
\citet{choudhury2009} find a two stage process starting 
with an ``inside-out'' topology in which high-density regions 
hosting ionizing sources are the first to be ionized.
From there, reionization proceeded directly into voids 
while dense regions which hosted no ionizing sources 
and had remained neutral to this point, 
were slowly ionized from the outside (``outside-in'').

After reionization is complete, the relative flux of 
ionizing radiation that is left over could trace the 
history of reionization of the large-scale structure.
In particular, the scatter in the intensity of the UV 
background is affected by the distribution of sources of 
the ionizing field \citep[e.g.][]{mesinger2009} and the 
evolution on the mean free path of ionizing photons 
\citep[e.g.][]{miralda-escude2003, bolton2007, faucher-giguere2008}.
\citet{mesinger2009} compare analytic, semi-numeric and 
numeric calculations of inhomogeneous flux fields 
and found a highly variable ionization state 
predicted to persist after reionization at $z{=}5$--6.
For example, if reionization proceeded in a fully
inside-out geometry, a highly ionized IGM 
is expected to exist in regions where the density of 
matter is higher than the average \citep[e.g.][]{trac2008},
leading to a possible 
direct observational test for the challenging question 
of which regions were the first ones to be permanently 
ionized.

A highly ionized IGM can be detected through
metal absorption systems. A typical signature 
is the presence of strong triply ionized carbon
absorption (\civ, ionization energy${=}47.89$eV).
In the redshift range $5.3{<}z{<}6.2$, only four \civ\ absorption systems 
with column densities \nciv${>}10^{14}$\cm\ have been reported 
from a sample of 13 sight-lines towards high-redshift QSOs 
\citep{ryan-weber2009, simcoe2011b,dodorico2013}.
Interestingly, two of them 
lie at similar redshift ($z{\sim}5.72$--5.73).
One strong \civ\ absorption system at $z_{\rm abs}{=} 5.7244$ in the 
line-of-sight towards QSO J1030+0524
\citep{ryan-weber2009, simcoe2011b} is accompanied by a weaker system 
at $z_{\rm abs}{=}5.7440$ \citep[][]{simcoe2011b,dodorico2013}.
The second example is found at redshift $z_{\rm abs}{=}5.7383$ 
towards QSO J1137+3549 \citep{ryan-weber2009}.

The redshift of these three systems 
is in co-incidence with an atmospheric transmission window 
at $\lambda {\sim}8180$\AA\ 
($z_{\rm Ly\alpha}{\sim}5.73$) 
and sets the possibility to search for galaxies in their vicinity 
using ground based observations.
Conveniently located at $z{\sim}5.72$--5.73, these three \civ\ systems 
provide an opportunity to study 
the connection between the ionization state of the intergalactic 
medium shortly after the EoR and the population of galaxies 
in their environment on different scales. 

The evolution with time of the comoving mass density of \civ\ ($\Omega_{\text \civ}(z)$)
shows a rapid rise between $z{\sim}6$ and $z{\sim}5$ 
\citep{ryan-weber2009, becker2009,simcoe2011b}.
This is unlikely to be solely due to a sharp rise in 
the metal content of the Universe as the star formation 
rate density is reasonably smooth over this short period 
of time \citep[e.g.][]{bouwens2007,stark2009,stark2013}. 
More recently, \citet{dodorico2013} revisited the evolution 
of $\Omega_{\text \civ}(z)$ and find it to be smoothly rising
from $z{\sim}6$ to $z\sim1.5$,
as expected from a progressive accumulation of metals. 
Nevertheless, they also report that the column density 
distribution function of \civ\ is lower at $5.3{<}z{<}6.2$ 
than $1.5{<}z{<}5.3$, which suggests a change in 
the number density and/or the physical size of 
the \civ\ absorption systems.
Furthermore, the evolution of the \siiv /\civ\ column density
ratios towards lower redshift currently suggest a change in 
the ionization conditions of the absorbing gas at $z{<}5$ \citep{dodorico2013}.
Therefore, it is possible that the observed
evolution of the abundance of high ionization absorption systems 
is the result of a change in the ionization balance 
driven by the evolution of the ionizing UV background 
after the EoR.
If the distribution of sources dominates the ionizing
field at that time, then the environment of high redshift
highly ionized absorption systems contains information not only 
on the origin of the enrichment of the Universe but also 
on the nature of the ionizing sources.

Recent studies suggest that sub-L$^{\star}$ 
galaxies most likely dominate the ionizing photon 
budget at $z{\simgt}6$ \citep{cassata2011, dressler2011, kuhlen2012, jaacks2012, 
finkelstein2012b, ferrara2013, robertson2013, cai2014, fontanot2014}.
Moreover, many authors have observed an anti-correlation 
between UV luminosity and the strength of the \Lya\ emission line,
with fainter objects showing larger \Lya\ equivalent widths 
\citep[e.g.][]{shimasaku2006, ouchi2008, vanzella2009, ouchi2010};
and an anti-correlation between the UV luminosity and the fraction of 
galaxies with \Lya\ emission \citep{stark2010, stark2011}, with fainter objects
being more likely to show \Lya\ emission.
Because at high redshift the UV luminosity of star forming galaxies
correlates with the stellar mass \citep[e.g.][]{mclure2011, gonzalez2011},
the interpretation of these trends suggest that is possible to use LAEs
to preferentially select low-mass galaxies.

The two goals of this work are: a) to identify the galaxies associated
to the nearby environment of highly ionized \civ\ absorption systems 
shortly after the EoR; and b) to characterise the matter density
distribution at larger scales using galaxies as tracers of the large-scale structure.
First, if the change in $\Omega_{\text \civ}(z)$ is driven only by
the metal content of the IGM, meaning that at \zlbg\
the IGM is simply less enriched, then it is reasonable 
to expect that strong \civ\ systems are associated with 
regions of earlier star formation episodes where the IGM
was polluted first and for longer times. 
This is found at lower redshift, $2{\simlt} z {\simlt} 3$ 
\citep[e.g.:][]{adelberger2005b, steidel2010},
where galaxies in denser environment are more likely to have
a strong \civ\ system within 1${\it h}^{-1}$ comoving Mpc. 
Thus, if the absorbing gas at \zlbg\ 
is not affected by changes in the 
IGM ionization state, then 
we would expect \zlbg\ \civ\ absorption systems 
near over-densities of LBGs similar to that found 
at $z{\simlt} 3$.
Second, if the change in $\Omega_{\text \civ}(z)$ results
from the evolution in the ionizing flux density background fluctuations, 
then \civ\ systems at \zlbg\ would trace regions of high 
flux density of ionizing radiation. In this case, a simple
prediction from an inside-out reionization process is  
a positive correlation between mass distribution
and the ionization level of the IGM.  
Under this scenario, rare highly ionized strong 
absorption systems would be expected 
to reside in dense structures that collapsed earlier 
and were reionized first. 
Finally, a third scenario involving the reversal of 
the topology of reionization is also possible.
If young low-mass galaxies forming away from 
the main over-densities provide a final push for 
the cosmic reionization, then these will be the regions
that, at large scales, will present a higher 
ionizing flux density that favours the detection of \civ\
in absorption. 

We report that \civ\ absorption systems are found
in low-to-intermediate density environments
populated by low mass galaxies and are not associated with
over-densities of massive galaxies, which
is in tension with the expectation from a fully inside-out reionization,
but in agreement with an outside-in reionization during the last stage of the EoR.
This work supports the idea that, although reionization is complete
by $z{\sim}5.7$, the predicted inhomogeneous ionizing flux density 
of the IGM affects the detection of high ionization metal absorption systems.
In this case, the finding that the immediate environment of highly-ionized absorption 
systems at \zlbg\ is dominated by low-mass galaxies
is a new piece of evidence that these galaxies are 
an important sources of ionizing radiation at $z{\sim}6$.

This paper is organised as follows:
Section \ref{s:obs-red} describes the observations,
Section \ref{s:sel} explains the photometric selection of the 
galaxies for the study, Section \ref{s:results-colours}
presents the colours and magnitudes of the LBGs and LAEs candidates,
Section \ref{s:results-num} reports the number density of each sample,
and Section \ref{s:results-sd} describes in detail their surface density
distribution. 
The discussion on the origin of the \civ\ and the 
reionization of the IGM can be found in Section \ref{s:discussion}. 
Finally, a summary of the work and the conclusions 
are presented in Section \ref{s:conclusion}.
Throughout this work we use AB magnitudes and assume a 
flat universe with $H_{0}{=}70$\kms Mpc$^{-1}$ (${\it h}{=}0.7$),
$\Omega_{m}{=}0.3$ and $\Omega_{\lambda}{=}0.7$.  

\section[]{Observations and data reduction}\label{s:obs-red}

\subsection{Subaru data}

This section presents the observational data and
the reduction process.
The present work is based on broad-band 
and narrow-band photometry obtained with 
Suprime-Cam \citep{miyazaki2002} on the Subaru Telescope.
We use broad-band R$_c$, i' and 
z' filters covering the wavelength 
range ${\sim}5800$--$10 000$\AA\ 
and a custom-made narrow-band filter
to detect \Lya\ in emission at redshift $z{=}5.71\pm0.04$
(NB\civ , $\lambda_c{=}8162$\AA, FWHM${=}100$\AA).
Observations with the NB\civ, R$_c$ and i' band were acquired the
nights 07--08 March 2011
and images in the z' band were obtained 
31 March and 01 April 2011.
We observed two fields centred on 
QSOs SDSS J103027.01+052455.0 
($z_{em} {=} 6.309$, R.A.=10:30:27.01 , DEC.=05:24:55.0) and 
SDSS J113717.73+354956.9 
($z_{em}{=}6.01$, R.A.=11:37:17.73, DEC.=35:49:56.9) 
\citep{fan2006a}, here after J1030+0524 and J1137+3549.
The total exposure time of the final images 
and the full width half maximum (FWHM) of the 
point-spread function (PSF) are presented in Table \ref{t:obs}.
Note that the z' band images have
 the best resolution in both cases. 
These values were measured in science images resulting 
from the reduction process described next. 

The data were processed with the software SDFRED2  
\citep{yagi2002,ouchi2004a} designed to reduce Suprime-Cam data. 
The reduction process includes bias subtraction 
and overscan, flat field correction, distortion correction,
PSF-equalization of different exposures (when needed),
masking of bad regions (e.g. satellite tracks and saturated pixels), 
alignment of all frames and 
stacking to create the final combined image.
Detection and extraction of objects was 
carried out using the source extraction 
software \textsc{sextractor} 2.5.0 \citep{bertin1996}.

In order to have consistent photometry,
first we matched the PSF of 
the field J1030+0524 to PSF=0.87"
and the field J1137+3549 to PSF=1.13",
which equates to the PSF of the filter with the poorest seeing.
The catalogue of broad-band detected objects
was obtained running \textsc{sextractor} in dual mode using
the best resolution z' band image  
for detection and the  
PSF-matched images for aperture photometry. 
The two samples of LBGs analysed in this work 
(sections \ref{s:sel-lbg} and \ref{s:sel-idrop})
are extracted from this catalogue.
Similarly, the catalogue of narrow-band selected objects
used to identify LAEs (section \ref{s:sel-LAE})
was obtained running \textsc{sextractor} in dual mode
with the NB\civ\ image for detection and the  
PSF-matched images for aperture photometry. 

The following \textsc{sextractor} parameters that regulate 
the detection of sources were used:
DETECT\_MINAREA${=}5$, 
DETECT\_THRESHOLD${=}2.0$, ANALYSIS\_THRESHOLD${=}2.0$ 
and DEBLEND\_MINCON${=}0.005$. 
For both detection and measurement images, the corresponding
{\it rms-map} of the background was converted to a weight-map
($WEIGHT{=}1/RMS^2$) and provided to \textsc{sextractor}
using WEIGHT\_TYPE${=}$MAP\_WEIGHT.
Colours were computed from magnitudes
measured in a 2.0" aperture (MAG\_APER) in J1030+0524 and a 
2.4" aperture in J1137+3549, and MAG\_AUTO 
was used for the total magnitude of an object.
Considering the z' filter samples the UV continuum (rest frame 1350\AA) 
of galaxies at redshift $z{\sim}5.7$,
the continuum magnitude of an object is
measured from the best resolution z' band image.

\begin{table}
 \caption{Exposure time and PSF of science images.}
 \label{t:obs}
 \begin{tabular}{@{}lcrc}
  \hline
  \hline
  Field & Filter & Exposure & PSF \\
   &  & Time (min) & (") \\
    \hline
           & NB\civ\ & 240 & 0.79 \\
           & R$_c$ & 80 & 0.87 \\
J1030+0524 & i' & 90 & 0.81 \\
           & z' & 116 & 0.69 \\
\hline           
           & NB\civ\ & 226 & 1.31 \\
           & R$_c$ & 100 & 1.11 \\
J1137+3549 & i' & 90 & 1.13 \\
           & z' & 114 & 0.67 \\ 
\hline
\end{tabular}
\\
\end{table}

\subsection{Photometric calibrations}

\begin{figure}
\includegraphics[width=84mm]{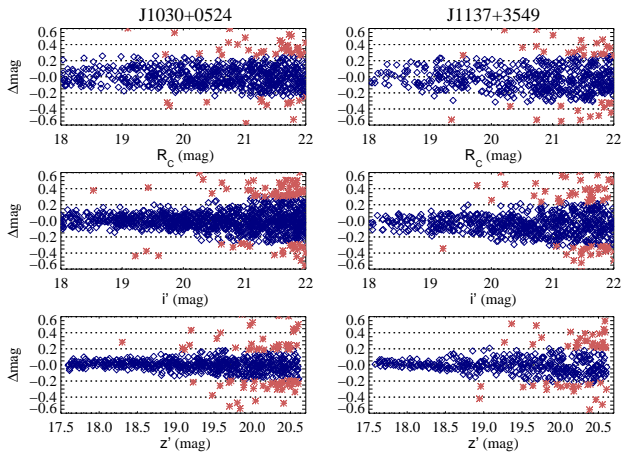}
 \caption{ 
Difference between zero-point corrected magnitudes 
and SDSS magnitudes of stars selected from SDSS.
From top to bottom, residuals in the R$_c$, i' 
and z' band photometry from Suprime-Cam. 
Diamond points are used to estimate the correction and
asterisks represent stars rejected after a $3\sigma$-clipping
used in the fitting process.}
  \label{f:zpoint}
\end{figure}

The zero-point magnitude 
in each broad-band image
was tested in each field
against point-like sources from the Sloan Digital Sky Survey (SDSS)
as both of our fields are covered by the survey.
Stars were selected with magnitudes in the range ${\sim}18$--22
and cross-matched with a total of 690 (545), 1034 (665) and 576 (385) 
point sources in R$_c$, i' and z' band 
in the J1030+0524 (J1137+3945) field.
The best fit to the Sloan magnitudes was found
after applying a three sigma-clipping process. 
Zero-point magnitudes in the J1030+0524 and J1137+3549 field are 
R$_{c\,zp}{=}34.57{\pm}0.11$ and R$_{c\,zp}{=}34.52{\pm}0.14$, 
i'$_{zp}{=}34.62{\pm}0.11$ and i'$_{zp}{=}34.50{\pm}0.11$, 
z'$_{zp}{=}33.52{\pm}0.07$ and z'$_{zp}{=}33.82{\pm}0.09$, 
respectively.
Figure \ref{f:zpoint} shows 
the residuals after the zero-point correction
of the stars selected from SDSS.
Magnitudes in all filters in both fields are in good 
agreement within ${\pm}0.2$--0.3 magnitudes 
with respect to the SDSS magnitudes.
Note the increment in the vertical scatter 
is due to the increase in the uncertainty in SDSS as we
approach its point source limiting magnitude. 
The zero-point magnitudes for the NB\civ\ band
were derived from photometric standard stars
observed the same nights. The stars are GD50 
for the field J1030+0524 and HZ44 for the field J1137+3549.
The zero-point magnitudes are NB\civ$_{zp}{=}32.28$ (J1030+0524)
and NB\civ$_{zp}{=}32.22$ (J1137+3549).

Galactic extinction from the dust map of 
\citet{schlegel1998} is $E(B-V){=}0.024$ in
the field J1030+0524
and 0.018 in the field J1137+3549.
We apply a correction of 0.064 (0.048) magnitudes in R$_c$, 
0.050 (0.038) magnitudes in i' and NB\civ , 
and 0.035 (0.027) magnitudes 
in z' for the field J1030+0524 (J1137+3549).

Aperture corrections in the four bands 
were estimated from the
flux of isolated point sources in 20 apertures 
from 0.4" (2.0 pixels) to 6.0" (29.7 pixels).
The measured fluxes level off in a 5.0" (5.5") 
aperture in the field J1030+0524 (J1137+3549).
Therefore, the fractional flux is estimated 
as the ratio between the flux in the aperture and the flux 
in a $5.0$" ($5.5$") aperture.
We then searched for an aperture with a fractional flux close to $90\%$
in all four bands and find good compromise
with a 2.0" (10 pixels) aperture in the field J1030+0524 and
a 2.4" (12 pixels) aperture in the field J1137+3549.
In particular,
we find that in the field J1030+0524
the fractional fluxes in an aperture of 2.0" in the R$_c$, i', NB\civ\ and z' bands
are $86.8\%$, $89.2\%$, $89.1\%$ and $89.6\%$, respectively,
implying aperture corrections of -0.15 mag, -0.12 mag, -0.11 mag and -0.11 mag. 
In the same way, in the field J1137+3549 the fractional fluxes 
in an aperture of 2.4" are $90.7\%$, $89.9\%$, $75.6\%$ and $94.7\%$.
The corrections for this case are  -0.10 mag, -0.11 mag, -0.24 mag and -0.05 mag.

\subsection{Limiting magnitudes} \label{s:limitmag}

Since the sources of interest are very faint, 
the magnitude error is dominated by the sky background level.
Therefore, the limiting magnitude of 
each PSF-matched image used for aperture photometry
is important to understand the limits of the photometric selection 
criteria described in Section \ref{s:sel}.
To estimate the $5\sigma$-limiting magnitude due to the sky level,
the background level is measured in 10\,000 apertures placed
randomly on the sky. 
The same diameter as for the aperture photometry 
was used in the {\it measurement} images: 2.0" in the field J1030+0524 and 2.4" 
in the field J1137+3549.
For z'-band {\it detection} images (best PSF), a 2.0" aperture was used in both fields.
Then, the FWHM of a Gaussian fit to the distribution of 
background counts is used to estimate $\sigma{=}$FWHM$/2.35482$
and finally obtain the $5\sigma$-limiting magnitude 
 $m_{5\sigma}{=}m_{zp}{-}2.5\log_{10}(5\sigma)$.
The resulting values and the aperture sizes used 
on the process are reported in Table \ref{t:mlim}.
Other ways to explore the detection limits of the data 
are presented and compared to $m_{5\sigma}($z'$)$ 
in Appendix \ref{app:limitmag}.

\begin{table}
 \caption{5$\sigma$-limiting magnitude of science images.}
 \label{t:mlim}
 \begin{tabular}{@{}llccc}
  \hline
  \hline

Field & Image & Filter &  m$_{5\sigma}\,$ & Aperture\\
 &         &   &  magnitudes & (``)\\
  \hline
J1030+0524 & Detection & NB\civ\ & 25.60$^a$ & 2.0 \\ 
           & Detection & z'  &  25.66$^a$ & 2.0 \\
           & Measurement  & NB\civ\ & 25.65$^b$ & 2.0 \\ 
           & Measurement & R$_c$  & 26.60$^b$ & 2.0 \\
           & Measurement & i' &   26.29$^b$ & 2.0\\
           & Measurement & z' &   25.74$^b$ & 2.0\\
  \hline
  \hline    
 J1137+3549 & Detection & NB\civ\  & 25.30 & 2.4   \\
           &  Detection & z'  &  25.85$^a$ & 2.0\\ 
          & Measurement  & NB\civ\ & 25.32 & 2.4  \\
          & Measurement & R$_c$ & 26.24$^b$ & 2.4  \\
          & Measurement & i'  & 25.87$^b$ & 2.4\\
          & Measurement & z'  & 25.64$^b$ & 2.4\\
\hline
\end{tabular}
\\
$^{\it (a)}${Before PSF-matching.}\\
$^{\it (b)}${After PSF-matching.}\\
\end{table}

\section{Photometric candidates for high redshift galaxies}\label{s:sel}

Broad-band photometry 
can be used to detect significant features in the 
observed spectral energy distribution of a galaxy.
In particular, the Lyman break technique
 \citep{steidel1995, steidel1996}
has been successfully  
applied to select high redshift galaxies
as proven by the high fraction of spectroscopic
confirmations (e.g. 82\% at $z {\sim}6$, \citealt{vanzella2009};
${\geq}71$\% at $6{<}z{<}6.5$, \citealt{curtis-lake2012}).
However, a galaxy sample will inevitably
include contamination from lower-redshift objects like 
reddened elliptical galaxies and Galactic stars. 
Therefore, defining a criterion to select galaxies based on broad-band colours
is a matter of compromise between a clean sample and a complete sample.

\begin{figure}
\includegraphics[width=84mm]{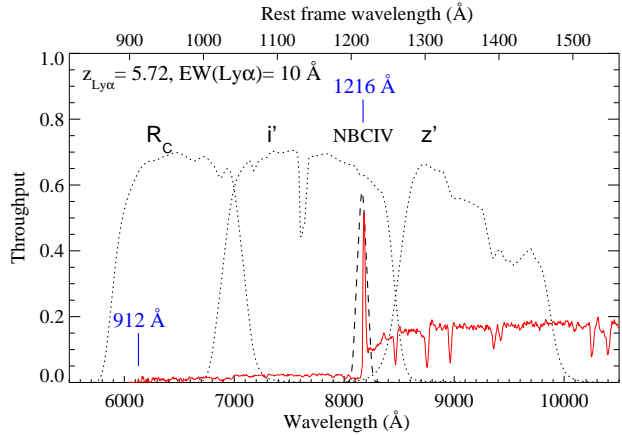}
\caption{Throughput of filters NB\civ\ (dashed line), R$_c$, i', and z'
(dotted lines) on SuprimeCam. 
A Lyman break galaxy template 
with EW(\Lya)=10\AA\, at $z{=}5.72$ is shown in red.
The rest frame wavelength of the Lyman limit and 
Lyman-$\alpha$ is indicated. For $z{>}5.7$,
the ``break'' at \Lya\  enters the z' band and can result in
a rapid evolution in the (i'-z') colour.}
\label{fig:spec}
\end{figure}

In this work, the filters R$_c$, i', and z'
on SuprimeCam were used to 
sample the rest frame 
UV continuum ($ \langle \lambda_{Rest} \rangle {\sim}1350$\AA)
and the hydrogen Lyman series 
(912\AA${<}\lambda_{Rest}{<}$1216\AA) of galaxies at redshift $z{>}5$,
as shown in Figure \ref{fig:spec}. 
High-redshift 
LBGs are identified by the break in their continuum flux at the
wavelength corresponding to the limit of the hydrogen Lyman series 
($\lambda_{Rest}{\simeq}912$\AA) and a significant flux attenuation 
at wavelengths shorter than Lyman-$\alpha$ ($\lambda_{Rest}{\simeq}1216$\AA) 
caused by the presence of neutral hydrogen in the IGM. 
For example, using similar filters
the most common colour
selection that aims to identify such features
at redshift $z {\sim}6$ is (i'-z')${>} 1.3$, and
no detection (e.g: R$_c{<}2\sigma$) in a bluer band 
when available \citep{stanway2004, bunker2004, stiavelli2005,  bouwens2007,  kim2009, vanzella2009, stark2010, pentericci2011}. 
We will refer to this criterion as ``standard" for
$z {\sim}6$ LBGs, or i'-dropout.
 
The redshift of interest for this work ($z {\sim}5.7$)
falls near the lower end of the redshift distribution probed 
by the i'-dropout criteria.
Therefore, to optimize the selection of galaxies at $z{\sim}5.7$ 
it is necessary to understand the limitations of the standard i'-dropout criteria.
Furthermore, some authors have noted that a 
hydrogen Lyman-$\alpha$ emission line 
redshifted to the long wavelength end of the ``dropout'' band
(i.e. i' band in this case, see Figure \ref{fig:spec})
can influence the colour of a galaxy \citep[e.g.][]{stanway2007, stanway2008}. 
Considering that recent observational evidence suggests that
the fraction of galaxies with strong \Lya\, in emission 
(\Lya\, equivalent width EW(\Lya)${>} 20$\AA)
is ${>}50$\% at redshift ${\sim}6$ \citep[e.g.][]{stark2011,curtis-lake2012},
the impact of the equivalent width of the \Lya\ line 
on the broad-band colours of LBGs is studied 
prior to defining an alternative colour-selection for $z{\sim}5.7$ LBGs.

\subsection{Selection of \protect\zlbg\ LBGs}\label{s:sel-lbg}

This section presents the strategy adopted to select 
LBGs at \zlbg\ based on broad-band colours. 
Firstly, we present the template spectra used 
to explore the colours of the targeted galaxy 
population using spectrophotometry.
Secondly, we discuss each of the explored parameters and the 
results from the spectrophotometry of the templates. 
Thirdly, the resulting selection criteria is presented in 
section \ref{s:selcriteria}. 
Finally, the simulated observations 
used to define the colour selection criteria 
are described in Appendix \ref{app:simulationobs}
and the contamination factors are 
discussed in Appendix \ref{app:contamination}.

\begin{figure*}
\includegraphics[width=160mm]{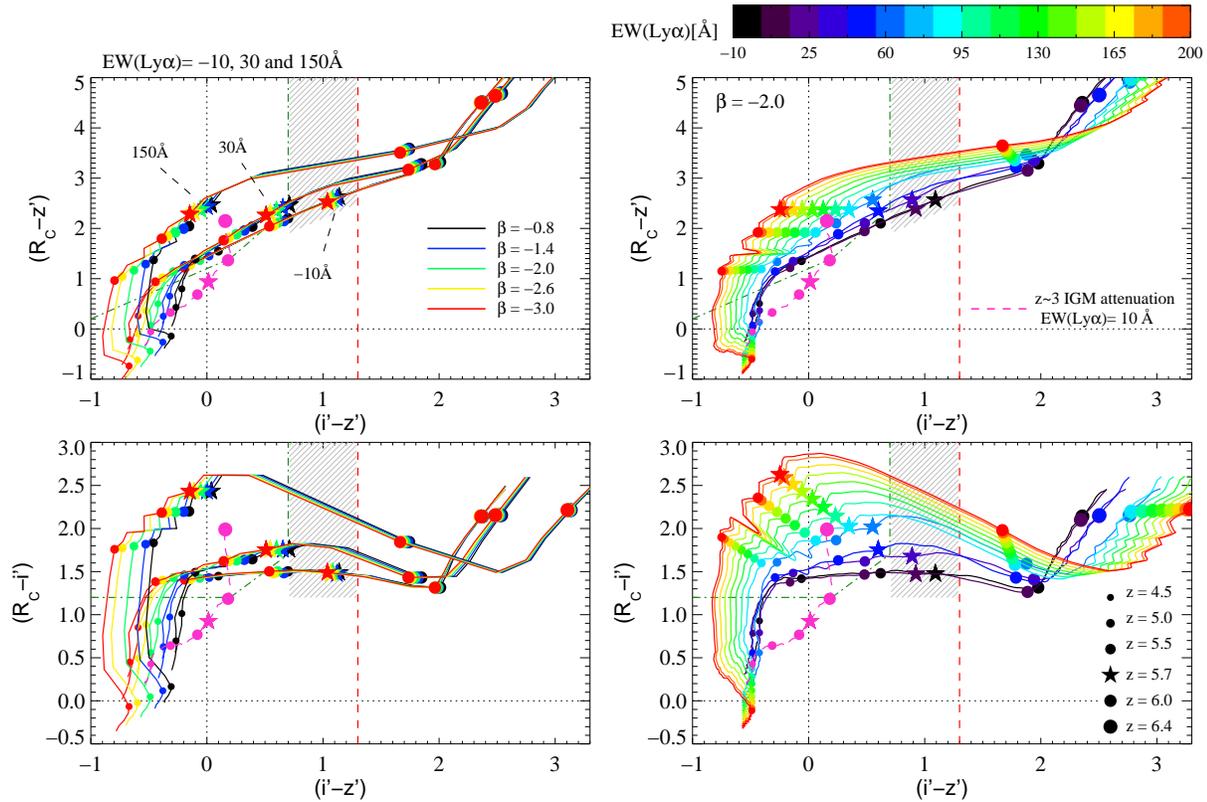}
\caption{Colour-colour diagrams showing the
evolution of LBGs from redshift z=4.1 to z=6.5.
The vertical red dashed line indicates the boundary for 
i'-dropouts selection (i'-z')=1.3.
The green dot-dashed line indicates the criteria
for $z \sim$ 5 objects: (R$_c$-i')${>} 1.2$, (i'-z')${<}0.7$
and (R$_c$-i')${>}$(i'-z')+1.0  \citep{ouchi2004a}.
The hashed region shows the window of interest 
where \zlbg\ LBGs with low EW(\Lya) are found.
{\it Left:}
Tracks are colour-coded according to the slope of the UV continuum $\beta$,
for three templates with EW(\Lya): -10, 30, 150\AA , respectively.
Templates with the same EW and different $\beta$ are almost 
indistinguishable from each other, particularly above $z{\sim}5$.
{\it Right:}
 Tracks are colour coded according to the equivalent width
of the \Lya\, line. The slope $\beta$ was fixed to -2.0.
The size of the solid circles indicate the redshift of the template.
All models meet the criterion (i'-z')${>}$1.3 for $z{\simgt}$6.
The star symbol indicates $z{=}5.7$ 
and shows that galaxies at $z{\simlt} 5.7$ do not meet 
the i'-dropout standard criteria.
The \Lya\, emission line acts to make galaxies bluer and
affects objects at $z {=} 5.7$  by moving them away from the
hashed region, which, therefore, contains only LBGs with EW(\Lya)${\simlt} 30$\AA .
The dashed magenta line is the colour-colour
evolution of a template with EW(\Lya)${=}10$\AA\ 
with the attenuation due to the \Lya\ forest for objects at $z{\sim}3$.}
\label{f:tracks}
\end{figure*}

The selection criteria were tested by adding 
artificial high-redshift LBGs to our images and 
recovering them, as described in Appendix \ref{app:simulationobs}.
The templates are based on
the composite spectrum of LBGs at redshift $z{\sim}3$
from \citet{shapley2003}. 
Firstly, they were modified to simulate UV spectral slopes 
($f_\nu \propto \lambda^{\beta}$)
covering the range of values
found in galaxies at $5{<}z{<}7$, i.e. $\beta{=}{-}0.8$ to ${-}3.0$ 
\citep{bouwens2012, finkelstein2012a},
using a step $\Delta\beta{=}0.2$.
Secondly, in order to sample the range of EW(\Lya) 
typically found at high-redshift 
\citep[e.g.][]{shapley2003,ouchi2008, stark2010,  kashikawa2011, cassata2011},
\Lya\ equivalent widths from -10\AA\ to 200\AA\ 
were simulated with a step of $\Delta$EW(\Lya)${=}10$\AA.
Thirdly, the effect of the \Lya\ forest absorption
in the wavelength region $\lambda_{rest}{<}$ 1216\AA\
was simulated applying a continuum attenuation
$D_A$ defined as
\begin{equation}
D_{\rm A}{=}1- \frac{FLUX_{\rm obs}}{FLUX_{\rm int}},
\label{eq:DA}
\end{equation}
where $FLUX_{\rm obs}$ is the observed flux level and
$FLUX_{\rm int}$ is the intrinsic flux level expected
without the absorption by the intervening \Lya\ forest.
The corresponding $D_A(z)$ was obtained from 
a fit to the values in the literature \citep{ giallongo1990, lu1994, reichart2001}.
Finally, each spectrum was redshifted from $z {=} 4.1$ to 6.5
using a step $\Delta z{=}0.1$, and magnitudes in the three 
filters, R$_c$, i', and z', were computed at each step.
Figure \ref{f:tracks} presents the redshift evolution 
of the colours of LBGs 
showing the effect of different EW(\Lya) and $\beta$ slopes. 
We discuss our findings in the following subsections.

\subsubsection{Template spectra for target galaxies} \label{s:templates}

This section presents and discusses the template
spectra used to 
compute the spectrophotometry of target galaxies.

Using stellar population synthesis models to simulate 
the spectrum of a galaxy implies the assumption of intrinsic 
properties like age, star formation rate (SFR), metallicity, etc.
In addition, they typically assume a single stellar population.
In this work we adopt a different approach by using
the four composite spectra of LBGs at redshift ${\sim}3$ 
representative of the quartiles of the EW(\Lya) distribution
from \citet{shapley2003} as initial templates for the analysis of 
the colour-colour diagram of high redshift galaxies.
Referring to the $z{\sim}4$ composite spectrum
of \citet{jones2012}, the changes in LBG spectra 
have a negligible effect on the broad-band filter colours.
Therefore, our results do not depend on any assumption
on the intrinsic properties of the galaxies.
In other words, the selection criteria introduced in Section \ref{s:selcriteria}
is based on templates that represent the average properties
of a well studied sample of observed galaxies.

The two spectroscopic features that could influence 
the UV colours the most are the \Lya\ line and the UV spectral slope.
The reported values in \citet{shapley2003}
are $\beta{=}{-}0.73$, ${-}0.88$, ${-}0.98$ and ${-}1.09$, and
EW(\Lya)${=}-14.9$\AA, ${-}1.1$\AA, 11.0\AA\ and 52.6\AA\ 
for the four templates denoted here as T1, T2, T3 and T4, respectively.
Using the templates ``as they are" and only accounting 
for IGM attenuation we find no significant difference 
in the redshift evolution of templates T1, T2 and T3 
in the colour-colour diagrams 
(i.e. differences in colour-colour tracks among the templates 
are smaller than photometric errors).
This is not surprising owing to the narrow range of $\beta$ slopes (-0.73 to -1.0)
and \Lya\ equivalent width (-15 to 11.0\AA) that they cover.
The only exception is T4 which departs from the general trend
in colour space for redshifts $5.5{<}z{<}5.9$
but this effect is driven by the \Lya\ emission and is
discussed in Section \ref{s:lyaEW}. 
Thus, it is safe to conclude that no dependence is found 
with the choice of initial LBG templates.
Because the templates represent the four quartiles
of the EW(\Lya) distribution of LBGs at $z{\sim}3$ , we use T1 and T2
as templates for LBGs with \Lya\ in absorption (net EW(\Lya) $\leq$ 0\AA),
T3 as template for LBGs with low \Lya\ emission (0\AA ${<}$net EW(\Lya)$\leq$30\AA)
and T4 as template for LBGs with strong \Lya\ emission (EW(\Lya)${>} 30$\AA).

\subsubsection{UV spectral slopes}

Figure \ref{f:tracks} shows the effect of the \Lya\ emission 
and the UV spectral slope $\beta$ in the evolution with redshift 
of the observed colours of template spectra.
The left side panels present templates with 
EW(\Lya)$ {=}{-}10$\AA, 30\AA\ and 150\AA,
colour coded according to $\beta$.
At $z {\geq} 5.5$, $\beta$ is the slope of the continuum 
that is covered by the z' band. 
Thus, it has practically no impact on the 
colour of the templates and the tracks are almost on top of each other.
Moreover, even at $z{=}5$ 
the smearing effect from different UV spectral slopes 
is comparable to the photometric errors.
Therefore, a single value $\beta{=}{-}2.0$ is adopted for subsequent analyses
because it is the typical value for M$_{\rm UV}{\leq}-19.0$ galaxies
at redshift $z{\sim}6$ \citep[e.g.][]{bouwens2012}. 

\subsubsection{\Lya\ equivalent width}\label{s:lyaEW}

It was noticed in previous works that the colours of an LBG
could be significantly affected by the \Lya\ 
emission line \citep[e.g.][]{stanway2008}.
The total effect depends on the set of filters,
redshift of the source and the equivalent width 
of the \Lya\ line. As shown in Figure \ref{f:tracks},
using SuprimeCam filters  R$_c$, i', and z'
we find that for $z{\sim}6$ (second to largest circles) 
all the explored EW(\Lya) values result in (i'-z') colours 
that agree with the standard i'{-}dropout criteria (i'-z')${>}1.3$ 
(vertical red dashed line). 
However, at $z{\simlt} 5.8$ galaxies no longer meet the 
criteria. Furthermore, \Lya\ emission results in an additional
effect on the (i'-z') colour.
To illustrate the significance of this effect, Figure \ref{f:tracks}
shows the position of the templates at $z{=}5.7$ with star symbols. 
The difference in (i'-z') between an LBG
with \Lya\ in absorption (EW(\Lya)${<}0$\AA)
and \Lya\ in emission (EW(\Lya)${>}0$\AA)
grows with EW(\Lya), reaching $\Delta \text{(i'-z')}{\sim}1$ mag between  
EW(\Lya)${=}{-}10$\AA\ and EW(\Lya)${=}150$\AA. 
The result is a region on the colour-colour diagram between
the $z{\sim}5$ and $z{\sim}6$ LBG selection criteria
which is devoid of galaxies with strong \Lya\ emission 
(shaded region in Figure \ref{f:tracks}).
We exploit this effect and develop criteria to select
\zlbg\ LBGs with little or no \Lya\ emission in this region.

In summary, the set of filters used by this work can produce
a segregation of LBGs with \Lya\ in emission and absorption
at the redshift of interest $ z {\sim}5.7$.
This effect is further explored in Appendix \ref{app:simulationobs}
and the selection criteria to generate a sample 
of bright LBGs with \Lya\ mainly in absorption
is presented in Section \ref{s:selcriteria}.

\subsubsection{\Lya\ forest attenuation} \label{s:da}

Neutral hydrogen absorbs radiation at wavelengths 
shorter than \Lya. Discrete absorption systems in the 
line-of-sight towards a background source produce 
a forest of narrow absorption lines that is called the \Lya\ forest.
The superposition of such absorption systems
can significantly reduce the flux observed at $\lambda_{rest}{<}1216$\AA.
Moreover, this attenuation will grow with redshift
as a result of the increasing fraction of neutral hydrogen 
in the IGM and the increasing density of the Universe.
This key feature is taken advantage of in the detection 
of high redshift galaxies.
For example, the magenta dashed line in Figure \ref{f:tracks}
corresponds to an LBG with EW(\Lya) $=10$\AA\ 
in which the average flux decrement ($D_{\rm A}$) was fixed at the 
average value at $z{\sim}3$.
When the expected $D_{\rm A}$ is considered 
(i.e. solid colour coded tracks) the points at  $z{=}6$ 
and $z{=}5$ are found in the regions corresponding to the selection 
for $z{\sim}6$ LBGs (vertical red dashed line) and for $z{\sim}5$ LBGs
(green dot-dashed lines).
Without accounting for the correct $D_{\rm A}$
(i.e. dashed magenta tracks)
the points do not meet their respective 
$z{\sim}6$ or $z{\sim}5$ colour
selection criteria.

\subsection{The selection criteria for $z{\sim}5.7$ LBGs}  \label{s:selcriteria}

Narrow-band imaging was designed to detect flux excess
from an emission line. As a result, star forming galaxies 
without dominant \Lya\ emission or with \Lya\ in absorption
will be missed by this technique.
Therefore, a reliable selection of LBGs from 
broad-band photometry that includes only objects 
with low \Lya\ emission plus objects with \Lya\ in 
absorption (or non-LAEs) will produce a valuable sample 
to compare with narrow-band selected LAEs 
(see discussion in Section \ref{s:disc-sel}).

Defining the optimal colour criteria based not only on
theoretical colour-colour tracks but also on 
simulated images of objects allows us to account for 
many observational uncertainties such as 
aperture corrections, sky background noise and
the choice of parameters for source extraction.
Moreover, to quantify the efficiency of a colour selection
in a particular set of images, it is necessary to
simulate the observation of the targeted objects, 
in this case high redshift LBGs. 
We used our analysis in the previous section
to compute magnitudes in the filters R$_c$, i', and z'
for templates covering a wide range in EW(\Lya) 
and a resolution in redshift of $\Delta z{=} 0.01$.
The magnitudes are used to generate
artificial objects in the science images
which are later extracted and reduced as 
a real source. The results of this exercise 
provides the basis to define 
colour selection criteria
for $z{\sim}5.7$ LBGs.
The details of our simulated observations 
and their results 
are presented in Appendix \ref{app:simulationobs}
and the possible sources of contamination are
reviewed in Appendix \ref{app:contamination}.

\citet{cooke2009} proposes a method based
on broad-band colours of galaxies at $z{\sim}3$
to select a sample of LBGs with \Lya\ predominantly 
in emission and a sample of LBGs with \Lya\ mainly
in absorption and find different environments
for each population \citep{cooke2013}.
In essence, we are applying
a similar procedure to 
higher redshift galaxies
and 
we aim to select LBGs that
occupy a narrow redshift range
around $z{\sim}5.7$. 
In particular, the $z{\sim}5.7$  LBG selection criteria 
for this work can be described as a
magnitude-colour-colour selection plus a size restriction.
The sample meets the following conditions:
\begin{enumerate}\itemsep2pt 
\item z' $ \geq 24$,   
\item S/N$_{\rm z\textrm'} \geq 5$,
\item 0.7 $\leq$ (i'-z') $\leq$ 1.3,
\item (R$_c$-z') $\geq$ (i'-z') + 1.2,
\item r$_{hl} \leq$ 0.45" (or 2.23 pixels), 
\item ISO\_AREA\_R$_c < 22$ pixel$^2$, and
\item $0.01 \leq$ S/G$_{\rm z\textrm'}< 0.9$,
\end{enumerate}
where r$_{hl}$, ISO\_AREA\_R$_c$, and S/G$_{\rm z\textrm'}$
are output values from \textsc{sextractor} for the 
half-light radius, the isophotal area in the R$_c$ band and
the stellar-to-galaxy coefficient, respectively.
The LBG candidates and their photometry are presented
in Tables \ref{t:J1030_LBGs} and \ref{t:J1137_LBGs} (Appendix \ref{app:tables}).

Several sources are affected by common undesired
artefacts of observations. 
For example, saturated columns and empty columns in the image
caused by bright stars can simulate 
a non-detection in a particular band.
These types of effects in R$_c$ or i' 
produce incorrectly selected objects.
Therefore, after applying the selection criteria, 
these and other types of ``bad column" objects
are removed from the sample by visual inspection.
Next, we present the selection of \zid\ galaxy 
candidates, also called i'-dropouts.

\subsection{Selection of \MakeLowercase{i'}-dropouts}\label{s:sel-idrop}

This section describes the criteria adopted
to select LBGs at $z{\sim}6$. 
The QSO in the field J1137+3549 is at $z_{em}{=}6.01$
and the QSO in J1030+0524 is at $z_{em}{=}6.309$.
These redshift values are within the expectations for a
sample of i'-dropouts from 
the spectrophotometry of the LBG 
templates and the sensitivity of the data.
Therefore, a sample of i'-dropout selected LBGs would be 
more likely associated with the environment of the background 
QSOs than the \civ\ systems.

Moreover, Figure \ref{f:tracks} shows that all LBG templates 
and possible combinations of $\beta$ and EW(\Lya) reach 
(i'-z') colours redder than 1.3 by \zid . 
Thus, we adopted the standard i'-dropout
criteria (i'-z')${>}1.3$, plus the additional restrictions 
described in the previous section, to sample \zid\ LBGs. 
In particular, i'-dropouts are selected using the following
conditions:

\begin{enumerate}\itemsep2pt 
\item z' $ \geq 24$,
\item S/N$_{\rm z\textrm'} \geq 5$,
\item (i'-z') $>$ 1.3,
\item S/N$_{\rm R_{c}}<2$,
\item r$_{hl} \leq$ 0.45" (or 2.23 pixels),
\item ISO\_AREA\_R$_c<22$ pixel$^2$, and
\item 0.01 $\leq$ S/G$_{\rm z\textrm'}{<}$ 0.9.
\end{enumerate}

As with the \zlbg\ LBGs sample,
individual i'-dropouts are inspected by eye to
remove objects that lie in bad columns of the image
and are wrongly measured as no-detections 
in the R$_c$ and i' bands.
The final candidates are presented
in Tables \ref{t:J1030_idrops} and \ref{t:J1137_idrops} (Appendix \ref{app:tables}).
Next, we present the strategy adopted to identify 
LAEs at redshift $z{\sim}5.7$.

\subsection{Selection of \protect\zlae\ LAE\MakeLowercase{s}}\label{s:sel-LAE}

This section presents the selection criteria 
that defines the LAE sample based on narrow-band 
photometry using the NB\civ\ filter. 
The filter detects the \Lya\ emission of
galaxies at redshift $z{\sim}5.71\pm0.04$,
which traces the more recent 
star formation episodes in the environment 
of the \civ\ absorption systems.
The LAEs selection is based on the detection 
of flux excess in the NB\civ\ band with respect to the i' band 
due to \Lya\ emission.
This excess is measured in the colour (i'-NB\civ).
Following the criteria defined by \citet{ouchi2008}, and 
accounting for the small filter difference, 
the strongest condition of the selection criteria for LAEs is
(i'-NB\civ)${>}1.335$. 

Considering the redshift of the target galaxies,
the neutral hydrogen in the line-of-sight
produces the same flux decrements at $\lambda{\simlt}912$ 
and $\lambda {\simlt} 1216$ that characterise \zlbg\ LBGs  
(Figure \ref{fig:spec}).
This feature can be used in the selection 
by including a condition in the broad-band colours.
Although many LAEs drop out the R$_c$ band, 
the (R$_c$-i') colour is commonly used, for example
(R$_c$-i')$>1.0$. 
However, we have found that some objects with significant 
flux excess in the NB\civ\ are not detected in i'.
In some of these cases, the alternative colour (R$_c$-z')
seems to be complementary. 
Therefore, we defined an alternative
condition $\text{(R}_c\text{-z')}{>} 1.3$. 
Finally, some objects have significant NB\civ\ flux (NB\civ${>}5\sigma$)
but are not detected in any of the broad-bands 
(i.e.: R$_c{<}1\sigma$, i'${<}1\sigma$ and z'${<}1\sigma$).
We also included these objects as part of our LAEs sample.

Finally, we note that only colour criteria are included
in the LAE selection as no additional conditions are applied.
The selection criteria are the following:

\begin{enumerate}\itemsep2pt 
\item S/N$_{\hbox{NBC\,{\sc iv}}} \geq 5$,
\item (i'-NB\civ) $>1.335$, and
\item $[$(R$_c$-i')$>1.0]$ $\cup$  $[$(R$_c$-z')${>}1.3]$ $\cup$  $[$(R$_c<1\sigma $) $\cap$ (i' $<1\sigma $) $\cap$ (z' $<1\sigma$)$]$.
\end{enumerate}

The LAEs sample was visually inspected
to remove sources in bad columns as for the other
two samples of photometric galaxy candidates.
Tables \ref{t:J1030_LAEs} and \ref{t:J1137_LAEs} (Appendix \ref{app:tables})
present the photometry of the LAE candidates in each field.

\section{Colours and magnitudes of \protect\zlbg\ galaxies} \label{s:results-colours}

This section presents colour-colour 
and colour-magnitude diagrams 
of the three photometric samples.
We find agreement with our expectations
from spectrophotometry described in Section \ref{s:templates}
and Appendix  \ref{s:recoveredobs}.
We report that LAEs are fainter and 
have bluer broad-band colours than LBGs selected 
at similar redshift ($z{\sim}5.7$). 
In addition, $z{\sim}5.7$ LBGs have low NB\civ\ brightness
and only one of them shows evidence of excess in the NB\civ .

It is important to keep in mind that
LBGs and LAEs are selected independently.
In the first case, photometric catalogues
were obtained for z' band detected sources and 
the selection is independent of the NB\civ\ magnitude.
In the second case, the catalogues were constructed
for sources detected in the NB\civ\ band and
the selection is defined by the excess in this band.

\begin{figure}
\includegraphics[width=84mm]{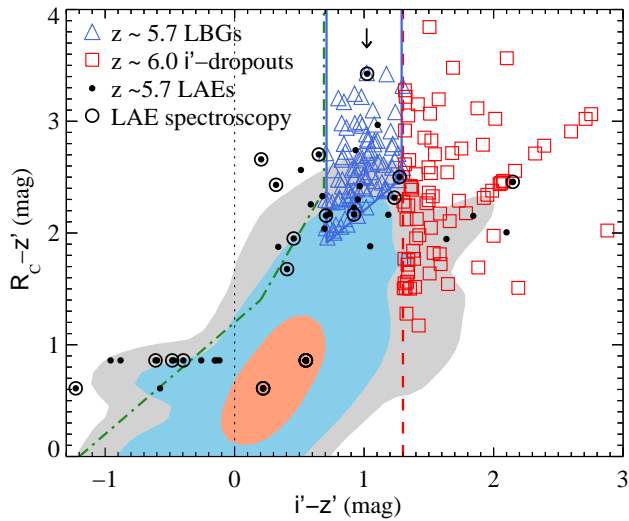}
\caption{Broad-band (R$_c$-z') vs. (i'-z') colours of 
the three samples from the two fields of this study: 
\zlbg\ LBGs (blue triangle), $z{\sim}6$ i'-dropouts (red squares) 
and \zlbg\ LAEs (black dots).
Open circles indicate LAEs in the spectroscopic sample,
which is presented in a forthcoming paper.
They all show a single emission line. 
The arrow indicates the object that is selected by
both the \zlbg\ LBGs and the LAEs criteria.
The red dashed vertical line is the (i'-z')${=} 1.3$ boundary
for i'-dropouts, the blue solid lines
indicate the colour criteria for \zlbg\ LBGs and the green
dot-dashed line shows a typical boundary adopted for
$z{\sim}5$ LBGs.
The contours correspond to the full catalogue of detections in the z' band
and contain 50\%, 90\% and 97\% of the sources in the catalogue.
}
\label{f:BBcol} 
\end{figure}

\subsection{Broad-band colours}

Figure \ref{f:BBcol} shows the (i'-z') vs. (R$_c$-z') colour diagram
of all sources and highlights the three different samples.
The contours contain 50\%, 90\% and 97\% of the total number of objects
including both fields.
The $1\sigma$ magnitude limit is assigned to an object 
when not detected in a particular band (i.e. S/N ${<}1$).
For example, many \zid\ i'-dropouts have S/N$_{\rm R_{c}}{<}1$ and
S/N$_{\rm i\textrm'}{<}1$, hence they form a diagonal line of red squares 
in Figure \ref{f:BBcol}.
Objects not detected in two bands, will have a constant colour
corresponding to the difference in the $1\sigma$ magnitude limit 
of the two bands involved. 
Due to the limitation of our broadband photometry, many LAEs  
are not detected in several broad-bands. For example, LAEs
with no detection in R$_c$ and z' form a horizontal line 
at (R$_c$-z')${=}0.86$ if the field is J1030+0524
and (R$_c$-z')${=}0.61$ if the field is J1137+3549. 
The (i'-z') colours of these objects are upper limits
only determined by their i' magnitude 
since the z' magnitude is set to the $1\sigma$ limit. 
LAEs with no detection in all three broad-bands
are overlapping each other in points $(0.55,0.86)$ 
if the field is J1030+0524
and $(0.22,0.61)$ if is J1137+3549.

The most important feature to notice from Figure \ref{f:BBcol} is that
LAEs have broad-band colours in agreement with our analysis
in Section \ref{s:sel-lbg}. 
First, LAEs detected in both i' and z' band
have colours bluer than the i'-dropout boundary $\text{(i'-z')}{=}1.3$, 
as expected due to the presence of the \Lya\ emission.
Second, the (i'-z') colours of LAEs with S/N$_{\rm z\textrm'}{<}1$ 
are upper limits and are 
consistent with our expectations, i.e. deeper
z' band photometry will result in lower values (bluer colours).
Third, although some LAEs 
occupy the colour space of \zid\ i'-dropouts, 
they are not selected as i'-dropouts because
their z' magnitudes have S/N$_{\rm z\textrm'} {\simlt} 2$. 
This means that their (i'-z') colours
have large uncertainties, and it is still possible that 
their true colour is bluer than our current estimate.
Similarly, regarding the LAEs situated 
in the colour region of \zlbg\ LBGs, all but one 
has S/N$_{\rm z\textrm'}{<}5$ and are not 
included in the LBG sample for this reason.
The one exception is the LAE with the highest 
(R$_c$-z') indicated with an arrow in Figure \ref{f:BBcol},
for which we have spectroscopic confirmation. 
This object is also selected as a \zlbg\ LBG
without using information from the narrow-band
and confirms that the selection criteria in Section \ref{s:selcriteria}
does target bright galaxies in the redshift range of interest.
The fact that only one \zlbg\ LBG shows NB excess
is evidence that the selection criteria introduced 
in Section \ref{s:selcriteria} 
certainly avoids most strong emitters.

The LBG samples (\zlbg\ and \zid ) are significantly 
more sensitive to photometric errors than the LAE sample.
The colour criteria for the two samples of LBGs
include part of the grey and the blue contours 
shown in Figure \ref{f:BBcol}. These contours 
contain 90\% and 97\% of the total number of 
detections in both fields.
As a result, the boundaries of the colour criteria
are defined in well populated regions of 
the colour diagram. This has implications on the 
effect that photometric errors have in the sample.
The reason is that many objects outside the boundaries
have colours that, within $1\sigma$ error, are consistent 
with the colour criteria. Similarly, several 
objects within the colour criteria have photometric errors 
that may place them outside the colour boundaries. 
In other words, the level of contamination expected
in a sample would increase with the density of objects
around the boundaries of the colour selection window.
Therefore, the LAEs sample is more robust to photometric errors
because, as it is shown below, the LAEs colour criteria
is in a less populated region of the NB colour-colour diagram.

In summary, the few LAEs in the i'-dropout colour region are significantly
fainter than the LBG magnitude limit (S/N$_{\rm z\textrm'}{<}5$) and have 
large uncertainties in their broad-band colours. 
LAEs in the colour window of \zlbg\ LBGs
are also too faint to be selected from the z' band as LBGs.
The only exception has S/N$_{\rm z\textrm'}{>}5$ and it confirms the redshift 
window aimed at by the \zlbg\ LBG selection criteria introduced in Section \ref{s:selcriteria}.
In general, the distribution of LAEs in the broad-band colour-colour diagram
is in agreement with expectations from spectrophotometry of typical LBGs:
that strong \Lya\ emission at \zlbg\ can move the object 
away from the i'-dropout selection window.
Photometric uncertainty affects the number of candidates
in the broad-band selected samples because the colour
boundaries are densely populated and photometric errors
can modify the position of an object in the colour-colour diagram.
We include this source of uncertainty in all our results
using Monte-Carlo simulations of its effect, 
as explained in Section \ref{s:results-num}.

\begin{figure*}
\begin{minipage}{150mm}
\centering 
\mbox{
\includegraphics[width=160mm]{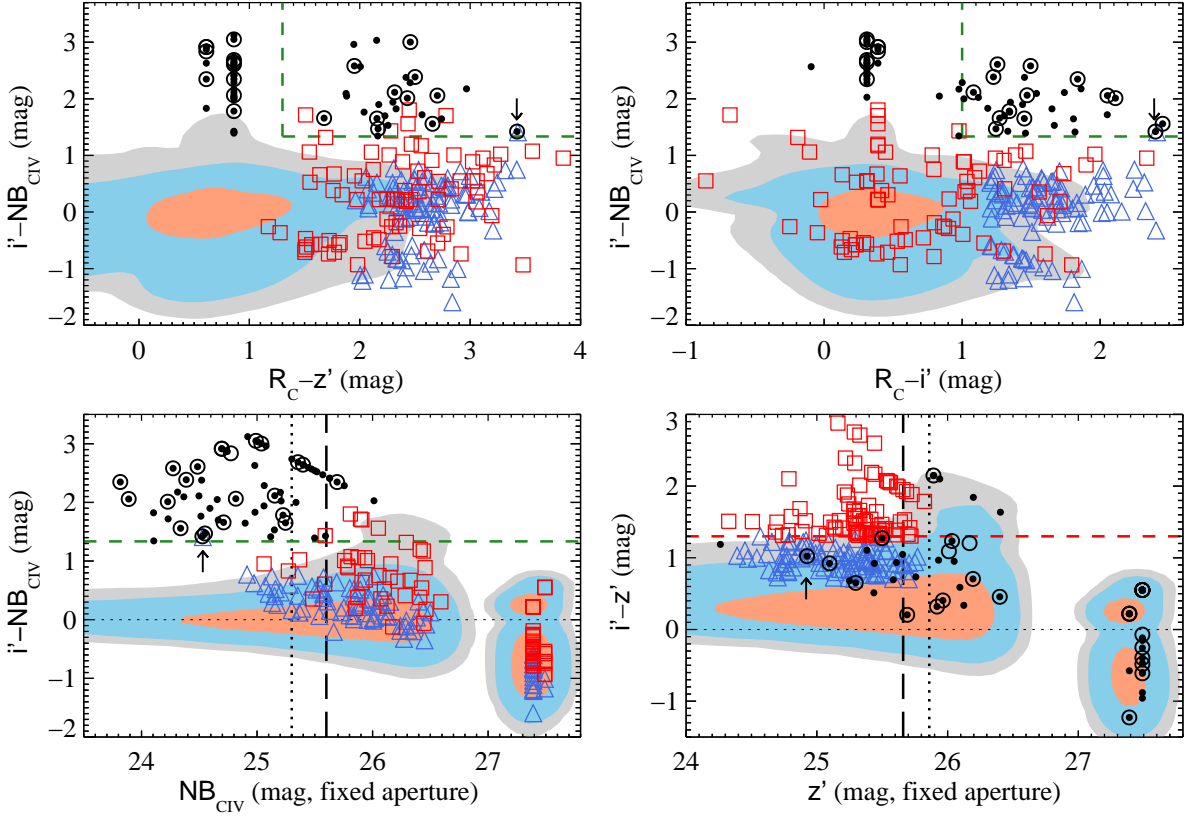}
}
\caption{ {\it Top:} Broad-band colours and the excess in the NB\civ :
(i'-NB\civ) vs. (R$_c$-z') ({\it left}) and (i'-NB\civ) vs. (R$_c$-i') ({\it right}).
Colours, symbols and contours are the same as Figure \ref{f:BBcol}.
The green dashed lines indicate the boundaries of the LAE colour selection criteria.
{\it Bottom left:} (i'-NB\civ) vs. NB\civ\ colour-magnitude diagram.
Vertical dashed and dotted lines indicate the NB\civ\ $5\sigma$
limiting magnitudes of field J1030+0524 and field J1137+3549, respectively.
Many broad-band selected sources are not detected in
the NB\civ\ band and are assigned the corresponding $1\sigma$ magnitude limit 
(NB\civ$_{1\sigma}\sim27.4$--27.5 mag).
{\it Bottom right:} 
(i'-z') vs. z' colour-magnitude diagram.
Vertical dashed and dotted lines indicate the z' $5\sigma$
limiting magnitudes of field J1030+0524 and field J1137+3549, respectively. 
The horizontal dashed line is the i'-dropout boundary (i'-z')${=}1.3$.
LAEs have fainter broad-band UV magnitude than LBGs.}
\label{f:NBcol}
\end{minipage}
\end{figure*}
\subsection{Narrow-band colours}\label{s:NBcol}

The top panels of Figure \ref{f:NBcol} present 
the colour space where LAEs are selected
as described in Section \ref{s:sel-LAE}. 
The green dashed lines show the windows
(i'-NB\civ) ${>}1.335$ \& $\text{(R}_c\text{-z')}{>}1.3$ (left panel), and 
(i'-NB\civ) ${>}1.335$ \& $\text{(R}_c\text{-i')}{>}1.0$ (right panel).
These boundaries are at ${>}0.3$ magnitudes from the 
grey contour that contains 97\% of the sources 
in both fields. As a result of a very low 
density of objects near the boundaries,
the LAEs sample is very stable to photometric uncertainty.

The condition of S/N$_{\text{NBC\,{\sc iv}}}{\geq}5$ 
implies that all selected objects have NB\civ\ errors ${\leq} 0.198$.
This means that photometric errors in the NB\civ\ 
has negligible impact in the selection of the LAEs sample.
Moreover, the condition (iii) on the broad-band colours of an 
LAE candidate (Section \ref{s:sel-LAE}) 
allows the inclusion of objects with no detection
in the three broad-bands, which is supported by
spectroscopic confirmation of some of these objects.
Therefore, contrary to the two LBG samples, photometric uncertainty
in the broad-band does not dominate the errors in the LAEs samples.
Objects not detected in at least two broad-bands 
form two columns in each panel 
aligned with (R$_c$-z')${=}0.86$ and 0.61, and
(R$_c$-i')${=}0.31$ and 0.39, respectively. 
Most of them are indeed not detected in the three broad-bands,
however, we have spectroscopic detection of an emission line
from all the LAEs selected for spectroscopic follow-up (open circles).

We note that five i'-dropouts have (i'-NB\civ) ${>}1.335$. 
However, because they are not detected in i' and have S/N$_{\text{NBC\,{\sc iv}}}{<}3$,
their (i'-NB\civ) colours have large uncertainties.
This is more obvious in the bottom left panel of Figure \ref{f:NBcol} which shows 
the $5\sigma$ detection limit in the NB\civ\ band with vertical lines,
dashed for the field J1030+0524 and dotted for the field J1137+3549.
In this figure we can see how i'-dropouts with significant NB\civ\ excess
are actually not detected in i' and their (i'-NB\civ) colours are 
dominated by their faint NB\civ\ magnitude.
Moreover, these objects belong to the field J1137+3549 
which has a brighter NB\civ\ magnitude limit than the field J1030+0524. 
Finally, the black arrow in Figure \ref{f:NBcol} shows the only
\zlbg\ LBG that has significant (i'-NB\civ) flux excess, while 
all the other \zlbg\ LBGs have low NB brightness. 

\subsection{Colour magnitude diagram} \label{s:colmag}

Several studies of LAEs at different redshift
have found that these galaxies are
typically fainter and have average
UV colours bluer than LBGs.
Our results are in agreement with these findings.
The bottom right panel of Figure \ref{f:NBcol} shows that LAEs are dominated
by fainter z' magnitude than the two LBG populations.
Although the figure also shows that LAEs populate a region
of bluer (i'-z') colours, this effect
is expected from the presence of a strong \Lya\ emission.

\section{Number density} \label{s:results-num}

\subsection{Counts of LBGs} \label{s:numberLBG}

This section presents the sample
resulting from the i'-dropout and the 
$z{\sim}5.7$ LBG selection criteria. 
We compare the number of objects 
per magnitude bin and the total number of objects with 
expectations from the luminosity function of \citet{bouwens2007}.
In general, our results agree with the expected number 
counts of LBGs at $z{\sim}5.7$ 
and i'-dropouts at $z{\sim}6.0$.

\begin{figure}
\includegraphics[width=84mm]{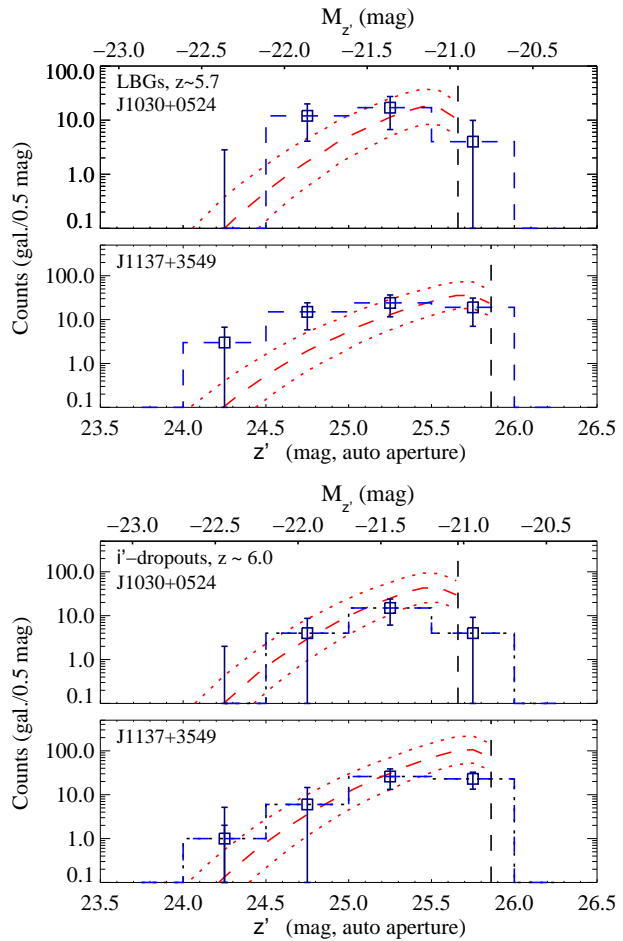}
\caption{{\it Top:} Number of LBGs per z' magnitude bin.
The blue dashed histogram corresponds to $z{\sim}5.7$ LBGs
in each field. The error bars include Poisson and photometric errors.
The red dashed line is the prediction from \citet{bouwens2007} luminosity function
and the red dotted lines include the error reported in that work.
The vertical black dashed lines indicate the $5\sigma$-limit
magnitude of each field.
{\it Bottom:} Number of i'-dropouts per z' magnitude bin.
The blue dashed histogram corresponds to i'-dropouts ($z{\sim}6$ LBGs)
in each field. 
As for the plot above, the red lines (dotted and dashed) are the expected 
number of galaxies ${\pm}1\sigma$, according to \citet{bouwens2007} luminosity function.}
\label{f:counts-mag-LBGs}
\end{figure}

Figure \ref{f:counts-mag-LBGs} shows the
number of objects detected in the z' band
that meet our selection criteria for \zlbg\ LBGs (section \ref{s:selcriteria}). 
The error bars include Poisson and photometric errors ($1\sigma$). 
To estimate the errors induced by photometric uncertainties
we run the following Monte-Carlo simulation.
For all the objects in the catalog of detections produced by \textsc{sextractor},
the magnitudes in each of the three broad-bands of each object
are simulated by randomly selecting a new value from
the interval [$m{-}\sigma_m$,$m{+}\sigma_m$],
where $\sigma_m$ is the error in the magnitude $m$ of each detection.
In other words: we assume that the real magnitude 
$m$ of each detection is contained within $m{\pm}\sigma_m$, 
with uniform probability. 
Once a new catalog of detections is created,
the selection criteria of Section \ref{s:selcriteria} is applied 
and a new sample of \zlbg\ LBGs is obtained. 
Then, the number of objects per magnitude bin is recalculated.
After $10\,000$ iterations, the standard deviation of the number 
counts in each magnitude bin is obtained as the error induced
by the uncertainty in the photometry of the sources.
Over-plotted with red dashed lines are the predicted 
number of galaxies from equation \ref{eq:counts} (Appendix \ref{s:pmz})
using \citet{bouwens2007} luminosity function of $z{\sim}6$ galaxies.
The dotted lines enclose one standard deviation in the three parameters
of the Schechter luminosity function \citep{schechter1976}
reported in \citet{bouwens2007}.
Our sample is in general agreement 
with the currently known luminosity function of high redshift LBGs. 

In the field J1030+0524 we select
33 sources as potential $z{\sim}5.7$ LBGs
and in the J1137+3549 field the number of sources is 61.
The expected total number of LBGs 
from equation \ref{eq:counts} (Appendix \ref{s:pmz}) and 
the luminosity function of high-redshift LBGs
is 18$^{+22}_{-10}$ in the J1030+0524 field and 39$^{+44}_{-21}$ in 
the J1137+3549 field. 
Although the number of LBG candidates in each field are within the errors of 
the expected number of galaxies from the $z{\sim}6$ 
luminosity function, we note that both samples are larger than 
the predicted mean number of \zlbg\ LBGs.
This is not a surprise since 
the colour selection window is heavily populated 
by Galactic cool dwarf stars (Appendix \ref{app:contamination-stars}) and, 
despite our attempt to minimise their fraction,
the contamination 
is expected to be higher than in other LBG colour-selection criteria.

The z' magnitude distribution of i'-dropouts
is shown in the bottom panel of Figure \ref{f:counts-mag-LBGs}.
In this case, 
we select 23 i'-dropouts
in the field J1030+0524, where 42$^{+53}_{-24}$
are expected, and 56 i'-dropouts in the field J1137+3549,
where the expectation is 103$^{+122}_{-55}$. 
Although the size of both samples are smaller than the 
mean predicted number of galaxies for each field,
they are within the uncertainty of the luminosity function.

There are several sources of error that can lead to 
an over-prediction in the number of high-redshift galaxies.
First, the volume kernel used in equation \ref{eq:counts} 
is over-estimated because we did not account for the area of the sky
that is covered by foreground galaxies in the field, which reduce the area 
of the sky in which high redshift galaxies can be detected.
Second, the effect is driven by the faintest magnitude bin (z'$>$25.5, bottom panel
of Figure \ref{f:counts-mag-LBGs}).
We are aware that the simulated observations (Appendix \ref{app:simulationobs})
used to estimate the selection probability function $P(m,z)$
can lead to an overestimation of the number of faint detections because
they are based on point-like sources. 
Using the same detection parameters in \textsc{sextractor}, at the faintest magnitudes, 
high redshift galaxies with relatively more extended light profiles
are more likely to be missed than galaxies with the same magnitude 
but with concentrated light profiles.
Therefore, the fainter end of real observations do sample a smaller 
volume than the one predicted by simulated observations of point-like sources.
Third, the condition (vii) of the i'-dropout criteria (Section \ref{s:sel-idrop}) that
aims to avoid stellar contamination 
will also remove small galaxies with concentrated light profiles. 
Considering that bright LBGs are typically more extended than faint LBGs,
this condition could also reduce the counts of objects in the faintest magnitude bin.
However, the removal of faint ``suspicious'' sources does not modify our results
because we aim to select the UV bright LBGs that trace the most massive haloes.

\subsection{Counts of LAEs}\label{s:numberLAE}

\begin{figure}
\includegraphics[width=84mm]{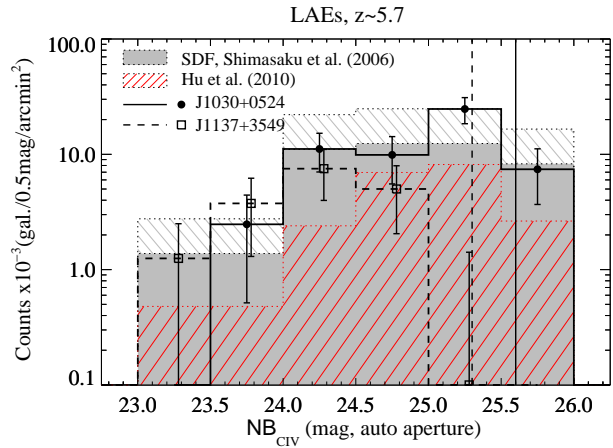}
\caption{
Surface density of LAEs per NB\civ\ magnitude.
Solid line histogram and circles correspond to
the field J1030+0524, and
dashed line histogram and open squares
correspond to the field J1137+3549 
(squares have been artificially shifted 0.03mag).
The vertical lines indicate the $5\sigma$
limiting NB\civ\ magnitude of each field.
For comparison, the grey filled histogram
corresponds to confirmed $z\sim5.7$ LAEs in SDF from a similar 
survey area (725 arcmin$^2$)
and the red line filled histogram
are confirmed $z\sim5.7$ LAEs from \citet{hu2010}
from a larger survey area (4168 arcmin$^2$).}
\label{f:counts-mag-lae}
\end{figure}

We report 45 narrow-band selected 
LAEs in the field J1030+0524 
(total area: 809 arcmin$^2$),
and 14 in the field J1137+3549 
(total area: 799 arcmin$^2$).
Although the samples have different number of objects,
there is very good agreement
in the surface density of LAEs per 0.5 NB\civ\ magnitude
detected in each field,
as shown in Figure \ref{f:counts-mag-lae}.

The two black line histograms (solid line for J1030+0524 and dashed line for J1137+3549) 
agree within the error bars, 
which include Poisson and photometric errors ($2\sigma$).
The errors are obtained with the procedure 
described in the previous section.
Nevertheless, the Monte-Carlo simulation of the photometric errors
for narrow-band selected objects 
allows for an interval [$m-2\sigma_m$,$m+2\sigma_m$],
from which magnitude values are pulled with uniform probability.
As a result, the error bars indicate two times the standard deviation
expected from photometric uncertainties.
This modification was needed because
the LAEs selection criteria is
not very sensitive to the photometric errors 
in the broad-band.\footnote{The reason is that 
many objects are not detected in at least one broad-band
and are set to the $1\sigma$ limit,
which means that their colours depend on one band only.
The Monte-Carlo simulation of the photometric errors
does not randomise the magnitude in the band where
the object is not detected. In other words, if a source
is not detected in one band, the magnitude assigned 
is the $1\sigma$ limit in every random realisation.}
Moreover, the LAEs selection requires S/N$_{\text{NBC\,{\sc iv}}}{\geq}5$
which makes the excess in the NB\civ\ to be almost
insensitive to the photometric uncertainty.
As discussed in Section \ref{s:NBcol},
the contour that contains 97\% of the sources 
in Figure \ref{f:NBcol} (grey contour)
is at ${>}0.3$ magnitudes from the colour selection
of LAEs.
Therefore, hardly any of the $5\sigma$ detections 
from the grey contour could reach the colour criteria
in our Monte-Carlo experiment,
suggesting that the criteria is very stable.

Figure \ref{f:counts-mag-lae} also shows the observed 
surface density per NB magnitude (i.e. no completeness correction is applied) 
of 34 confirmed $z{\sim}5.7$ LAEs in the Subaru Deep Field 
(SDF, total area: 725 arcmin$^2$) from \citet{shimasaku2006} (grey filled histogram).
The spectroscopic sample from which these objects were obtained
contains almost half the SDF LAE photometric sample.
Thus, we estimated the predicted surface density assuming
the same confirmation fraction for the complete SDF LAE photometric sample.
The grey line-filled region shows the possible range of LAE counts in 
each magnitude bin and suggest that our two samples are in good 
agreement within the errors with the $z{\sim}5.7$ LAEs in the SDF.

The red line-filled histogram correspond to the catalog of 88 confirmed 
$z{\sim}5.7$ LAEs from \citet[][total area: 4168 arcmin$^2$]{hu2010}.
It is slightly below the histograms corresponding to the samples of 
photometric LAE candidates in the two fields of our study.
Such effect is realistic considering that some level of contamination 
is present in our photometric samples.
Moreover, this catalog combines seven fields (with some overlap) observed 
with slightly different exposure times and seeing.
Thus, since the surface density is obtained using 
the total area of the survey, shallower fields will 
contribute to the total area but not to the fainter end of the distribution
in Figure \ref{f:counts-mag-lae}.
In addition, two of their fields are in the direction to a massive 
foreground cluster ($z{\sim}0.37$) which will affect the ``effective'' 
observable area due to a higher number of foreground galaxies,
and the brightness of the objects due to gravitational lensing.
Both of these effects depend on the position in the field-of-view,
thus the impact on the counts of LAEs is not very clear.
Finally, \citet{hu2010} use a more flexible selection condition in (i'-NB) 
but almost all the objects were targeted for spectroscopic follow-up.
They confirm that a higher (i'-NB) threshold provides a less contaminated 
sample, but the effect in the total number of confirmations is not significant.

In summary, the surface density per NB magnitude of the photometric samples 
in the fields of this study is in good agreement with samples from other studies.
Next, we present the projected surface density distribution
of the galaxies in the environment of the \civ\ systems (\zlbg\ LBGs and LAEs)
and the environment of the background QSOs (\zid\ i'-dropouts).

\section{Surface density distribution} \label{s:results-sd}

In order to characterise the environment of the \civ\ absorption systems
within the large-scale distribution of galaxies in each field, 
we start with a qualitative analysis of the morphology of the density
field traced by the three populations of galaxies.
We describe the projected distribution of sources at scales  
${\sim}80{\times}60h^{-1}$ comoving Mpc
and the environment of the \civ\ systems
at a scale of $10h^{-1}$ comoving Mpc.
Then, we quantify the surface density of galaxies
within $20h^{-1}$ comoving Mpc radius
centred in the lines-of-sight to the \civ\ systems
and we compare our results with expectations from 
a non-clustered distribution of sources (random distribution).

\subsection{$z{\sim}5.7$ LBGs}

In both fields we find that:
a) the \civ\ systems are in 
a low density region of LBGs, and 
b) the two-dimensional distribution of LBGs
shows a clumpy structure.

\begin{figure*} 
\begin{minipage}{170mm}
\centering 
\mbox{
\includegraphics[width=130mm]{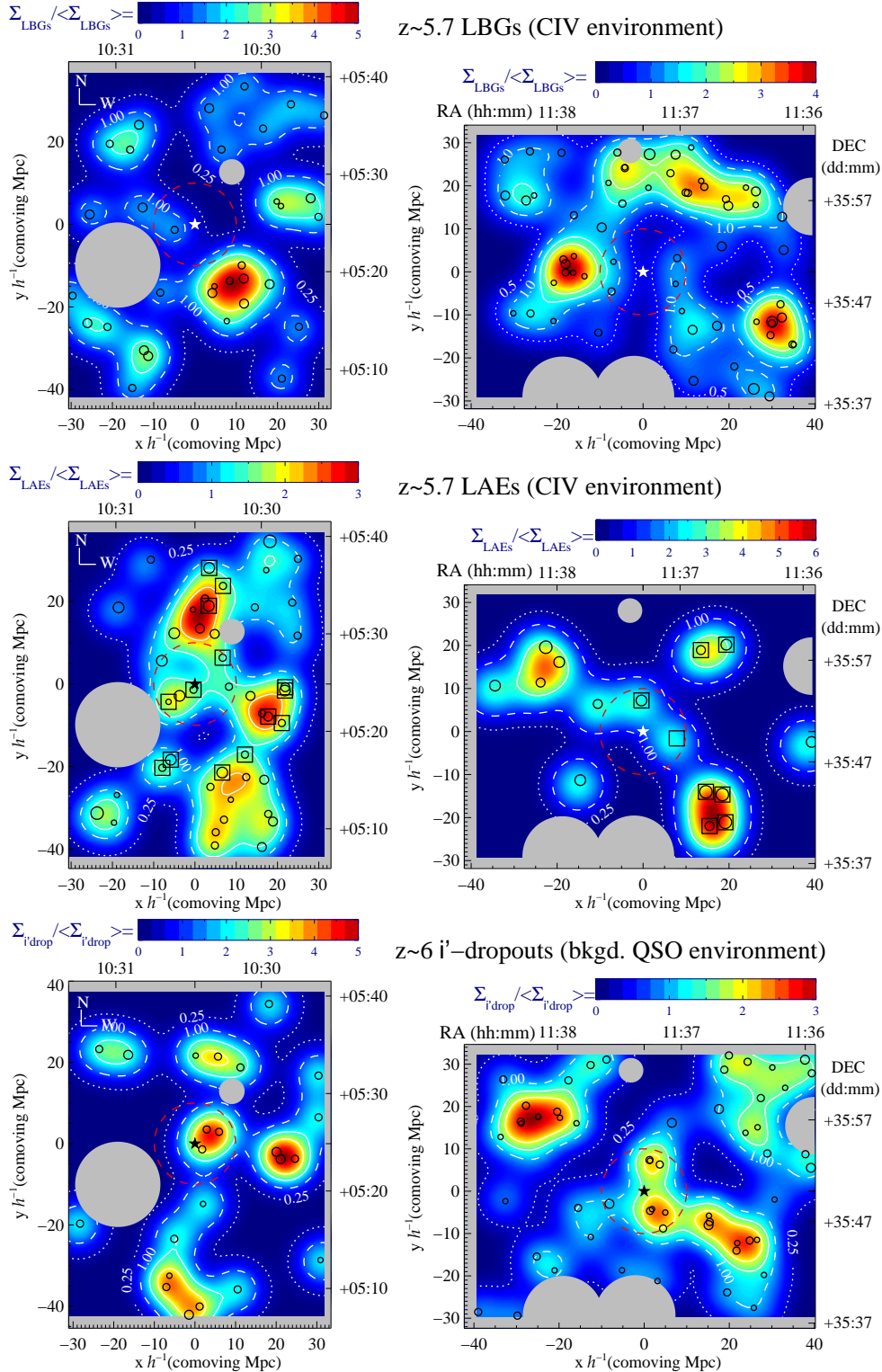} 
}
\caption{Distribution of \zlbg\ LBGs ({\it top}),
\zlbg\ LAEs ({\it middle}) and $z{\sim}6$
i'-dropouts ({\it bottom}) in the field J1030+0524 ({\it left column})
and J1137+3549 ({\it right column}). 
The size of each circle indicates the 
apparent magnitude bin 
with larger circles representing brighter magnitudes. 
The star symbol indicates the line-of-sight to the \civ\ 
systems. Masked areas of the field (bright Galactic stars and edges of 
the CCD) are shaded grey.
The area is colour-coded according to the 
surface density contrast $\Sigma/ \langle \Sigma \rangle$ 
obtained using 1${\it h}^{-1}$ comoving Mpc bin size 
and a Gaussian smoothing kernel with FWHM of 10 ${\it h}^{-1}$ 
comoving Mpc. The colour bar of each plot is on the top
of each panel. In general, blue colours represent less 
than the mean surface density of the field and red colours
correspond to the highest density of each sample.
Yellow and green are intermediate densities.
Dotted contours correspond to under-dense 
regions ($\Sigma / \langle \Sigma \rangle{<}1$), 
solid contours correspond to over-dense regions 
($\Sigma / \langle \Sigma \rangle{>}1$), 
and dashed contours correspond to mean 
density regions ($\Sigma / \langle \Sigma \rangle = 1$)
The red dashed circle centred on the star symbol 
has a radius of $10{\it h}^{-1}$ comoving Mpc.
}
\label{f:maps}
\end{minipage}
\end{figure*}

The top panels of Figure \ref{f:maps} presents the position 
of $z{\sim}5.7$ LBGs (circles) in comoving Mpc 
with respect to the \civ\ lines-of-sight (white stars). 
The size of the circle represents the
apparent magnitude according to the 
bins of the histogram in Figure \ref{f:counts-mag-LBGs}, 
with smaller circles for fainter objects.
The contours indicate constant levels of density 
contrast quantified by $\Sigma_{LBG}/\langle \Sigma \rangle_{LBG}$,
where $\langle \Sigma \rangle_{LBG}$, is the mean 
surface density of LBGs averaged over the size of
the field and $\Sigma_{LBG}$ is 
the surface density of LBGs obtained using 1${\it h}^{-1}$ 
comoving Mpc bin size and a Gaussian smoothing kernel 
with FWHM${=}$10${\it h}^{-1}$ comoving Mpc. 
Dotted contours correspond to under-dense regions, 
dashed contours correspond to mean density regions
and solid contours correspond to over-dense regions.

We report that, in both fields, 
the lines-of-sight to the \civ\ systems 
are in regions with a surface density of LBGs lower than the mean of the field,
at more than 10${\it h}^{-1}$ comoving Mpc 
from the main concentrations of bright LBGs. 
The top panels of Figure \ref{f:maps} show that LBG candidates
form ``islands'' of size ${\geq}10h^{-1}$ comoving Mpc
surrounded by large regions where the density
is below the mean of the field (blue areas).
In both sight-lines, a circle of 10${\it h}^{-1}$ comoving Mpc 
radius (red dashed circle) is easily contained
between over-densities of LBGs.
In other words: at a scale of 10${\it h}^{-1}$ comoving Mpc
radius, the environment of the \civ\ looks empty of LBGs
in comparison to the rest of the field.

The surface density within a circle of 10${\it h}^{-1}$ comoving Mpc 
radius from the \civ\ line-of-sight in the J1030+0524 field (top left panel) 
is $\Sigma_{LBG}(10){=}18\pm16{\times}10^{-3}$ gal./arcmin$^2$, 
which is ${=}0.4^{{+}0.5}_{{-}0.4}$ times the mean surface density of the field
$\langle \Sigma \rangle_{LBG}{=}41\pm5 {\times}10^{-3}$ gal./arcmin$^2$.
In other words:  $\Sigma_{LBG}(10)$ is lower than 
the mean surface density of the field by a factor of $\sim 2$. 
The errors include photometric uncertainties and
the uncertainty introduced by the masked areas.

In the J1137+3549 field, the surface density within 10${\it h}^{-1}$ 
comoving Mpc radius is $\Sigma_{LBG}(10){=}71{\pm}31{\times}10^{-3}$ gal./arcmin$^2$
and the total surface density is $\langle \Sigma \rangle_{LBG}{=}76{\pm}9 {\times}10^{-3}$ gal./arcmin$^2$, 
which results in a density contrast of ${=}0.9^{{+}0.6}_{{-}0.4}$.
Thus, $\Sigma_{LBG}(10)/\langle \Sigma \rangle_{LBG}\simlt 1$
indicates that, at scales of 10${\it h}^{-1}$ comoving Mpc, 
the surface density of LBGs in the environment of the \civ\ system is 
in agreement with the mean surface density of the field.
At 8${\it h}^{-1}$ comoving Mpc radius $\Sigma_{LBG}(8)=28^{{+}37}_{{-}28}{\times}10^{-3}$ gal./arcmin$^2$,
which results in a density contrast of ${=}0.4^{{+}0.6}_{{-}0.4}$, similar
to the field J1030+0524 at 10${\it h}^{-1}$ comoving Mpc scales.

In the field J1030+0524, the closest LBG brighter 
than z' = 25.50 magnitude is found at 5.1${\it h}^{-1}$ comoving Mpc
(760.8${\it h}^{-1}$ kpc physical, 2.16 arcmin) projected distance
from the \civ\ system line-of-sight.
In the field J1137+3549, the two closest LBGs brighter than 
z'${=}25.5$ are found at ${\sim}8.6{\it h}^{-1}$ comoving Mpc 
(${\sim}$1.28${\it h}^{-1}$ physical Mpc, ${\sim}$3.6 arcmin) projected distance 
from the line-of-sight to the \civ\ system.
There are two fainter LBGs (z'${\sim}25.6-25.7$) 
at slightly closer distances 
(7.28 and 8.08${\it h}^{-1}$ comoving Mpc),
but they also are far enough from the \civ\ to rule out
a physical origin.

In summary, the environment of the \civ\ systems 
can be described as deficient of bright LBGs
with a surface density at a scale of 
10${\it h}^{-1}$ comoving Mpc that is lower than 
the mean surface density of LBGs in the entire field of view.
Next, we describe the distribution of LAEs and 
discuss the differences in structure from the LBGs.

\begin{figure*}
\begin{minipage}{150mm}
\centering 
\mbox{
\includegraphics[width=140mm]{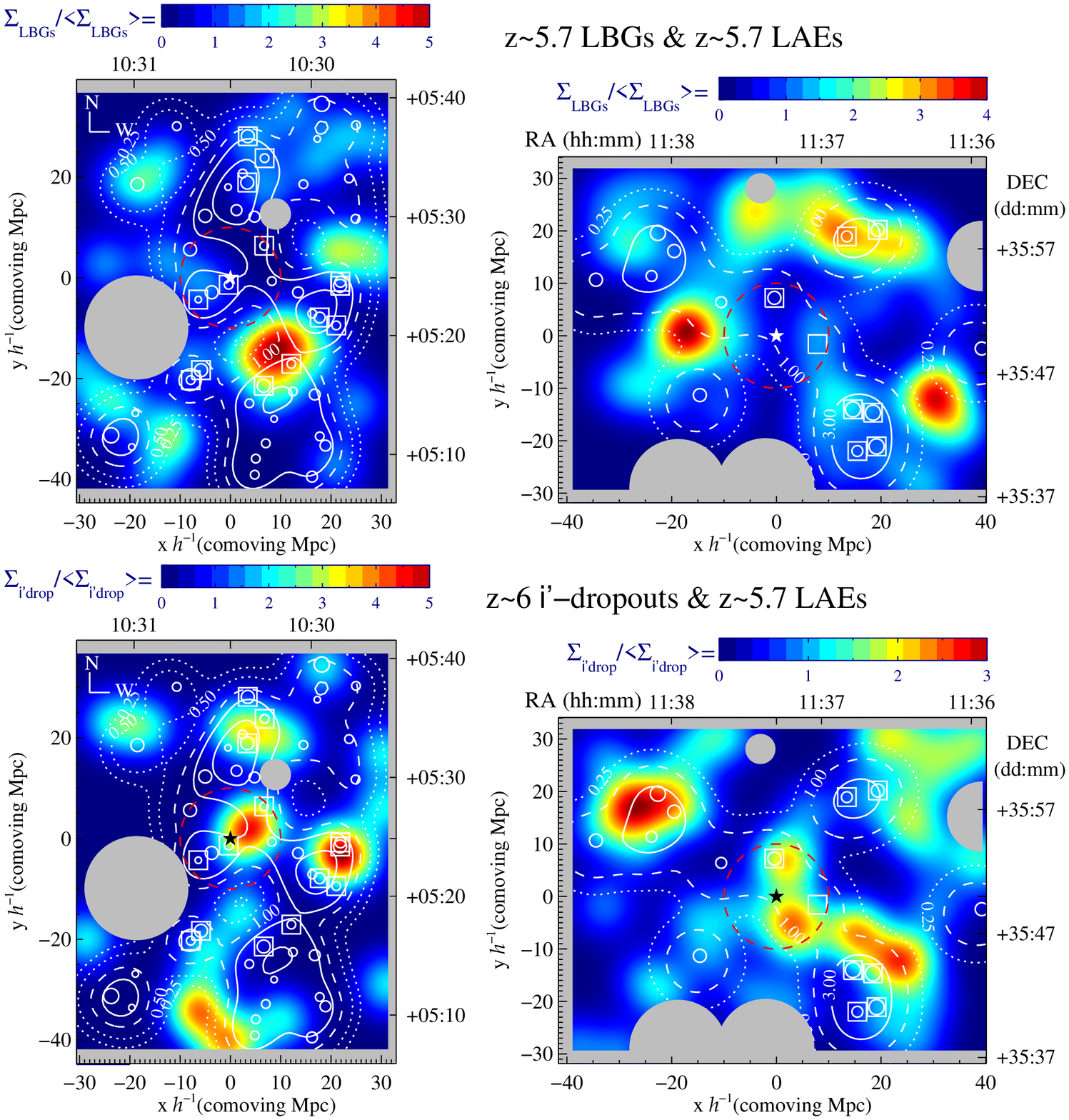} 
}
\caption{{\it Top:}
Density contrast
of \zlbg\ LBGs (colour-coded background) and \zlae\ LAEs 
(line contours) showing that over-densities of 
the two samples are not aligned:
the solid line contours of both panels are shifted from
the yellow and red areas.
In general, LAEs (open circles) are found around or between
the areas with over-densities of LBGs.
The exceptions are: one object in the north-west corner 
of the field J1030+0524 ({\it left panel})
and the the north-east corner 
of the field J1137+3549 ({\it right panel}).
{\it Bottom:} Density contrast of \zid\ i'-dropouts 
(colour-coded background) and \zlae\ LAEs (line contours).
Although the two samples are at different redshifts, 
the distributions are collapsed in projection and show some agreement
and avoidance.
The north side and the centre of the field J1030+0524 ({\it left panel})
show good agreement between the position of over-densities
of both populations. However, in the south of the field
the LAEs are avoiding the positions of the i'-dropouts.
In the field J1137+3549 ({\it right panel}), 
agreement is found in the east side and the centre.
However, in the west of the field, the over-densities
of each sample do not match.
Symbol key is as per Figure \ref{f:maps}.}
\label{f:comp}
\end{minipage}
\end{figure*}

\subsection{\zlbg\ LAEs}

In both fields we find that: a) the \civ\ systems are in regions with
more LAEs than the average of the field, and 
b) LAEs are less clustered 
than LBGs and surround the concentrations of LBGs.

The distribution of narrow-band selected 
LAEs is presented in the middle panels 
of Figure \ref{f:maps}.
The size of the circles
represents the NB\civ\ magnitude of the objects.
Spectroscopic detections of emission lines
are indicated with open squares. 
Since the spectroscopic data will be presented 
in a follow-up paper, it is beyond the scope of this work 
to report our findings from the spectroscopic campaign.
Nevertheless, one of the eight spectroscopic confirmations
in the field J1137+3549 (Figure \ref{f:maps}, middle right panel),
is not bright enough in the NB\civ\ band (S/N$_{\hbox{NBC\,{\sc iv}}}{<}5$)
to be in the photometric sample, but is close enough 
to the \civ\ line-of-sight that is considered relevant 
for the characterisation of the environment of the \civ\ system.
Therefore, this object is included in the results
presented in this section and Section \ref{s:compare-with-rnd}.

Contrary to the LBGs result, we report a high surface 
density of LAEs within 10${\it h}^{-1}$ comoving Mpc 
from the \civ\ systems.
In particular, the mean surface density of LAEs in the field J1030+0524 
is $\langle \Sigma \rangle_{LAE}{=}56 {\pm}3 {\times}10^{-3}$ 
gal./arcmin$^2$ and the surface density within 
10${\it h}^{-1}$ comoving Mpc is 
$\Sigma_{LAE}(10){=}107{\pm}4{\times}10^{-3}$ gal./arcmin$^2$, 
which corresponds to a density contrast of $\sim1.9{\pm}0.2$.
The errors are small because the uncertainty 
in the photometry has a negligible impact
in the selection of LAEs (Section \ref{s:NBcol}) and the 
very small effect of masked regions is only noticed at scales
${>}10{\it h}^{-1}$ comoving Mpc. 
The three closest LAEs are found 
at projected distances of 1.43${\it h}^{-1}$ comoving Mpc 
(212${\it h}^{-1}$ kpc physical, 0.60 arcmin), 4.70${\it h}^{-1}$ comoving 
Mpc (700${\it h}^{-1}$ kpc physical, 1.99 arcmin), and 7.72 ${\it h}^{-1}$ 
comoving Mpc (1.15${\it h}^{-1}$ Mpc physical, 3.27 arcmin).
We expand on the closest LAE at 212${\it h}^{-1}$ physical kpc 
in Section \ref{s:disc-closest}.

In the field J1137+3549, 
the density of LAEs towards the \civ\ system 
is also higher than the average of the field.
The surface density within 
10${\it h}^{-1}$ comoving Mpc radius is
$\Sigma_{LAE}(10){=}36{\times}10^{-3}$ gal./arcmin$^2$,
and compared to the mean of field 
$\langle \Sigma \rangle_{LAE}{=}19{\pm}1 {\times}10^{-3}$ gal./arcmin$^2$
represents a density contrast of $1.9{\pm}0.1$.
The two closest LAEs are at 7.24${\it h}^{-1}$ comoving Mpc 
(1.08${\it h}^{-1}$ Mpc physical, 3.06 arcmin) and 8.02${\it h}^{-1}$ comoving Mpc 
(1.19${\it h}^{-1}$ Mpc physical, 3.39 arcmin) 
from the line-of-sight to the \civ\ system. 

At scales of ${\sim}80{\times}60{\it h}^{{-}1}$ comoving Mpc,
LAEs appear to form a filamentary structure,
similar to previous findings in other fields from wide-field 
imaging at similar redshift \citep[e.g.][]{ouchi2005a}.
Furthermore, the particular distribution of LAEs occupies 
the space between the LBG ``clumps''. 
The main concentrations of each population
do not share the same position in the field of view
which is further enhanced by the fact that the two
distributions are in projection.
The projected over-densities of LBGs and LAEs
(solid line contours, Figure \ref{f:maps}) 
are in different regions of the field.
This pattern is more obvious in the top panels
of Figure \ref{f:comp} that shows the density contrast 
of both populations using colour-coded 
contours for \zlbg\ LBGs and line contours for \zlae\ LAEs,
where over-densities of one sample fill the space 
of under-densities of the other.
Interestingly, except for the north-west corner of
field J1137+3549, both fields show 
\zlae\ LAEs scattered around more compact \zlbg\ LBGs groupings.
This type of behaviour is observed at $z{\sim}3$ \citep{cooke2013}.

These two galaxy populations 
likely inhabit different large-scale environments.
According to the selection probability function 
(Appendix \ref{s:pmz}), the depth of the 
volume sampled by the LBGs ($\Delta z {\sim}0.2$) 
is ${\sim}90{\it h}^{-1}$ comoving Mpc, 
which is similar to the vertical size of the left side panels in Figure \ref{f:maps}
(${\sim}80{\it h}^{-1}$ comoving Mpc). 
In addition, the depth of the LAEs sample ($\Delta z {\sim}0.08$) 
is ${\sim}36{\it h}^{-1}$ comoving Mpc, almost half the vertical
size of the field in the left side panels.
Then, LAEs are contained in the volume sampled by the LBGs. 
As a result, the projected distribution of the over-densities
of each populations suggest that, at this redshift, LAEs and LBGs are not 
evenly distributed (or mixed) in real space.

\begin{figure*}
\begin{minipage}{150mm}
\centering 
\mbox{
\includegraphics[width=140mm]{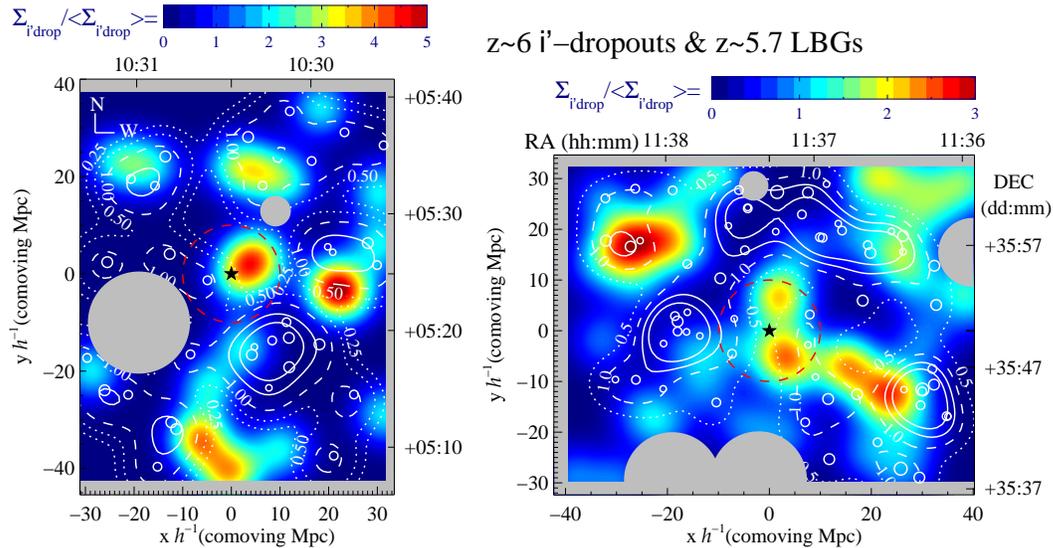} 
}
\caption{Density contrast of \zid\ i'-dropouts (colour-coded background) 
and \zlbg\ LBGs (line contours) in the field J1030+0524 ({\it left}) 
and the field J1137+3549 ({\it right}).
In general, both panels show that over-densities of the two samples are not aligned.
Symbol key is as per Figure \ref{f:maps}.}
\label{f:comp-b}
\end{minipage}
\end{figure*}

\subsection{$z{\sim}6$ i'-dropouts}

In both fields we find that: a) the line-of-sight towards the 
QSO intercepts an over-density of i'-dropouts, and b) the projected distribution
of i'-dropouts is partially aligned with the distribution of \zlae\ LAEs but
not aligned with the projected distribution of \zlbg\ LBGs. 

The distribution of i'-dropouts in the field J1030+0524 
(bottom left panel of Figure \ref{f:maps}) shows
three galaxy candidates 
at projected distances of 2.26${\it h}^{-1}$ comoving Mpc 
(323${\it h}^{-1}$ kpc physical, 0.94 arcmin),
4.54${\it h}^{-1}$ comoving Mpc (649${\it h}^{-1}$ kpc physical, 1.89 arcmin) and 
6.60${\it h}^{-1}$ comoving Mpc (942${\it h}^{-1}$ kpc physical, 2.75 arcmin) 
from the QSO's line-of-sight.
We find a surface density of i'-dropouts 
within 10${\it h}^{-1}$ comoving Mpc of 
$\Sigma_{\rm i\textrm{'}\rm{drop}}(10){=}55{\pm}13{\times}10^{-3}$ gal./arcmin$^2$, 
which is twice the mean surface density averaged over the entire field
$\langle \Sigma \rangle_{\rm i\textrm{'}\rm{drop}}{=}28{\pm}4 {\times}10^{-3}$ gal./arcmin$^2$
(density contrast ${=}2.0^{{+}0.8}_{{-}0.7}$) and represents 
an over-density in the position of the QSO.

A similar over-density of i'-dropouts is seen 
in the line-of-sight towards the QSO in the field J1137+3549.
In particular, the two closest i'-dropout to the \civ\ line-of-sight
are at projected distances of 4.62${\it h}^{-1}$ comoving Mpc 
(660${\it h}^{-1}$ kpc physical, 1.92 arcmin) and
4.84${\it h}^{-1}$ comoving Mpc (693${\it h}^{-1}$ kpc physical, 2.02 arcmin),
with z'${=}25.7$ and z'${=}25.4$, respectively.
The next four closest objects are at
${\sim}7.3{\it h}^{-1}$ comoving Mpc and have z'${\sim}25.5$.
The surface density within 
10${\it h}^{-1}$ comoving Mpc radius is
$\Sigma_{\rm i\textrm{'}\rm{drop}}(10)=146\pm35{\times}10^{-3}$ gal./arcmin$^2$,
which, in agreement with the field J1030+0524, is twice the mean surface density of field 
$\langle \Sigma \rangle_{\rm i\textrm{'}\rm{drop}}{=}70{\pm}8 {\times}10^{-3}$ gal./arcmin$^2$
(density contrast $=2.1^{{+}0.8}_{{-}0.7}$).

The over-densities of i'-dropouts in the direction to each QSO
are very likely associated to the environment of the QSOs, in particular
in the field of QSO J113717.73+354956.9 which is at $z_{em}{=}6.01$ \citep{fan2006a}.
The i'-dropout criteria applied to the broad-bands used in this work
is found to recover LBGs at $z{\simgt}5.87$, 
which is $\simgt64.1h^{{-}1}$ comoving Mpc ($\simgt6500$ km s$^{{-}1}$) 
from the strong \civ\ absorption system at $z_{abs}{=}5.7242$
and $\simgt57.8h^{{-}1}$ comoving Mpc (${\simgt}5860$ km s$^{{-}1}$) from the
\civ\ absorption system at $z_{abs}{=}5.7383$.
Thus, our simulated observations
described in Appendix \ref{app:simulationobs} predicts that 
the i'-dropout sample is too far from these \civ\ systems
to be physically associated. 
Although it is true that photometric errors could introduce 
objects from lower redshift in the i'-dropout colour selection,
this effect is more significant for objects close to 
the faint end of the selection function, and 
the three objects within 10$h^{{-}1}$ projected comoving Mpc of 
the QSO J103027.01+052455.0 ($z_{em}{=}6.309$) are bright sources (z'${\sim}25.2$ mag).
Moreover, we have spectroscopic confirmation of one of them
at $z{=}5.973{\pm}0.002$ \citep{diaz2011},
originally reported by \citet{stiavelli2005}.
In the field J1137+3549, the i'-dropouts in the over-density towards the QSO 
are fainter than the field J1030+0524. Hence, photometric errors
could explain that some LBGs at redshift $z{\sim}5.7$ might 
end up in the i'-dropout sample. However, the large number
of candidates in a small projected area implies that this over-density 
is extended over a significant fraction of the volume probed by the i'-dropout sample.
Therefore, the over-density very likely covers the environment of the QSO.

Since LBGs are star-forming galaxies,
the UV-brighter examples are also more massive \citep{mclure2011, gonzalez2011}.
Therefore, although the number of objects is low 
(23 and 56 i'-dropouts in each field, Section \ref{s:numberLBG}),
the sample contains the more luminous objects, 
which in comparison to the LAEs, correspond to
older and more massive haloes in the field of view. 
In the J1030+0524 field, there is good agreement in the position of
the projected over-densities 
of LAEs and i'-dropouts in the centre and the 
north half but not in the south half.
This is shown in the bottom left panel of Figure \ref{f:comp} 
that compares the surface density of i'-dropouts 
(colour coded contours) and LAEs (line contours).
Similarly, in the field J1137+3549 
the bottom right panel of Figure \ref{f:comp} shows
good agreement between LAEs and i'-dropouts 
in the centre and east of the field. 
However, the west of the field shows 
the two populations not aligned. 
Because the contours represent the distribution
collapsed in the line-of-sight direction,
this behaviour is consistent with the expectation
that LBGs trace group over-densities (nodes) and LAEs
trace filamentary structures.
A filament oriented in the line-of-sight direction
will be detected in both samples whereas
filaments tangential to the observer will show 
no counterpart in the sample at higher (or lower) redshift.
Moreover, shifted positions of ``nodes'' in the two distributions
correspond to structures with intermediate orientations.

Finally, i'-dropouts are compared to \zlbg\ LBGs (Figure \ref{f:comp-b}),
and their clustering mismatch is more significant.
In particular, both fields show that 
the positions of all the concentrations of galaxies 
of one sample are shifted from the over-densities of 
the other sample. The exception is the north-east corner
of field J1137+0524 (right panel).  
This result has important implications for 
the observed surface density contrast 
within $10{\it h}^{{-}1}$ comoving Mpc
from the line-of-sight to each QSO.
Previously, we showed that while both \zlbg\ LBG 
samples are under-dense in the line-of-sight 
towards the background QSO, the i'-dropout samples
show the opposite picture with both lines-of-sight 
intercepting an over-density. 
Therefore, the fact that both fields show the same 
characteristic density distribution pattern of ``avoidance''
across the entire field of view strongly suggest that 
the structure is real. 

In summary, the over-densities of i'-dropouts are likely
associated to the environment of the QSOs in the background and
the mismatch between the distribution of i'-dropout and 
\zlbg\ LBGs is found across the two fields of view
which strengthens the significance of the 
under-density of bright \zlbg\ LBGs in the environment 
of the \civ\ absorption systems.
Moreover, the distribution of LAEs show partial agreement 
with i'-dropouts and supports the 
idea that LAEs trace filamentary structure between nodes of LBGs 
\citep[e.g.][]{cooke2013}.
In the following section, we compare our results with 
expectations from a random distribution of sources.

\begin{figure}
\includegraphics[width=80mm]{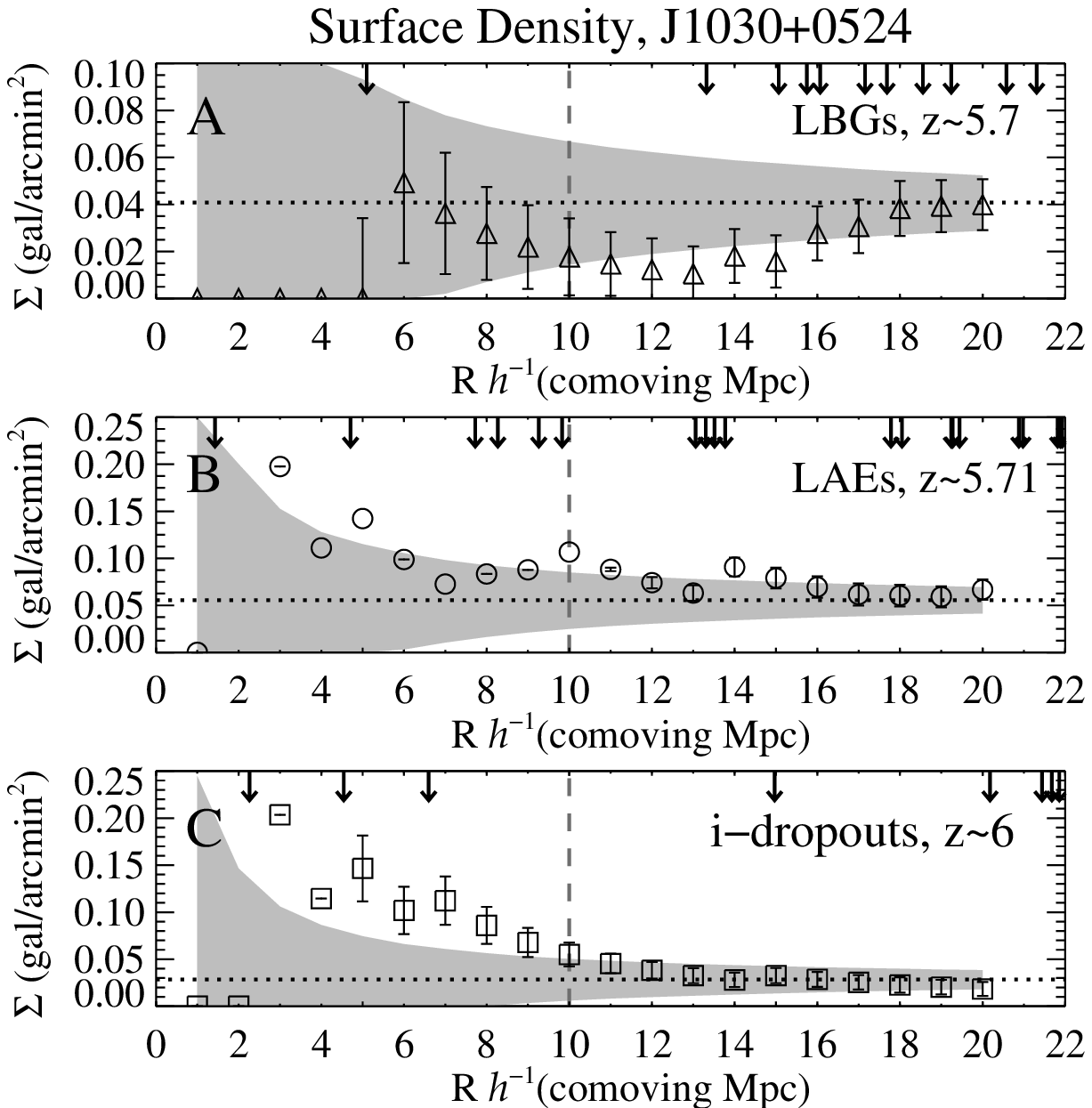} 
\caption{ Surface density of galaxies 
as a function of radius from the line-of-sight to the \civ\ systems
in the field J1030+0524.
The grey area contains ${\pm}1$ standard deviation
from  the average of 10$^5$ random realisations of galaxy distribution,
which were created with the same total surface density of \zlbg\ LBGs 
(Panel A), \zlae\ LAEs (Panel B) and \zid\ i'-dropouts (Panel C).
The black dotted horizontal line indicates 
the total surface density of the corresponding 
sample in the field J1030+0524. 
The arrows in the top axis of each panel indicate the 
radial position of objects within 22${\it h}^{-1}$ comoving Mpc radius.
Panel A shows that the surface density of \zlbg\ LBGs (triangles) 
in the J1030+0524 field is lower than the mean of the field
at most scales out to ${\sim}18 h^{-1}$ comoving Mpc.
Panel B corresponds to \zlae\ LAEs (circles) and shows the 
opposite to \zlbg\ LBGs. With a stronger signal, the number of LAEs is 
higher than the mean of the field at all scales out to ${\sim}17 h^{-1}$ comoving Mpc.
In panel C, the surface density of i'-dropouts is higher than 
the mean of the field at scales ${\simlt}10 h^{-1}$ comoving Mpc.
The error bars account for the uncertainty introduced 
by photometric errors and masked areas.
As discussed in Section \ref{s:NBcol}, the error bars
in Panel B are very small because the LAE sample
is very stable and the Monte-Carlo simulation of the errors
very rarely recover new objects or lose selected LAEs.}
\label{f:sdradius-1030}
\end{figure}

\begin{figure}
\includegraphics[width=80mm]{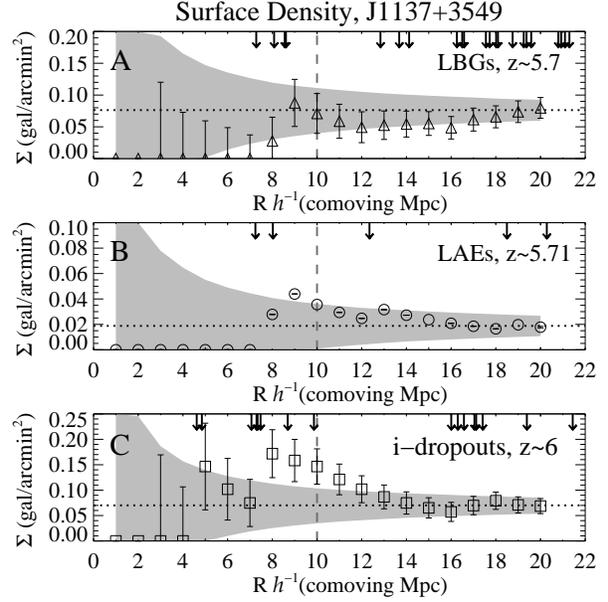} 
\caption{ Surface density of galaxies 
as a function of radius from the line-of-sight to the \civ\ system
in the field J1137+3549.
Symbols and lines are as per Figure \ref{f:sdradius-1030}.
Panel A shows that the surface density of \zlbg\ LBGs 
is lower than the mean of the field at all scales out to 
${\sim}19 h^{-1}$ comoving Mpc, in agreement with the field J1030+0524.
Panel B shows that the surface density of \zlae\ LAEs 
is higher than the mean of the field on scales of ${\sim}8$--$16 h^{-1}$ comoving Mpc,
which is similar to our findings in the field J1030+0524 (Figure \ref{f:sdradius-1030}).
In panel C, i'-dropouts are significantly in excess 
(more than one standard deviation)
from the mean of the field, at scales of 8--11${\it h}^{-1}$ comoving Mpc
and probe the environment of the QSO.}
\label{f:sdradius-1137}
\end{figure}

 \subsection{Expectations from non-clustered sources}\label{s:compare-with-rnd}

This section presents the surface density of each sample 
of galaxies in the environment of the \civ\ systems
measured as a function of radius
and the probability of the observed number of sources
assuming a random distribution.

\subsubsection{Surface density profiles}\label{s:compare-with-rnd-SDP}

Figures \ref{f:sdradius-1030} and  \ref{f:sdradius-1137} 
present the surface density profile of the three samples  
out to ${\sim}20{\it h}^{-1}$ comoving Mpc 
from the lines-of-sight to the \civ\ systems in the each field.
The error bars include Poisson and photometric errors.
In order to compare with a non-clustered distribution,
$10^5$ random galaxy distributions
with the same total surface 
density of each sample were simulated,
from which the surface density was measured radially 
from the \civ\ line-of-sight, after applying the same
masks used in the data.
The grey shaded areas in Figures \ref{f:sdradius-1030} 
and \ref{f:sdradius-1137} indicate one standard deviation 
of the radial surface density of the random sample.

The surface density profile of \zlbg\ LBGs (triangles, top panels)
is different from \zlae\ LAEs (circles, middle panels) in both fields of view.
In the first case, the surface density of LBGs in the environment of 
the \civ\ is below the total surface density
of the corresponding field (the triangles are mainly below the dotted line out to 
${\sim}18{\it h}^{-1}$ comoving Mpc).
This under-density is more significant in the field J1030+0524
where the largest deficit of LBGs with respect
to a non-clustered distribution is at scales of 13${\it h}^{-1}$ comoving Mpc.
The opposite is observed in the LAEs population:
in the middle panels of Figures \ref{f:sdradius-1030} and  \ref{f:sdradius-1137},
the circles are above the dotted line out to ${\sim}17{\it h}^{-1}$ comoving Mpc.
Interestingly, the LAEs show the most significant excess 
at a scale of 10${\it h}^{-1}$ comoving Mpc, close
to the scale of maximum deficit of LBGs in the field J1030+0524.
Moreover, in Figure \ref{f:sdradius-1137}, the number of LAEs
around the \civ\ system in the field J1137+3549 
is also higher than the total surface density  of the field
at similar scales (${\sim}10{\it h}^{-1}$ comoving Mpc).
In summary, both lines-of-sight intercept
regions of the Universe that --at scales of
${\sim}10{\it h}^{-1}$ comoving Mpc--
have two times more LAEs, and almost half the number of LBGs, 
than the average in the surrounding 
${\sim}80{\times}60{\it h}^{-2}$ comoving Mpc$^2$.

The i'-dropout samples of the two fields (bottom panels)
present similar profiles.
First, in the field J1030+0524 (Figure \ref{f:sdradius-1030}) 
the QSO line-of-sight is close to three galaxy candidates
and we have spectroscopic confirmation of the closest one.
In comparison with the expectation from a uniformly distributed 
sample (grey shaded area) the surface density is 
higher than one standard deviation from the mean
surface density of i'-dropouts in the field
at ${\simlt}10{\it h}^{-1}$ comoving Mpc scales.
At ${>}10{\it h}^{-1}$ comoving Mpc the surface density is
in agreement within the errors with the average of the field.
Second, seven i'-dropouts in the field J1137+3549 (Figure \ref{f:sdradius-1137}) 
are within $10{\it h}^{-1}$ comoving Mpc from the line-of-sight to the QSO. 
Assuming no clustering, the excess in the surface density 
of i'-dropouts is higher than one standard deviation 
from the mean of the field at 8--10${\it h}^{-1}$ comoving Mpc scales,
which was also identified as an excess scale
in the LAE samples of both fields.

\subsubsection{Probability of the observed number of sources.}

The $10^5$ random realisations described in Section \ref{s:compare-with-rnd-SDP}
were used to estimate the probability of the results 
under the assumption of no clustering.
We calculate the probability, for a random distribution,
of the number of galaxies in a circle of 6, 8 and 
10${\it h}^{-1}$ comoving Mpc radius
centred on the \civ\ lines-of-sight.
We present each field in turn.

Table \ref{t:prob} presents the number of 
objects detected in each sample
at scales of 6, 8 and 10${\it h}^{-1}$ comoving Mpc
centred on the \civ\ line-of-sight, 
and the probability obtained for a random distribution 
with the same mean surface density of LBGs, 
LAEs and i'-dropouts.

Starting with the field J1030+0524,
the probability of having three or more LAEs in the field
at 8${\it h}^{-1}$ comoving Mpc is $P_{8,\,LAE}({\geq}3){=}0.32$.
At 10${\it h}^{-1}$ comoving Mpc the probability of having six LAEs 
or more is $P_{10,\,LAE}({\geq}6){=}0.09$.
Thus, we find more LAEs than expected from a random distribution
which suggests that the distribution of LAEs is not random
and the \civ\ line-of-sight intercepts 
an environment with significant excess of LAEs.
This concentration of objects at large scales is not
seen in the \zlbg\ LBGs.
The random distribution easily predicts a high number of cases
with zero or one source for the number of \zlbg\ LBGs.
Therefore, \zlbg\ LBGs are found to be consistent with 
a random distribution.
This can also be seen in Figure \ref{f:sdradius-1030}
where points at 6, 8 and 10${\it h}^{-1}$ comoving Mpc
are within the shaded regions.
In particular at 10${\it h}^{-1}$ comoving Mpc,
the probability for a random distribution
of finding one or less \zlbg\ LBGs is $P_{10,\,LBG}({\leq}1){=}0.32$. 

In the field J1137+3549, at 8${\it h}^{-1}$ and 10${\it h}^{-1}$ 
comoving Mpc scales, \zlae\ LAEs are consistent with the expectation 
from no clustering since the probabilities 
are $P_{8,\,LAE}({\geq}1){=}0.50$ and $P_{10,\,LAE}({\geq}2){=}0.29$.
Similarly, the \zlbg\ LBGs are also consistent with a non-clustered sample.

For the i'-dropouts in the field J1030+0524, 
the over-density is more significant
at scales of 8${\it h}^{-1}$ comoving Mpc at which
the probability that three objects result from a random distribution
is P$_{8,\,i'{\textrm-}drop}(\geq 3){=}0.07$,
while the probability at 10${\it h}^{-1}$ comoving Mpc 
radius is P$_{10,\,i'{\textrm-}drop}(\geq 3){=}0.20$.
This shows that, around the \civ\
line-of-sight, the excess of LAEs and i'-dropouts 
is peculiar whereas the low density of LBGs could 
be simply random.
Similarly, i'-dropouts in the field J1137+0524 have 
a probability of six or more sources
P$_{8,\,i'{\textrm-}drop}({\geq}6){=}0.03$.
Furthermore, at 10${\it h}^{-1}$ comoving Mpc radius,
the probability of eight or more i'-dropouts is also 0.03.

Our comparison with non-clustered
distributions is a good reference point 
yielding conservative results, since
accounting for clustering of sources in 
our simulated distributions will 
increase the predicted probability of zero detection,
which would enhance the significance of the observed 
LAE and i'-dropout over-densities.
Overall, the results in the field J1137+3549
are in agreement with the field J1030+0524:
the excess of i'-dropouts is robustly determined, 
the deficit of \zlbg\ LBGs is consistent with a random distribution 
and the over-density of LAEs is detectable (lower significance in the J1137+3549 field).

\begin{table*}
\begin{minipage}{150mm}
\centering
 \caption{Probability of the number of galaxies within a radius of 6, 8 and 10$h^{{-}1}$ comoving Mpc
centred on the \civ\ line-of-sight in the corresponding field,
obtained from 10$^5$ random realisations of galaxy distribution (no clustering is assumed) 
with the same mean surface density of \zlbg\ LBGs, LAEs and i'-dropouts.
The errors include photometric errors.
For LAEs and i'-dropouts, we present the probability of finding an excess
of sources $P_{r}({\geq}N)$, except for the case of LAEs in the J1137+3549 field at a radius of 6${\it h}^{-1}$ comoving Mpc.
For LBGs, we present the probabilities for a deficit of sources $P_{r}({\leq}N)$.}
 \label{t:prob}
 \begin{tabular}{@{}lccccccc}
  \hline
  \hline
  Field & Sample & N$_6$ & P$_6$ &N$_8$ & P$_8$ & N$_{10}$ & P$_{10}$ \\
    \hline
           & \zlae\ LAEs & 2 & 0.31$^a$& 3 & 0.32$^a$& 6 & 0.09$^a$ \\
J1030+0524 & \zlbg\ LBGs & $1{\pm}0.7$ & 0.80$^b$& $1{\pm}0.7$& 0.56$^b$ & $1{\pm}0.9$ & 0.32$^b$ \\
           & i'-dropouts & $2{\pm}0.5$ & 0.10$^a$ & $3{\pm}0.7$ & 0.07$^a$ & $3{\pm}0.7$ & 0.20$^a$ \\
\hline           
           & \zlae\ LAEs & 0 & 0.68$^b$ & 1 & 0.50$^a$ & 2 & 0.29$^a$ \\
J1137+3549 & \zlbg\ LBGs & $0^{{+}1}$ & 0.21$^b$ & $1^{{+}1.3}_{{-}1}$ & 0.24$^b$ & $4{\pm}1.8$ & 0.57$^b$  \\
           & i'-dropouts & $2{\pm}1.2$ &0.40$^a$ & $6{\pm}1.6$ & 0.03$^a$ & $8{\pm}1.9$ & 0.03$^a$  \\ 
\hline
$^{\it (a)}${$P_{r}({\geq}N_{r})$}\\
$^{\it (b)}${$P_{r}({\leq}N_{r})$}\\
\end{tabular}
\end{minipage}
\end{table*}

\section{Discussion} \label{s:discussion}

\subsection{The impact of the $z{\sim}5.7$ LBG selection criteria} \label{s:disc-sel}

Many studies on LBGs have found that EW(\Lya) 
correlates with other galaxy properties \citep[e.g.][]{shapley2003, vanzella2009, cooke2010, jones2012}. 
This section discusses the importance of the $z{\sim}5.7$
LBG selection criteria to directly test a
possible correlation between environment and EW(\Lya).
First, we summarise the trends 
previously reported between the UV luminosity 
and EW(\Lya), to show the differences 
between the two populations of galaxies 
presented in this work. 
The LBG sample comprises only the
bright end of the population and is dominated
by galaxies with EW(\Lya) ${\simlt}25$\AA.
In contrast, our LAE sample includes only 
galaxies with EW(\Lya) ${>}20$\AA, and is dominated
by galaxies with faint UV continuum.
Second, we review the dependence of 
UV luminosity and EW(\Lya) with other observables.
Finally, we discuss the implications from different 
clustering properties and the potential dependence 
between EW(\Lya) and environment. 

\subsubsection{Two different samples: UV bright LBGs and \Lya\ bright LAEs} \label{s:two-samp}

In Section \ref{s:colmag} we showed that
the LAEs sample is fainter in the UV (${\sim}1350$\AA) than
the LBGs sample. 
This result is in agreement with studies
of LAEs at $z{\sim}5.7$ \citep[e.g.][]{shimasaku2006}
that find emitters with fainter far UV magnitudes 
to have stronger \Lya\ emission.
Therefore, a sample of bright LAEs will be dominated by objects
with faint UV magnitudes. 

Moreover, from a large spectroscopic campaign of $z{\sim}3$ LBGs, 
\citet{shapley2003} reported that the average luminosity in the UV 
increases with decreasing EW(\Lya).  
The same trend is observed at 
higher redshifts \citep[e.g.][]{stanway2007, ouchi2008, vanzella2009, stark2010, stark2011}.
As a result, a sample containing the brightest LBGs (M$_{UV}{<}-21.0$ mag)
will be biased towards low EW(\Lya). 
Thus, based only on the UV luminosities of the LBGs in our samples 
it is reasonable to expect a low fraction of them with \Lya\ in emission.

\citet{shapley2003} also reported that LBGs with stronger \Lya\ 
emission have bluer continua.
In addition, \citet{ouchi2008} find that LAEs from $z{=}3.1$ to 5.7
have bluer UV continuum colours than dropout LBGs. 
This is consistent with the picture at $z{\sim}3$
presented by \citet{cooke2009}, who showed 
a correlation between 
the position of an LBG in the colour-magnitude diagram 
and its EW(\Lya), such that
luminous LBGs are dominated by \Lya\ in absorption
with the associated redder UV continua while blue LBGs
typically present \Lya\ in emission.
In other words: the LAE population dominates the faint-blue
quadrant of the colour-magnitude diagram, while the opposite
bright-red quadrant is dominated by LBGs with \Lya\ 
in absorption. 

At $z{\sim}3$, \citet{shapley2003} finds weaker low-ionization interstellar 
absorption and smaller kinematic offset between \Lya\ and interstellar
absorption lines among strong \Lya\ emitters.
At $z{=}4$ and 5, \citet{vanzella2009} compare the stacked spectra
of LBGs with \Lya\ in absorption and LBGs with \Lya\ in emission 
and find stronger interstellar absorption in the composite with \Lya\ in absorption.
Moreover, \citet{vanzella2009} also compare different morphological 
parameters ($r_{hl}$, isophotal area, FHWM and Gini coefficient)
of LBGs at $\langle z \rangle {\sim}3.7$ and find that all parameters 
suggest that LBGs with \Lya\ in absorption are more extended and 
diffuse sources than LBG with \Lya\ in emission.
This \Lya --morphology relationship is also discussed in \citet{cooke2010}.
All these results suggest that strong \Lya\ emitters are younger,
less massive and less dusty than strong \Lya\ absorbers. 

Furthermore, following the growth of structure from 
density fluctuations, more massive objects are
expected in denser regions of the Universe.
Therefore, it would be expected that two samples of
galaxies statistically dominated by different
\Lya\ properties also present different spatial distribution.
If LAEs are less massive systems than bright
LBGs, then they would be less clustered. 
As a result, the LAE and LBG samples in this work 
are complementary tracers of the large-scale structure 
and the spatial density of these galaxies could be used 
to characterise different aspects of the large-scale environment.

\subsubsection{Distribution of galaxies: is EW(\Lya) an environment indicator?}

There is abundant evidence that LBGs are biased tracers of matter,
as they inhabit over-dense regions of the universe 
\citep[e.g.][]{steidel1998, ouchi2004b, adelberger2005a,cooke2006, hildebrandt2009, bielby2013}.
Furthermore, it has been reported that at redshifts $z{=}3$--5.5
more luminous LBGs are more strongly clustered 
\citep[e.g.][]{ouchi2005b, kashikawa2006a, lee2006, hildebrandt2009}.
Recently, \citet{cooke2013} analysed ${\sim}55\,000$ LBGs
at redshift $z{\sim}3$ from the Canada-France-Hawaii 
Telescope Legacy Survey and found that
the subpopulation of LBGs with \Lya\ in absorption (aLBGs)
are preferentially in groups, clustered in massive dark 
matter haloes (${\sim}10^{13}$M$_{\sun}$). Interestingly, 
the subpopulation of LBGs with \Lya\ in emission (eLBGs) 
is more evenly spread in space, located in group outskirts 
and the field. Such an environment--EW(\Lya) connection 
suggests that the \Lya\ line could be a statistical 
indicator of environment. Therefore, under this picture, 
if the bright LBG population and the LAE population
trace different matter density bias, their distributions
provide complementary information 
on the environment at all scales. 

Our results cannot rule out an environmental
dependence for EW(\Lya).
We find UV bright LBGs and \Lya\ bright LAEs
preferentially avoiding each other, with bright LBGs
forming ``clumps'' and LAEs extending and
clustering around them.
This is in agreement with the expectation that our 
sample of \zlbg\ LBGs traces the most massive 
haloes while LAEs 
inhabit their surroundings areas.
As mentioned above, a mismatch between the 
distribution of LBGs with different EW(\Lya) is inferred 
at redshift $z{\sim}3$ by \citet{cooke2013}.
If the processes responsible for this effect 
are already in place at \zlbg,
it is possible that a sample of galaxies selected from a narrower
redshift slice will show the ``shell''-like structure
tested in \citet{cooke2013}.
Figure \ref{f:comp} shows that, in both fields,
over-density contours of $z{\sim}5.7$ LBGs 
and \zlae\ LAEs seem to alternate, filling 
the under-density regions of each other.
We highlight that our selection criteria for $z{\sim}5.7$ LBGs
has the advantage of aiming for a population of galaxies 
with \Lya\ preferentially in absorption or low emission 
(EW(\Lya\ $\simlt25$\AA))
in a cosmic volume ($\Delta z{\sim}0.2$) that is
a good complement to the LAEs sampled volume ($\Delta z{\sim}0.08$).
Considering that the volumes of the samples are similar,
the suggested scenario 
where galaxies with low EW(\Lya) are more clustered
than galaxies with high EW(\Lya)
is a simple explanation for our results.
Furthermore, it supports the use of LAEs and LBGs 
as complementary tracers of environment.
We will review this topic with follow-up spectroscopy
of the samples of this work.

\subsection{Environment of \civ\ systems and the chemical enrichment of the Universe}\label{s:disc-environment}

In this section we discuss the environment
of the \civ\ systems and the implications
from the evolution of the \civ\
cosmic density $\Omega_{\text \civ}(z)$.
Firstly, we review the expectations 
on the environment of the systems at \zlbg\
from lower redshift observations 
and find disagreement. 
We discuss possible scenarios
and conclude that background ionizing 
flux density fluctuations 
are likely affecting the detection
of high ionization absorption systems.
Secondly, the origin of the \civ\ absorption
systems is considered.
The predictions from theoretical models that 
reproduce the evolution of the \civ\ cosmic density
are compared with observational results
to conclude that LAEs are the most 
favourable candidates for the physical origin
of the metals. 
Finally, we comment on previous studies 
in the J1030+0524 field.

\subsubsection{The strongest \civ\ absorption systems at $z{\sim}5.7$:
implication for the build up of cosmic metals}\label{s:dis-civ-oi}

The growth of $\Omega_{\text \civ}(z)$ with cosmic time
has been measured by several authors \citep{ryan-weber2009, becker2009,cooksey2010,simcoe2011b,dodorico2013}.
In this section we interpret our results as evidence that 
the ionization state of the IGM plays a role in the 
observed evolution of $\Omega_{\text \civ}(z)$.
We argue that detection of high ionization systems at $z\geq 5.7$ depends
on the fluctuations in the background ionizing flux density after the EoR.
In this picture, we conclude that 
higher levels of ionizing flux at \zlbg\
seem to be associated with an excess of LAEs.

\citet{adelberger2003} analysed the relative spatial distribution of \civ\ 
absorption systems and LBGs at redshift
$2{\simlt} z {\simlt} 3$ and reported significant evidence 
that strong systems (\nciv ${\geq}10^{14}$ \cm) and LBGs 
are found in similar parts of the Universe, suggesting that
they may be similar objects. 
Moreover, galaxies in denser environments
are more likely to have a \civ\ system with \nciv ${>}10^{13}$ \cm\
within 1${\it h}^{-1}$ comoving Mpc \citep{adelberger2005b}.
Another piece of the puzzle is the significant
evidence of galactic outflows in star-forming
galaxies at $z{\simlt}3.5$
\citep[e.g.][]{heckman2000, pettini2002, shapley2003, martin2005, rupke2005, weiner2009, dessauges-zavadsky2010}.
In particular, \citet{steidel2010} report strong evidence for 
outflowing enriched gas in their LBG sample
and find that \civ\ absorption systems at 
$\langle z \rangle {\sim}2.3$ are tracing
the circum-galactic medium (CGM) of LBGs. 
These findings show that the surroundings
of $z {\simlt} 3$ LBGs have been enriched.

If $\Omega_{\text \civ}(z)$ is controlled only by the metal content of the IGM,
then at $z{\sim}5.7$ the IGM is simply less enriched.
In this case, \civ\ systems should be found in regions 
of the IGM that were polluted first, i.e. regions of 
earlier star formation episodes.
Therefore, they would be associated 
with over-densities of galaxies, as
it is found at $2{\simlt} z {\simlt} 3$ \citep[e.g.:][]{adelberger2005b}.
Similarly, if \civ\ absorption 
systems trace the CGM of LBGs,
the evolution of $\Omega_{\text \civ}(z)$ would simply reflect 
an increase with cosmic time in the number of haloes 
hosting strong \civ\ systems.
In this case, for $z{\sim}5.7$ \civ\ absorption systems
we still expect a similar environment
as found at $z{\simlt} 3.5$. 

However, in this work we trace back the earliest
absorption systems known to-date in two lines-of-sight and
find them removed from the main over-densities. 
The strong \civ\ absorption systems at $z{\sim}5.7$ 
studied here are not found in over-dense 
regions of bright and massive LBGs ($L>L^{\star}$),
but in rather under-dense regions of the large-scale 
distribution of \zlbg\ LBGs (top panels of Figure \ref{f:maps}).
In the fields J1030+0524 and J1137+3549
the \civ\ absorption systems are distant 
(${>}10{\it h}^{-1}$ projected comoving Mpc)
from the main concentrations of LBGs.
Thus, the high column density \civ\ systems are not 
found associated with the earliest star-forming regions.

Furthermore, we report a significant excess of LAEs
within 10${\it h}^{-1}$ projected comoving Mpc
from the \civ\ absorption systems at $z{\sim}5.7$ in
both fields of view (higher in the field J1030+0524).
This result completes the picture that
\civ\ absorption systems are tracing different environments
at $z{\leq} 3.5$ and \zlbg . 
This is in agreement with \citet{dodorico2013}
who finds that \civ\ systems at $z{\sim} 5.7$ are better explain
by gas with an over-densities of $\delta {\sim} 10$
whereas \civ\ systems at $z{\sim} 3$ are better explain
by gas with $\delta {\sim} 100$.
Moreover, we propose that such a density contrast 
is linked to larger scales: 8--10$h^{{-}1}$ comoving Mpc.
Considering that the presence of \civ\ is determined not only by the 
amount of carbon but also by the ionization condition of the gas,
our results suggest that the detection of \civ\ systems, 
and therefore the rise in $\Omega_{\text \civ}$ with cosmic time, 
is likely affected by a heterogeneous ionizing flux density
distribution left over from cosmic reionization.
Therefore, whether the ionizing flux is dominated by local sources
or the large-scale distribution of ionizing sources, 
if the ionizing flux density fluctuations 
have survived at least to \zlbg ,
we conclude that higher levels of ionizing flux 
seem to be associated with excess of LAEs.

Finally, considering that in both fields
the closest LBG is at more than 5${\it h}^{-1}$ comoving Mpc
of the \civ\ system, our results also
imply that the progenitors 
of strong \civ\ systems at $z{\sim}5.7$ 
are not bright and massive LBGs.
We expand the discussion on the origin of
the metal absorption systems in the Section \ref{s:disc-originciv}.

\subsubsection{The closest galaxy to a $z{\sim}5.7$ \civ\ system}\label{s:disc-closest}

We report one galaxy candidate close enough to 
the J1030+0524 \civ\ line-of-sight to consider 
the possibility of a direct causal connection 
through active outflows.
The object is a faint LAE selected candidate with
NB\civ${=}25.11\pm0.15$ mag, barely detected in z' (z'${=}25.89{\pm}0.52$ mag,
M$_{UV}{\sim}$-20.48) and non-detected in R$_c$ and i'.
The object lies at 0.60 arcmin from the sight-line
towards QSO J1030+0524. Thus, if the object
is confirmed to be a $z{\sim}5.72$ galaxy, 
it will be at 212${\it h}^{-1}$ physical kpc from
the \civ\ system.
However, the line-of-sight to J1030+0524
contains two \civ\ systems at $z{\sim}5.72$--5.74
and opens a window to explore the origin of the high-redshift 
enrichment in better detail.
Precise spectroscopic redshift measurement
will be required to confirm the association
of this LAE with any of the two absorption systems.
Spectroscopic data of the object has been obtained with
\textsc{deimos} on the Keck Telescope and will be
presented in a forthcoming paper.

\subsubsection{The origin of $z{\sim}5.7$ \civ\ systems}\label{s:disc-originciv}

It is well recognised that feedback from galactic winds
plays a fundamental role setting the properties of 
metal absorption systems in the IGM 
\citep[e.g.][]{oppenheimer2006, oppenheimer2008, martin2010, tescari2011, cen2011},
and is important for galaxy evolution 
\citep[e.g.][]{rupke2005, martin2005, martin2006, steidel2010, bradshaw2013}. 
This is why we are interested in the properties 
of the galaxies near the \civ\ systems.
In this section we present LAEs as the 
most likely physical origin 
of strong \civ\ absorption
at this redshift. We review the theoretical
expectations and observational results
that suggest this scenario, and we present
our results as direct evidence to support it.

The evolution of the comoving density of \civ\ 
ions in the IGM is only well reproduced by models 
that include outflows from star-forming galaxies. 
\citet{oppenheimer2009} find that 
\civ\ at $z{=}5$--6 is a good indicator of 
metals in the IGM and not gas in galaxies.
They predict that strong \civ\ systems are ionized by the galaxy 
that produced the enrichment, which would have rest-frame
UV magnitudes ${\geq}-18.9$ (1350\AA) and 
stellar masses of $10^7$--$10^8$\msun. 
Current measurements of the stellar mass of LAEs are 
at the upper end of this range.
For example, \citet{gawiser2006} reported mean stellar masses 
of $5\times10^8{\it h}^{-2} M\sun$ in LAEs at $z=3.1$.
More recently, \citet{lidman2012} estimated stellar masses of 
10$^8$--10$^9{\it h}^{-2} M_{\sun}$ for four of the five bright 
LAEs at $z{\sim}5.7$ they analysed.

When considering dark matter haloes, 
\citet{porciani2005} predicted that galaxies in
haloes of ${\sim}10^{8}$ -- $10^{10}M_{\sun}$ at $z{\sim}6$ could 
have produced large bubbles of metals that evolve
into the environment of $z{\sim}3$ LBGs.
Interestingly, \citet{ouchi2010} estimated that LAEs at $z{=}5.7$ and $z{=}6.6$ 
lie in dark matter halos of 10$^{10}$ -- 10$^{11} M_{\sun}$. 
Moreover, this result holds for LAEs at $z{=}3.1$--5.7.
Therefore, faint LAEs are
the best candidates for the galaxies responsible for the 
enrichment of the IGM at $z{\geq}6$.
This idea is supported by the excess of LAEs
around the \civ\ systems 
in the two fields of view of this study.

Simulations also predict that at redshift $z{\sim}6$ there should 
be a closer connection between \civ\ absorption systems and their 
parent galaxy as the gas would be in their first journey 
out of the galaxy \citep{oppenheimer2009}. 
Although the LBG sample seems
to be too distant to be associated with the \civ\ systems,
in section \ref{s:disc-closest} we report
a narrow-band selected LAE as the closest
galaxy candidate to a \civ\ system at $z{\sim}5.7$.
This particular example in the field J1030+0524
at ${\sim}212 h^{-1}$ projected physical kpc from the line-of-sight
to the \civ\ is in very good agreement with
expectations from cosmological simulations.

In conclusion, our results suggest that the origin of the \civ\ absorption 
systems at $z{\sim}5.7$ is more likely associated with a low-mass
LAE-type of galaxy than a massive LBG-type of galaxy.
This argument is strengthened 
by, first, the best candidate for a galactic 
outflow--absorption system causal connection is a faint
LAE at 212${\it h}^{-1}$ physical kpc in the field J1030+0524 
that awaits spectroscopic confirmation, and second,
the tentative detection of a \civ\
absorption system at $z{=}5.9757$ reported by \citet{dodorico2013}
which is at $\Delta v{\sim}110{\pm}85$km s$^{{-}1}$
and 325${\it h}^{-1}$ physical kpc transversal
from the i'-dropout with strong \Lya\ emission 
J103024.08+052420.41, at $z{\simeq}5.973\pm0.002$ \citep{stiavelli2005, diaz2011}.
However, the possibility that undetected low mass young 
galaxies are responsible for the enrichment of the IGM
cannot be ruled out.

\subsubsection{Previous environmental studies in the field J1030+0524}

Some evidence of a possible over-density
of galaxies in the direction to the 
QSO SDSS J1030+0524 has been provided by
\citet{stiavelli2005}. 
This result was later confirmed by \citet{kim2009} 
based on an excess of galaxies
photometrically classified as i'-dropouts 
(i.e. with $i_{775}{-}z_{850}{>}1.3$),
and hence potentially located at $z {\simgt} 5.5$.
They used the Advanced Camera for Surveys (ACS)
on the \textit{Hubble Space Telescope} (HST)
as part of a large study of i'-dropouts around 
five QSOs at $z{\sim}6$.
In particular, $14{\pm}4$ $i$-dropouts were found 
in the field of J1030+0524,
compared with $8 {\pm}3$ expected from a 
random distribution of galaxies
at this redshift \citep{giavalisco2004}.

In this study, we analyse the same field at significantly 
larger scales, we use an independent method to quantify 
the density contrast in the field and, although the current 
photometry is ground-based (seeing limited),
we include information from a bluer band (R$_c$) which was 
not available to previous studies.
We find an over-density of i'-dropouts relative to the projected
surface density averaged over scales of ${\sim}80{\times}60h^{{-}1}$ comoving Mpc,
in agreement with previous studies.

More recently, \citet{diaz2011} analysed the spectra of a small 
sample of three objects in the field J1030+3549 
selected as i'-dropouts in \citet{stiavelli2005}.
In particular, \citet{diaz2011} reported positive detection of 
a spectroscopic feature that at the rest wavelength 
of the \Lya\ line would position the objects 
at $z_{T1}{=}5.973$,  $z_{T2}{=}5.676$ and $z_{T3}{=} 5.719$ 
(Targets 1, 2 and 3 of \citealt{diaz2011}). 
However, our $z{\sim}5.7$ LBG selection does not recover them.
Although in our current study the photometry of Target 1 and 2  
are contaminated by a close object,
both targets are detected with S/N$_{\rm z\textrm'}{\geq}5$ and
the measured (i'-z') colours are in agreement with the values reported in \citet{diaz2011}. 
Moreover, Target 1 ($z_{T1}{=}5.973$) has colours expected for $z{\sim}6$ LBGs 
and is part of the i'-dropout sample, and Target 2 ($z_{T2}{=}5.676$) has colours 
close to the $z{\sim}5.7$ LBG criteria.
Finally, Target 3 is only detected with S/N$_{\rm z\textrm'} {\sim}2$ 
and has bluer colours than reported by \citet{diaz2011}.
More importantly, we have a $4\sigma$ detection in the 
R$_c$ band (R$_c{=}26.6{\pm}0.3$) which implies that, 
contrary to the other two objects, Target 3 is more likely a 
low-redshift object than a high redshift LBG.
We also note this object lies very close to the edge of
the ACS CCD camera, which can affect the 
photometry estimated from ACS data.
New spectroscopic data has been acquired and
the nature of these objects will be reviewed in a 
future work.

\subsection{The IGM and the sources of cosmic reionization}\label{s:dis-eor}

Metal absorption systems provide information 
of the IGM that could be used to test the topology of reionization.
Regardless of whether the \civ\ systems are IGM or CGM, 
if the change in $\Omega_{\text \civ}(z)$ is the result
of ionizing flux density fluctuations then
\civ\ systems close to the EoR 
trace regions of high flux density 
of ionizing radiation. In this case, a simple
prediction from an inside-out reionization is 
a positive correlation between mass distribution
and the ionization level of the IGM.
Under this scenario, rare 
high ionization strong absorption systems would be expected 
to reside in dense structures that collapsed earlier and were
reionized first. 
However, we do not find that \civ\ systems are associated with
over-densities of massive galaxies. 

Young massive stars provide the ionizing flux
that caused reionization. 
This means that the generation of stars that started the process
is not the same that finished it.
Under this scenario, it is more natural to expect the
ionization conditions that allow the detection of strong
\civ\ in regions of more recent star formation.
Indeed, our results support a connection between 
\civ\ absorption systems and recent star 
formation at $z{\sim}5.7$.
Moreover, the association with an excess of LAEs 
favours dark matter haloes 
of $M{\simlt}10^9$ -- $10^{10} M_{\sun}$
as progenitors of the \civ\ systems,
consistent with predictions from simulations \citep[e.g.][]{oppenheimer2009}.
As a result, young haloes in intermediate density 
regions at \zlbg\ could host the ionization 
conditions found in over-densities at $z{\sim}2$ -- 3.

Finally, there is general agreement that fainter galaxies (sub-L$^{\star}$)
dominate the ionizing photon budget. 
In particular at redshift $z{\sim}6$, \citet{finkelstein2012b}
report that the contribution from galaxies with L${>}$L$^{\star}$ 
to the total specific UV luminosity density is almost
the same as galaxies with 0.2L$^{\star}{<}$L${<}$0.5L$^{\star}$.
In addition, it has been found that LAEs are statistically 
fainter and less massive than LBGs, which makes them
good candidates for sources of ionizing radiation.
For example, current evidence suggest that the contribution
of LAEs to the ionizing photon budget increases with fainter 
magnitudes \citep[e.g.][]{kashikawa2011, cassata2011}.
Our results suggest a connection between highly ionized IGM
and LAEs, which is easily explained if LAEs provide a larger
amount of ionizing radiation than UV-brighter/more massive LBGs.
Therefore, our result is in good agreement with current findings
that sub-L$^{\star}$ galaxies play a central role in the reionization of the Universe,
at least at the final stages of the EoR.

In summary, this work supports the idea that
UV faint galaxies are the key to the reionization of the Universe,
and is in agreement with 
an inversion in the topology of cosmic reionization, 
meaning that in the latest stages of the EoR, the ionizing
flux density was higher in lower densities environments.
Hence, the simplest picture for the environment of the 
strong \civ\ absorptions at the highest redshift
is a ``filamentary'' structure populated by low-mass haloes
rather than an early over-density.

\section{Summary and conclusions}\label{s:conclusion}

This is a study of the environment of high column density
\civ\ systems (\nciv${>}10^{14}$\cm )
at $z{\sim}5.72$--5.74, in the fields J1030+0524 and J1137+3549.
Using wide field photometry in the R$_c$, i' and z' bands,
and a narrow-band filter NB\civ\ 
we select LBGs using broad-band colours,
and LAEs using narrow-band excess and colours.
Our results and conclusions are summarised 
as follows:

\begin{itemize}
\renewcommand{\labelitemi}{$\bullet$}

\item We find the selection criteria for LAEs
to be reliable and stable within the photometric uncertainty. 
We confirm that strong \Lya\ emission
can affect the broad-band colours
of the LBG population at the redshift of interest. 

\item We have tested a selection
criteria based on broad-band colours
that aims for a sample statistically dominated 
by bright star-forming galaxies with 
EW(\Lya) $\simlt 25$\AA\ (see Section \ref{s:sel}),
in a redshift slice $\Delta z {\sim}0.2$.
This $z{\sim}5.7$ LBG selection criteria is predicted
to be more effective at selecting galaxies 
at the redshift of the \civ\ systems
than the $z{\sim}6$ i'-dropout criteria. 
In general, the number of objects detected is consistent
with expectations from the $z{\sim}6$ luminosity function 
(Section \ref{s:results-num}).

\item We compare the projected distribution of LBGs ($\Delta z {\sim}0.2$)
with narrow-band selected LAEs ($\Delta z {\sim}0.08$).
A direct comparison of the clustering of the sources
is possible thanks to the narrow volume that is sampled.
In both fields-of-view, we find bright UV LBGs 
in clustered associations and 
LAEs distributed in the surroundings areas.
The structure detected in the distribution 
of galaxies is consistent, although not definite, 
with a possible dependence of the EW(\Lya)
with environment. 

\item The local environment of \civ\ absorption 
systems in the fields J1030+0524 and J1137+3549  
presents an excess of LAEs and a deficit of \zlbg\ 
LBGs per arcmin$^2$ on several scales. 
In the field J1137+3549, LAEs are low in number
but the surface density at a scale of $10{\it h}^{-1}$ comoving Mpc
is higher than the mean of the field.

\item i'-dropouts are found in excess towards both
QSO lines-of-sight and are likely related to the background ($z{\simgt}6$) 
QSO environment instead of the foreground ($z{\sim}5.72$--5.74) absorption systems of interest. 
This is in agreement with the expectation that 
QSOs inhabit massive haloes in the centre of large-scale
over-densities, and that large-scale over-densities can be 
traced by LBGs.

\item Our results suggest that strong \civ\ absorption 
systems at the end of the EoR are not related 
to over-densities of bright and massive 
(L${>}$L$^\star$) LBGs with low or 
no \Lya\ emission.
Therefore, the environment of 
strong $z{\sim}5.7$ \civ\ systems 
is different to the examples 
at lower redshift ($z{\leq}3.5$). 
Instead, \civ\ absorption systems 
are found in regions dominated by LAEs,
which are younger (recent star formation),
fainter and lower mass systems than LBGs.
This would imply that $z{\sim}5.7$ \civ\ systems trace 
low-to-intermediate density environments and 
are distant from the oldest star-forming 
regions of each field.

\item We report one LAE that lies 
at ${\sim}212{\it h}^{-1}$ physical kpc from the 
line-of-sight in the J1030+0524 field.
The close proximity suggests the 
faint galaxy is the progenitor of one of 
the two \civ\ systems in this sight-line.
This result supports the idea that LAEs are the most 
favourable candidates for the physical origin of 
the \civ\ systems at $z{\sim}5.7$.
Spectroscopic redshift determination 
is required to test a galaxy-absorption 
system connection and we will address 
this topic in more detail in a forthcoming paper.

\item The results support
the idea that the detection of high ionization 
absorption systems after the EoR
depends on the fluctuations of the ionizing
flux density.
In this case, the excess of LAEs 
found in the field J1030+0524 is
related to high levels of ionizing flux
that allow the detection of \civ .
This result implies that faint galaxies
are important sources of ionizing radiation,
in agreement with many other findings in the literature
that propose that faint galaxies are the
primary sources driving the end of 
cosmic hydrogen reionization.

\end{itemize}

More work on the environment of metal 
absorption lines in the EoR is needed. 
Recently, \oi\ absorption systems
have been proposed as probes 
of the physical state of neutral filamentary
over-densities in the later stages of the EoR
\citep{keating2014}.
If this is the case,
at $z{\geq} 5.7$ strong \civ\ 
absorption system with no low ionization
metal lines are the complement 
to \oi\ absorption systems.
It could be possible 
to combine low ionization and high ionization
metal absorption lines to study the EoR,
because \oi\ are expected to trace
the haloes of the galaxies that produced 
the cosmic reionization \citep[e.g.][]{finlator2013},
whereas \civ\ are likely tracing 
diffuse recently ionized IGM.

\section*{Acknowledgments}

We are grateful to the staff of the Subaru Telescope
for their help with the observations.
We thank the referee for her/his constructive comments.
Many thanks to Greg Poole and Vincenzo Pota for 
their valuable comments that helped to improve 
the presentation of this work.
CGD acknowledge the support of the 
funds by the Victorian Government.
YK acknowledges the support from the 
Japan Society for the Promotion of Science (JSPS) 
through JSPS research fellowships for young scientists.
ERW acknowledges the support of Australian 
Research Council grant DP1095600.
JC acknowledges support from Australian Research Council 
grant FT130101219.
Support for this work was provided by NASA through awards, 
RSA1345996 and RSA1367870, issued by JPL/Caltech.

\appendix

\section{\\Detection limits} \label{app:limitmag}

This section presents the limiting magnitude of the z' 
band images used for detection of sources, which
represents the detection limit in the \zlbg\ LBG
and \zid\ i'-dropout samples.
The sky background level dominates the error
in the magnitude of the faint sources analysed 
in this work.
To measure the background in individual exposures 
before the sky subtraction was applied, 
a map of the background was 
generated running \textsc{sextractor} 
in each single-exposure frame and asking for
CHECKIMAGE\_TYPE = ``BACKGROUND".
Then, the average and the
standard deviation in each background frame
was estimated with \textsc{imstatistic} in \textsc{iraf}.
No significant variations were found within each exposure
but differences of 10 -- 20$\%$ are found
in the mean number of background counts among 
individual exposures that 
form a single science frame.
The median of the distribution of the mean values 
was adopted as
the background level in the corresponding science frame. 
This source of noise was included in the corrected error
\begin{equation}
f^c_{\rm err}{=}\sqrt{FLUX\_ERR^2+\pi\left(\frac{FWHM}{2}\right)^2
\frac{BACKGROUND}{GAIN}},
\label{eq:flerr}
\end{equation}
where $FLUX\_ERR$ is the flux error estimated by \textsc{sextractor},
$BACKGROUND$ is the median background level in ADUs, 
$GAIN$ is the average detector gain ($GAIN=3.32$) and $FWHM$
is the full with half maximum of the source.
The corrected flux error is used to estimate photometric
errors  
\begin{equation}
MAG_{\rm err}{=} 2.5 \log(1+\frac{f^c_{\rm err}}{FLUX}),
\label{eq:merr}
\end{equation}
where $FLUX$ is the value measured by \textsc{sextractor}.

\begin{figure}
\includegraphics[width=84mm]{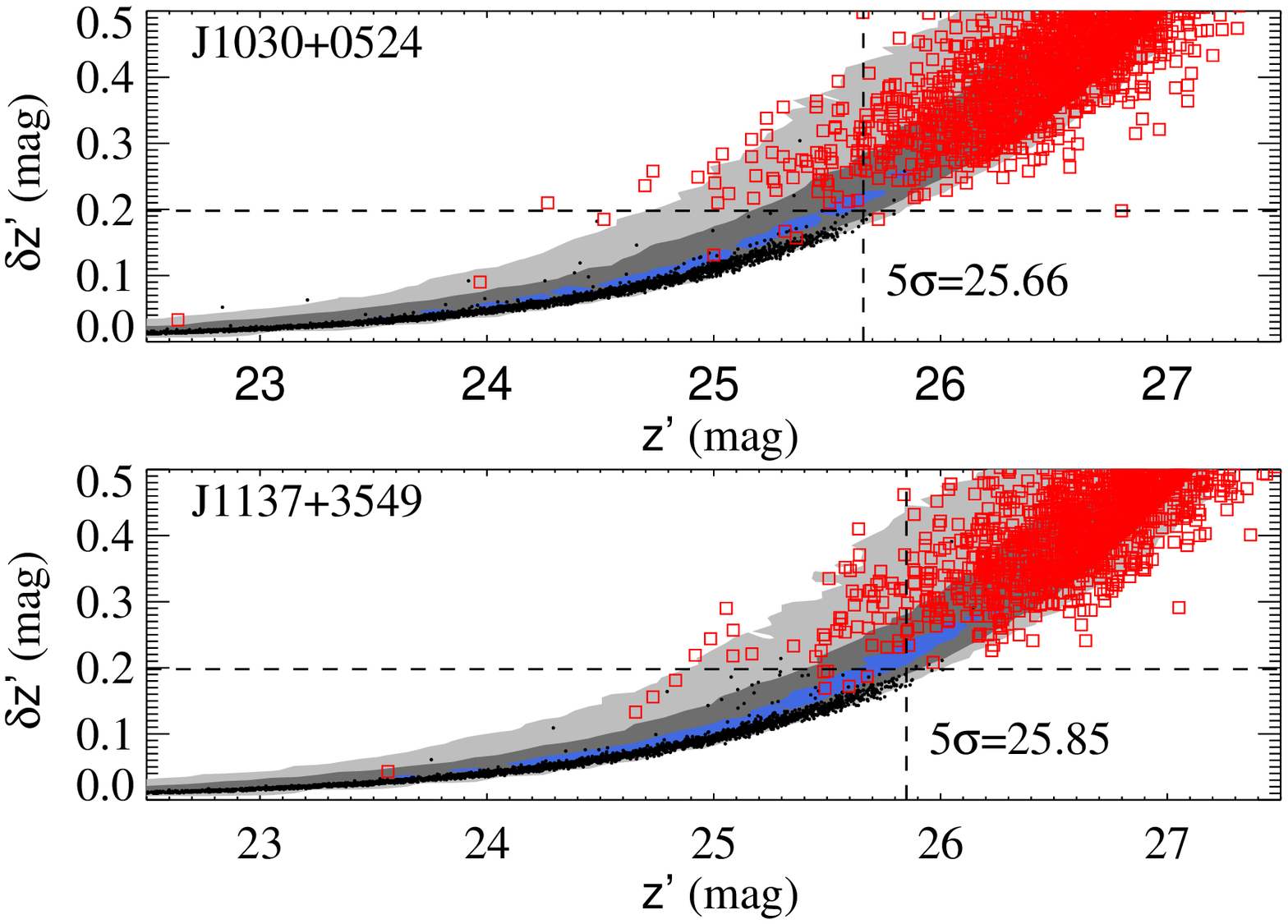}
 \caption{ Magnitude - Error relation
 from the z' band in the fields J1030+0524 ({\it top})
 and J1137+3549 ({\it bottom}).
 The grey, dark grey and blue contours
 indicate more than 20, 100 and 250 sources
 respectively. 
 Black dots are point sources and red squares are 
 detections in the negative image.
The horizontal dotted line indicates the error of a $5\sigma$
detection, i.e. 0.198 mag, and the vertical dotted line indicates the
$5\sigma$ limit in Table \ref{t:mlim}, showing that
an error-limited sample is cleaner than a magnitude-limited
sample.}
  \label{f:magera}
\end{figure}

One approach to estimate the $5\sigma$-limiting magnitude is 
by searching for the faintest object detected with
a S/N = 5, which is a $5\sigma$-detection.
This is equivalent to searching
for the faintest point 
source with an error of $0.198$ magnitudes.
Figure \ref{f:magera} 
shows the z' band magnitude of each detected
source plotted against the corresponding magnitude error.
The grey area contains more than 20 objects per bin,
the dark grey area contains more than 100 objects per bin, and
the blue area contains more than 250 objects per bin.
Point sources are plotted as black points and 
detections in the negative image obtained by multiplying the 
original science image by $-1$, are red squares.
The horizontal line indicates
an error of 0.198 magnitudes.
We find that the faintest point source 
in the J1030+0524 field with a magnitude error close to 0.198 mag has
a magnitude z' = 25.79 mag and a magnitude z' = 26.01 
 mag in the J1137+3549 field. 
They both are in agreement with 
our previous estimation of the $5\sigma$-limiting magnitude
using random apertures (Table \ref{t:mlim}) which are
indicated by the vertical lines.

\begin{figure}
\includegraphics[width=84mm]{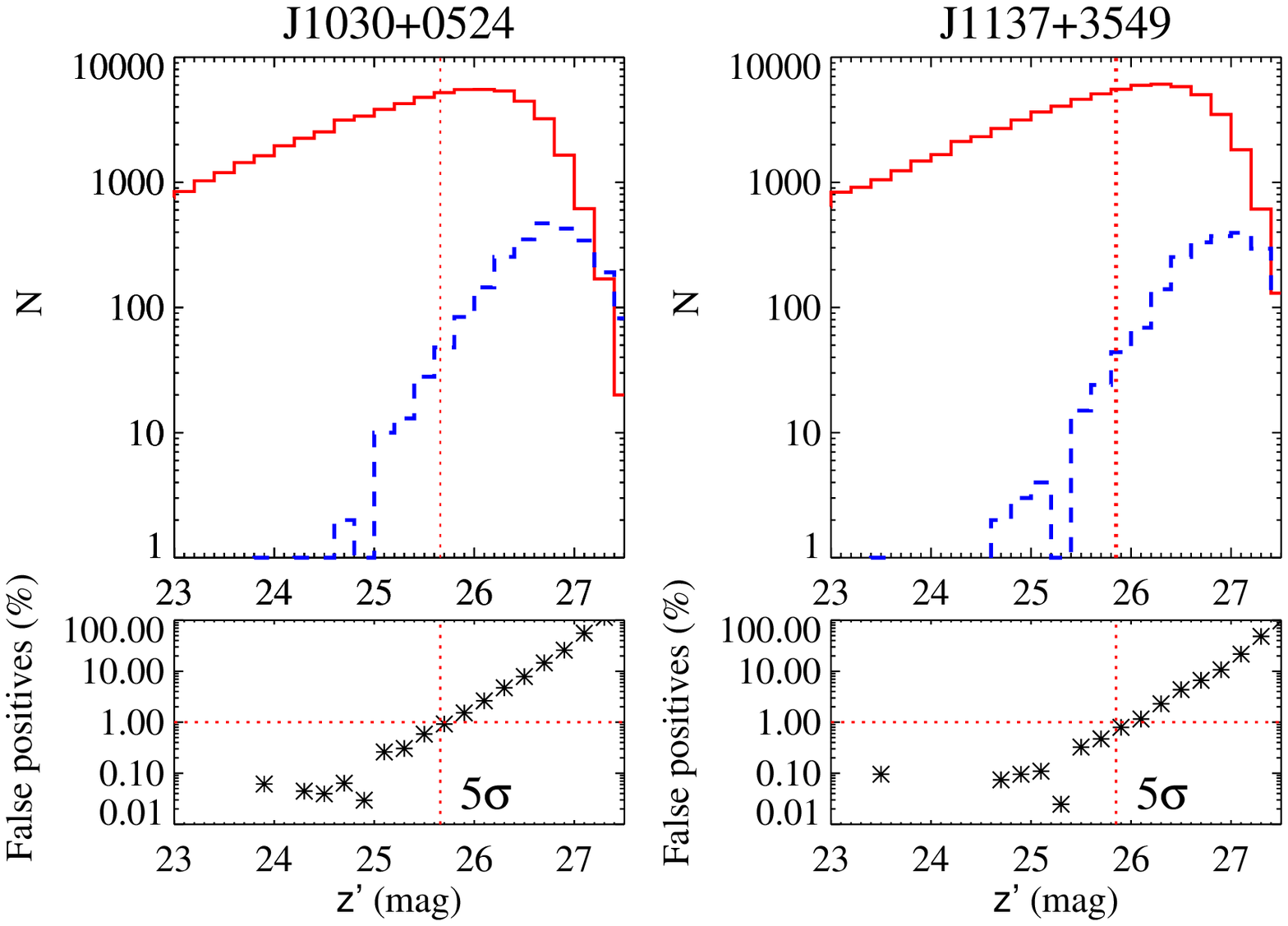}
 \caption{  {\it Top:} Distribution of $z'$-band magnitudes 
 of sources in the field J1030+0524 ({\it left})
 and J1137+3549 ({\it right}). The red solid line 
 corresponds to the positive image and the blue dashed
 line corresponds to the negative image. The vertical dotted line is the 
$5\sigma$-limit from Table \ref{t:mlim}.
{\it Bottom:} Percentage of false detection per magnitude bin
obtained as the ratio of detections in the negative image over 
detections in the positive image. The horizontal dotted line
indicates the 1\% contamination level.
In both cases, a magnitude limited sample ($m_{z'} \leq m_{5\sigma}$)
has less than 1$\%$ of contamination from false positive detections.}
  \label{f:zhista}
\end{figure}

The histogram of magnitudes provides 
another rough estimate of the limiting 
magnitude from the turning point
in which the number of objects starts to decrease. 
Figure \ref{f:zhista} 
shows the histogram of 
z' band magnitudes in the fields of interest. 
The $5\sigma$-limit seems to be in the range 25.8 -- 26.0 mag
for the field J1030+0524 and in the range 26.0 -- 26.2 mag
for the field J1137+3549, which is equivalent to or
slightly fainter than our previous estimates 
(Figure \ref{f:zhista}, vertical dotted line).
Also plotted in Figure \ref{f:zhista} 
is the histogram of magnitudes obtained from the corresponding 
negative image (dashed line).
Since the median background per exposure was subtracted during 
the reduction to bring the sky-level to zero, 
every detection in the negative image results from fluctuations 
in the background counts or from bad pixels. 
Therefore, the number of detections in negative images 
provide a true measurement of the number of false-positive 
detections in the original images. 
The bottom plot in Figure \ref{f:zhista} 
presents the fraction of false-positive detections 
obtained as the ratio between the number of detections 
in the original image and the number of detections in the 
negative image, per magnitude bin.
The horizontal line shows that in both cases 
the fraction of false detection is less than $1\%$ 
for magnitudes brighter than $m_{5\sigma}$ 
(vertical line). 
The contamination from false-positive detections 
is addressed in Appendix \ref{app:contamination}.

\section{\\ Simulated observations of LBGs} \label{app:simulationobs}

Section \ref{s:sel-lbg} shows that,  for this study, 
the influence of the UV spectral slope on
the colours of galaxies is negligible.
We also show that the \Lya\ forest attenuation 
and \Lya\ emission can significantly 
affect the colours of galaxies
in the redshift range $z{=}5.5$--5.9, using
the filters in this work.
Therefore, we adopt a value $\beta$ = -2.0 for the 
LBG templates (see Section \ref{s:templates} for 
a description of the templates) 
and we simulate the observations of objects
with \Lya\ equivalent widths from -10\AA\ to 200\AA\
using 10\AA\ steps.
Each spectrum was redshifted 
from $z =$ 4 to 6.5 using a step $\Delta z=$ 0.01.
At each step, the spectra was scaled to simulate
M$_{UV}$= -24.0 -- -19.5 mag using a step $\Delta $M$_{UV}{=}$ 0.1 mag.
Then, magnitudes in the three filters R$_c$, i', and z'
were computed using the throughput function of each filter,
including filter transmission, atmospheric transmission 
at airmass = 1.2, quantum efficiency of the detector, 
transmission of the primary focus unit and primary focus reflectivity.

The typical size of bright LBGs at $z>5$, defined by the average half-light radius,
 is $r_{hl}\sim$ 0.15"  \citep[${\simlt}$ 1 kpc][]{bouwens2004, oesch2010, malhotra2012, huang2013}.
However, the FWHM of the PSF in the PSF-equalized science images 
is 0.87" (1.13") in the field J1030+0524 (J1137+3549),
and not enough to spatially resolve most of these objects.
Therefore, we use point sources to simulate the observations.
For each realization described above, 100 point sources were added 
randomly distributed across the images, 
but avoiding superposition with real sources. 
The resulting number of simulated sources was $11\,700$ 
at each step of \Lya\ equivalent width.
After applying the same extraction process and calibration
as to the real objects, observed magnitudes were obtained
and  used to define an optimal selection of targets 
based on broad-band colours (see Section \ref{s:selcriteria}).

\begin{figure}
\includegraphics[width=84mm]{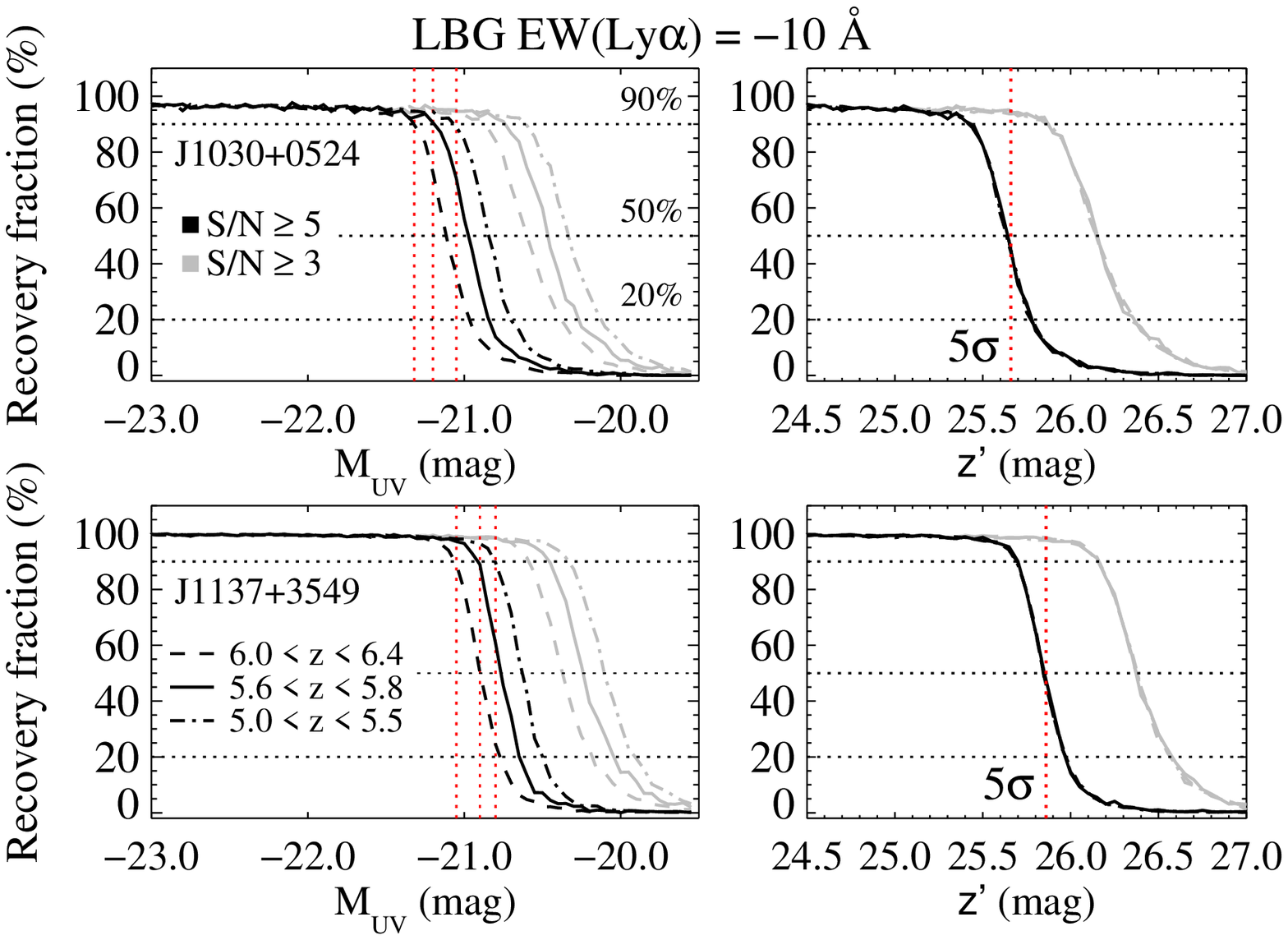}
\caption{Fraction of recovered artificial LBGs
as a function of absolute UV magnitude ($\lambda_{rest} {\sim}1350$\AA)
({\it left}) and apparent magnitude z' ({\it right}). 
The black lines correspond to 
a sample detected with S/N${\geq}5$ and 
the grey lines correspond to S/N${\geq}3$.
The dashed, solid and dot-dashed lines are the distribution
of objects at different redshift bins.
In the {\it left} panels,
the horizontal dotted lines indicate the
90\%, 50\%\ and 20\%\ levels of completeness 
and the vertical dotted lines
correspond to the UV magnitudes with 90\%\
level of completeness.
For the J1030+0524 field ({\it top}) 
the 90\%\ completeness of each redshift sample 
is M$_{UV}{=}-21.32$ (dashed), $-21.20$ (solid) and
$-21.05$ (dot-dashed), and for the J1137+3549 field ({\it bottom})
the corresponding values are
M$_{UV} = -21.05$ (dashed), $-20.9$ (solid) and
$-20.8$ (dot-dashed).
In the {\it right} panels,
the vertical dotted line
is the $5\sigma$-limiting magnitude from Table \ref{t:mlim}.
The recovery fraction of the sample with 
S/N $\geq$ 5 is 50\%\ at
a magnitude z'$_{5\sigma}$,
which means that a $5\sigma$-limited 
sample recovers 50\%\
of the objects with magnitudes z' = z'$_{5\sigma}$.} 
\label{f:compl}
\end{figure}

\subsection{Recovered objects and detection completeness} \label{s:recoveredobs}

This section explores the limits imposed by the 
quality of the data. First, we present the precision
of the magnitudes recovered from the simulations.
Second, we show the range in redshift and 
M$_{UV}$ that is sampled with the observational
data.

We compare the distribution of the residual 
magnitude measured in a 2" aperture $m_{input} - m_{recovered}$, 
in each of the three photometric bands in the field J1030+0524,
for objects in the redshift range $z = 5.2$ -- 6.2 
(i.e. $\langle z\rangle = 5.7, \Delta z = 1$). 
Considering the model with \Lya\
in absorption, 
90\%\ of the objects with S/N $\geq 5$ are 
recovered with a precision of ${\pm}0.22$ magnitudes
in the R$_c$ band, 
$\pm$0.14 in the i' band
and $\pm$0.06 magnitudes in the z' band.
Since the input magnitudes were calculated after calibrating 
the z' band magnitude, the spread in i' and R$_c$
is dominated by the higher end of the redshift 
range considered ($5.2{<}z{<}6.2$),  
where \Lya\ has been
redshifted to the z' band and the \Lya\ forest and 
Lyman break have been redshifted to 
the i' band and R$_c$ band respectively.
In contrast, the spread in z' is imposed by the 
quality of the data.

Figure \ref{f:compl} presents the fraction of
recovered objects per M$_{UV}$ bin and per
z' band magnitude bin as measured in an
automatic aperture (MAG\_AUTO) from the best seeing images.
Black and grey lines correspond to $5\sigma$ and $3\sigma$ 
detections respectively, of the model with \Lya\ in absorption.
Three redshift ranges are plotted for comparison.
In Figure \ref{f:compl} more than 90\% of the galaxies 
with M$_{UV} \leq -21.35$ mag (${-}21.05$ mag) are 
detected in the J1030+0524 (J1137+3549) field with S/N $\geq5$, 
at the three redshift bins.
If objects with a S/N $\geq3$ are included, the 90\% recovery
limit moves to M$_{UV} = -20.9$ mag (${-}20.60$ mag), i.e. half a magnitude fainter.
The three redshift ranges have different depths because the
limit is imposed by the apparent magnitude in the z' band,
which correspond to brighter M$_{UV}$ for higher redshift.
This effect is shown in the right panels of Figure \ref{f:compl} where the
three redshift ranges show the same distribution
of the recovery fraction as a function of z' magnitude.
In this plot, the vertical red dotted line indicates
the $m_{5\sigma}$ limiting magnitude estimated in 
section \ref{s:limitmag} (Table \ref{t:mlim}).
The figure shows that z'$_{5\sigma}$ 
represents the magnitude at which the fraction of
objects with S/N $\geq5$ drops to 50\%. 
If detections with S/N $\geq3$ are included, the fraction
of recovered objects remains $\geq$ 90\% 
for magnitudes z' $\geq$ z'$_{5\sigma}$. 
However, in the Section \ref{s:selcriteria}
only objects with S/N $\geq5$ are used
to define the colour selection.

\begin{figure}
\includegraphics[width=84mm]{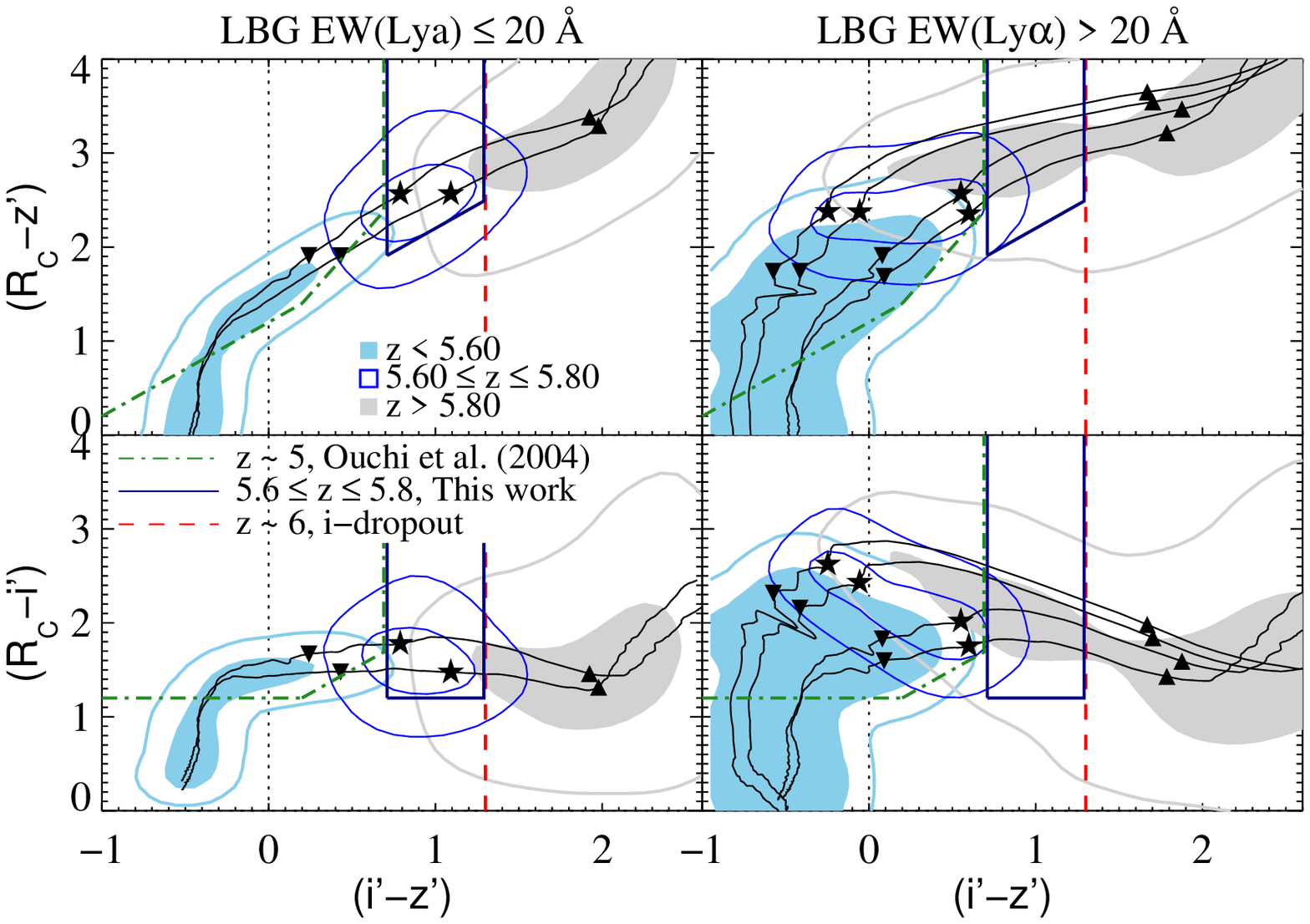}
\caption{
Colour-colour diagrams of recovered LBGs.
Left panels are LBGs with EW(\Lya) $\leq 20$\AA\
and right panels are LBGs with EW(\Lya) $>20$\AA.
The contours show the distribution of
the colours according to the redshift of the
input source. Light blue filled contours show $z{<}5.68$,
grey filled contours show $z{>}5.78$ and blue line
contours show the range $5.68 \leq z \leq 5.78$.
Colour-colour space selection criteria 
are delimited:
$z{\sim}6$ i'-dropouts (red dashed lines),
$z{\sim}5$ Ri'z'-LBGs (green dot-dashed lines)
and $z{\sim}5.7$ LBGs of this work (blue solid lines).
The black tracks correspond to 
the LBG templates used to generate input 
fake sources. The tracks in left panels have
EW(\Lya) = -10 and 20\AA\
and the tracks in the right panels have 
EW(\Lya) = 30, 50, 150 and 200\AA.
The plots show that among galaxies with 
EW(\Lya) ${\leq}20$\AA\ (left panels)
those in the narrow redshift range of 
interest (blue line contours)
are expected to dominate the colour window 
adopted for this work (blue solid lines) whereas
the number of galaxies at lower/higher redshift 
in this window is expected to be low.
Moreover, for galaxies with 
EW(\Lya) ${>}20$\AA\ (right panel)
the number of objects in the redshift range
of interest (blue line contours) found in the 
colour window drops significantly. 
Thus, the colour window shown by the solid lines 
can select a sample of galaxies
with EW(\Lya) ${\leq}20$\AA , thus
complementary to $z{\sim}5.7$ LAEs 
which by definition have EW(\Lya) ${>} 20$\AA.}
\label{f:color-sel}
\end{figure}

\begin{figure}
\includegraphics[width=84mm]{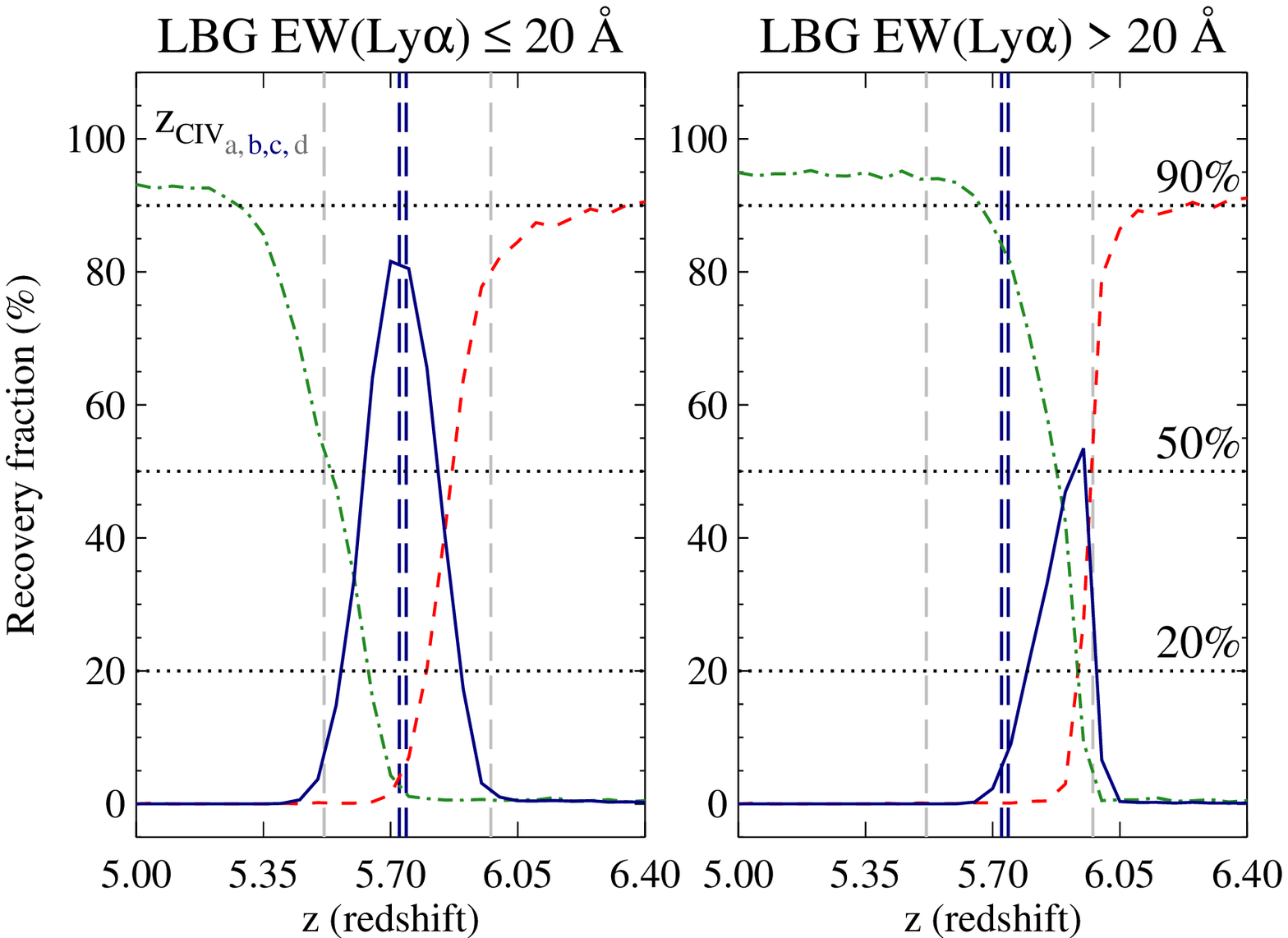}
\caption{ 
Fraction of recovered artificial LBGs
as a function of redshift.
This figure shows that the selection criteria adopted
for this work covers the window in redshift
that is left by the selection criteria of LBGs
at $z{\sim}5$ and $z{\sim}6$.
Moreover, this redshift windows correspond to the redshift 
of the two $z{\sim}5.7$ \civ\ absorption systems in the field.
In order to consistently compare the evolution with redshift,
only objects with  M$_{UV} \leq -21.5$ are included.
Green dot-dashed, red dashed and dark blue solid curves
correspond to objects selected with the $z{\sim}5$ Ri'z'-LBG 
selection of \citet{ouchi2004a}, the $z{\sim}6$ i-dropout criteria
and the $z{\sim}5.7$ LBG selection from this work, respectively.
The left panel shows that the fraction
of LBGs with EW(\Lya) $\leq 20$\AA\
selected with the colour window described 
in Section \ref{s:selcriteria} is maximum 
at the redshift of the \civ\ systems.
The right panel shows that LBGs 
with EW(\Lya) ${>}20$\AA\ 
selected with the colour criteria
for this work lie at higher redshift ($z{\sim}5.9$).
This source of contaminants, however,
is low and can be easy to identify from spectroscopy
due to the \Lya\ emission line.}
\label{f:red-sel}
\end{figure}

Figure \ref{f:color-sel} shows the (R$_c$-z') -- (i'-z') and (R$_c$-i') -- $\text{(i'-z')}$
colour-colour diagrams of artificial LBGs with different EW(\Lya).
The left panels correspond to EW(\Lya) $\leq 20$\AA\ 
and show that the contours of the population
of LBGs at $5.68 \leq z \leq 5.78$ (blue solid lines) are
concentrated outside the standard selection criteria 
for $z{\sim}6$ i'-dropouts (red dashed line)
and outside the selection criteria for $z{\sim}5$ Ri'z'-LBGs 
from \citet{ouchi2004a} (green dot-dashed line).
On the other hand, objects at lower and higher redshift 
(light blue and grey contours respectively) are mainly 
contained within either of these two selection criteria.
The result is a window in 
colour-colour space that is contained by the limits
\begin{equation}
0.7 \leq (\rm{i'} -\rm{z'}) \leq 1.3
\label{eq:colour-selA}
\end{equation}
and
\begin{equation}
(\rm{R}_{\rm c} -\rm{z'}) \geq (\rm{i'} -\rm{z'}) +1.2.
\label{eq:colour-selB}
\end{equation}

Interestingly,
the right panels show that the population of LBGs 
with EW(\Lya) ${>}20$\AA\ at  $5.68 \leq z \leq 5.78$ 
is shifted from the mentioned colour window. In addition, the
contamination from higher redshift objects remains low.
Therefore, the window defined by equations \ref{eq:colour-selA}
and \ref{eq:colour-selB} opens the opportunity to
explore a selection criteria that targets LBGs with 
EW(\Lya) $\leq 20$\AA\ (or non-LAEs)
in the redshift of interest $z{\sim}5.7$. 
We will refer to this colour-colour window as 
$z{\sim}5.7$ LBG colour criteria.

Figure \ref{f:red-sel}
shows in better detail the gap in redshift space 
left by the i'-dropout (red dashed line) and the 
Ri'z'-LBG selection (green dot-dashed line)
in the field J1030+0524.
In the left panel, the fraction of recovered 
LBGs with EW(\Lya) $\leq 20$\AA\ using  
equations \ref{eq:colour-selA} and \ref{eq:colour-selB}
(blue solid line) reaches 80\%\ at the redshift of the 
\civ\ systems (blue long-dashed vertical lines).
Coincidentally, the minimum recovery fraction of the other
selection criteria is reached at the redshift of the \civ\ systems.
The recovery fraction is $\geq 50\%$ in the redshift range
5.627 -- 5.830 ($\Delta z \sim0.2$), which correspond
to a comoving line-of-sight length of $\sim90 {\it h}^{-1}$ comoving Mpc.
Hence, the $z{\sim}5.7$ LBG colour criteria
selects LBGs in a narrow redshift range centred in the
redshift of the \civ\ absorption systems of interest. 
Now that the colour-colour region of interest has been
identified, it is important in understanding the nature of the
contaminant sources in the colour 
selection window to explore additional constraints 
in the selection to produce a cleaner sample.
The next section presents the contamination
sources and their expected impact in the 
selection of LBGs, as well as the additional
conditions used to select high-redshift galaxies.

\section{\\Contamination in the \protect\zlbg\ LBG sample}\label{app:contamination}

The contamination in the colour selection for
$z{\sim}5.7$ LBGs (eq. \ref{eq:colour-selA} and \ref{eq:colour-selB}) 
comes from two main sources: Galactic M stars and 
elliptical galaxies in the redshift range $z{=} 1.2 - 1.8$. 
In this section we analyse the contamination expected 
from each of these sources and we present
additional conditions to select $z{\sim}5.7$ LBGs.
We close the section with the contamination from 
background fluctuations.

\begin{figure}
\includegraphics[width=84mm]{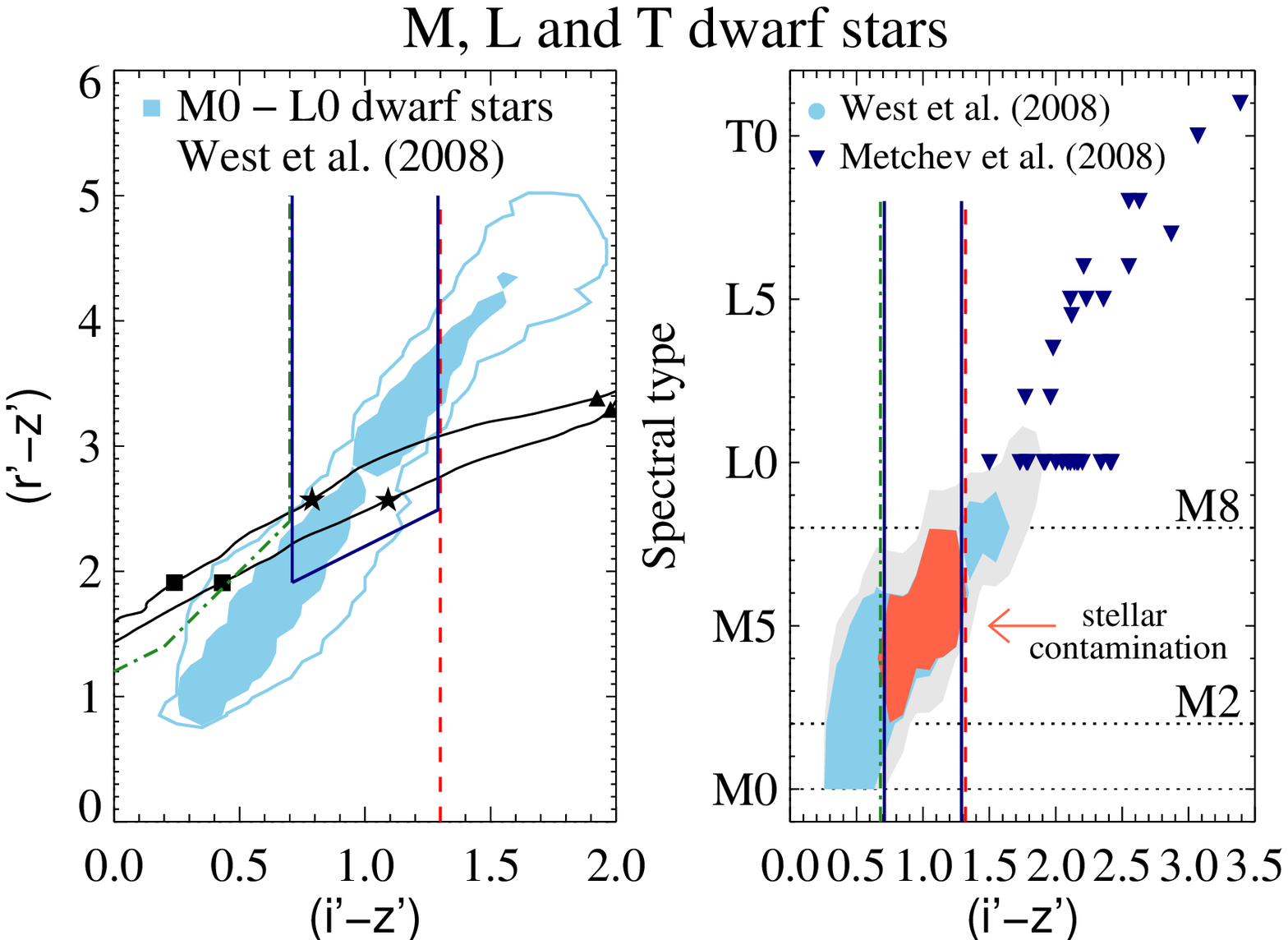} 
\caption{{\it Left:} Colour-colour diagrams 
(r'-z') -- (i'-z') and ${\text{(r'-i')}}$ -- (i'-z') for M stars
from \citet{west2008} (contours). Green dot-dashed, 
red dashed and dark blue solid lines
are the colour criteria described in Figure \ref{f:color-sel}.
This figure shows that stars cross through the window in 
colour-colour space occupied by the targeted LBG population 
(filled stars in black solid tracks, same as Figure \ref{f:color-sel}).
{\it Right:} Spectral type as a function of (i'-z'). Light blue
filled contours are M0 -- L0 stars from \citet{west2008} and
dark blue triangles are L and T dwarfs from 
\citet{metchev2008}.
All stars in the $z{\sim}5.7$ LBG colour selection 
window are indicated with red filled contours.
The stellar contamination comes from
stars in the spectral type range M2 -- M8.}
\label{f:colour-stars}
\end{figure}

\subsection{Galactic stars}\label{app:contamination-stars}
\begin{figure}
\includegraphics[width=84mm]{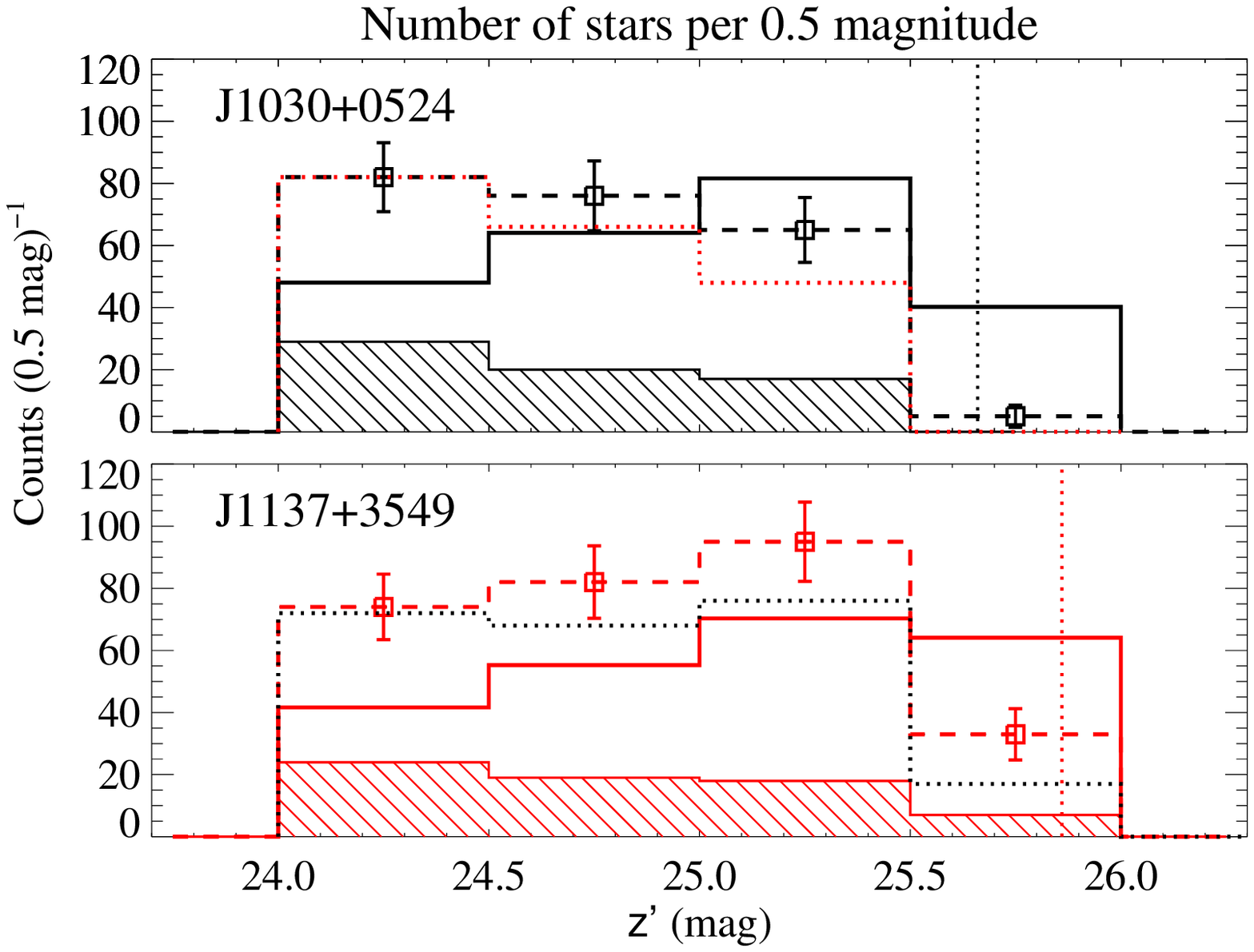} 
\caption{Number of stars per apparent magnitude 
bin estimated in each field using models for 
the distribution of stars from \citet{bochanski2010} 
and \citet{juric2008}.
The solid line histogram correspond
to the predicted number of stars corrected
for the sensitivity of our data.
The dashed line histogram is the total
number of observed sources in 
the colour selection window for $z{\sim}5.7$ LBGs.
The error bars correspond to Poisson and photometric
errors. The line-filled histogram corresponds to 
observed objects with S/G$_z'{>}0.9$
and shows a significant difference from the predicted
number of stars.
Finally, the dotted line histogram
corresponds to all the sources rejected
by our $z{\sim}5.7$ LBG selection criteria (section \ref{s:selcriteria}).
}
\label{f:num-stars}
\end{figure}

Given the colours and small sizes of $z{\sim}5$ -- 6
LBGs, stars can be an abundant source of contamination.
Fortunately, they occupy a well defined region 
in typical colour-colour diagrams used to select $z\sim6 $ 
galaxies (Figure \ref{f:colour-stars}). 
Because we are
selecting sources with colours that are typical
of late type stars, the contamination from Galactic 
low-mass stars can be high.
Figure \ref{f:colour-stars} presents the colours of 
cool low-mass stars from \citet{west2008} obtained
from the SDSS. 
They occupy a narrow stripe
that runs through the colour window
of  $z{\sim}5.7$ LBGs (blue solid line).
The black solid lines are the same
LBG tracks in the left panels of Figure \ref{f:color-sel} 
and the filled stars indicate $z{=}5.7$.
The Sloan r' band is plotted
instead of the R$_c$ band used in this work. 
However, the number of contaminants estimated
below is for the z' band magnitude.
In particular, the right panel of 
Figure \ref{f:colour-stars} presents 
(i'-z') vs. spectral type and shows that the 
stars inside the selection window (red filled contours)
have spectral types in the range M2 -- M8.
The figure also shows the position of L and T dwarfs from 
\citet{metchev2008} (inverted triangles). 
All stars colder than M8 have redder 
(i'-z') colours and, therefore, are not sources of 
contamination for the $z{\sim}5.7$ LBGs.

In order to estimate the number
of cool stars expected to contaminate our sample, 
we identify the absolute magnitude range
and distance range of the contaminant stars.
First, the absolute magnitude in the z' band
of the M stars with colours $0.7 \leq$ i'-z' $\leq 1.3$ 
was obtained from the colour-absolute magnitude
relation of \citet{bochanski2010} (Table 4 of their paper, 
but see the {\it Erratum} \citealt{bochanski2012}).
Second, from the absolute magnitudes of the contaminant 
stars we estimate the range of distances
$d=2.38$ kpc -- 22.64 kpc to meet the
apparent magnitude range of interest,
for our $z{\sim}5.7$ galaxies.
The galactic coordinates of
fields J1030+0524 and J1137+3549
are $l=239.4549$ $b= 50.0511$ and 
$l=179.4436$ $b= 71.9995$, respectively. 
Thus, both fields are above the 45 degrees 
of galactic latitude (well above the Galactic plane) 
and would include mainly 
the inner halo of the Milky Way
and the high end of the thick disc \citep{brown2004, carollo2007}.

A rough estimate of the number density of stars was
obtained from a simple model of the Galaxy.
We adopted a thin disc $+$ thick disc model 
from \citet{bochanski2010}, 
for distances smaller than 2000 pc above 
the plane of the galaxy.   
Then, for heights $Z{>}2000 $ pc, the model adopted was the
best-fit thin disc $+$ thick disc model of \citet{juric2008}
for faint and red stars ($1.3<$ (r'-i') ${<}1.4$, Table 4 of \citealt{juric2008})
with the additional halo distribution they obtained for 
a slightly bluer sample ($0.9<$ (r'-i') $<1.0$).
Finally, the density of stars in each distance bin is obtained 
from the models. The volume contained in the field of view
at each distance bin is used to estimate the total number 
of stars expected in each absolute magnitude bin. 
Then, the distance is used to convert it to apparent z' 
magnitude bin.

Figure \ref{f:num-stars} shows the average number of 
Galactic contaminant stars expected 
in each field (solid lines) per 0.5 apparent magnitude in the range of 
magnitudes of interest. 
To generate the histogram, we corrected the expected
number of stars described above
for the sensitivity in the z' band by multiplying for
the recovery fraction of objects as a function of z' (Figure \ref{f:compl}).
Hence, we obtain the equivalent of a magnitude-limited sample.
Over-plotted with dashed lines are the total number of
objects detected with S/N $\geq$ 5 (error-limited sample) 
and selected with the $z{\sim}5.7$ LBG criteria.
The dotted vertical lines indicate the 5$\sigma$ detection
limits of each field.
The comparison with this toy model shows that the 
number of detection is larger than the expected
average stellar contaminations, meaning that 
other type of sources also inhabit the colour region.
The exceptions are the faintest bins, which
show the effect of the comparison
between a magnitude-limited sample
expected from a toy model
and an error-limited observed sample like the \zlbg\ LBGs of this work.
In a magnitude-limited sample, the 5$\sigma$-limiting
magnitude is the magnitude at which 50\% of the objects are recovered
whereas in an error-limited sample, the 5$\sigma$-limiting magnitude
is the faintest average magnitude with a S/N $=5$, i.e. the magnitude
of the faintest object.

Comparing the dashed line histogram in each panel 
of Figure \ref{f:num-stars},
the number of objects selected 
in the field J1030+0524 drops towards 
the 5$\sigma$ detection limit 
whereas the opposite is observed 
in the field J1137+3549.
The reason is that the i' band image of the field J1030+0524
is 0.42 magnitudes deeper than the i' band image 
of the J1137+3549 field.
However, the {\it detection} image 
of the field J1030+0524 is 0.2 
magnitudes shallower than
that of J1137+3549
(Table \ref{t:mlim}). 
As a result, the excess of sources in 
the field J1137+3549
is dominated by photometric errors. 
Many 5$\sigma$ detections in this field have 
colours with lower signal-to-noise.

One approach to effectively minimise the number of 
Galactic stars is to simply avoid the well defined 
stellar locus in our colour-colour plots
but this would also remove a significantly large
number of $z\sim$ 5.7 galaxies.
Therefore, to minimise the contamination from stars, 
we opted for the standard approach and
included a condition in the 
star-to-galaxy (S/G) classification from \textsc{sextractor}
in the detection image: S/G$_{\rm z\textrm'}<0.9$.
This condition aims to remove objects with the highest 
probability of being a star.
However, the filled histograms in Figure \ref{f:num-stars} 
correspond to objects with S/G$_{\rm z\textrm'}\geq0.9$ and, in both fields,
is significantly lower than the predicted
number of stars (solid line histogram).
This indicates that many stars have S/G$_{\rm z\textrm'}<0.9$ and therefore
additional conditions are needed to remove further stars.
We added a restriction on the isophotal area in the R$_c$ band
(ISO\_AREA\_R$_c{<}22$ pixel$^2$)
which also reduces the contamination from low redshift red galaxies.
We discuss galaxy contamination in the following subsection.

\subsection{Red and evolved galaxies}

\begin{figure}
\includegraphics[width=84mm]{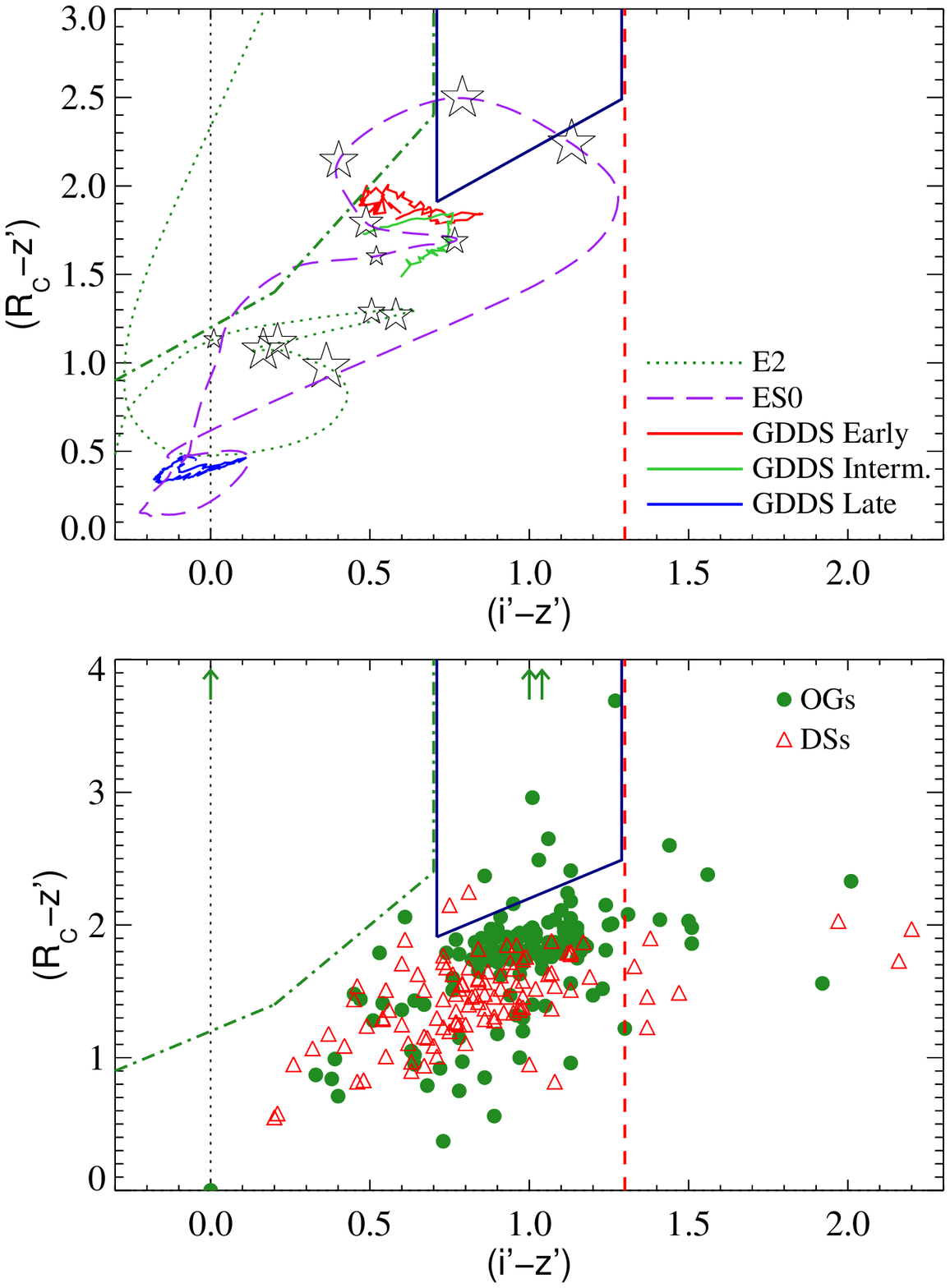}
\caption{ (R$_c$-z')--(i'-z') colour-colour diagram of red galaxies.
The green dot-dashed, 
red dashed and dark blue solid lines
are the colour criteria described in Figure \ref{f:color-sel}.
{\it Top:} 
Redshift evolution of the colours of the 
early type (red solid track), intermediate type (green solid track) 
and late type (blue solid track)
galaxy composite spectra of GDDS. 
The colours of synthetic model spectra of a typical ES0 (purple dashed track)
and an old elliptical (green dotted track) are shown for comparison.
The redshift range probed by the GDDS templates
is highlighted with stars.
Open stars from small to large indicate redshifts
$z = 0.9, 1.1, 1.3, 1.5, 1.7$ and 1.9.
The ES0 track approaches the red and green
tracks at the redshift of the galaxies used 
in the composites ($z\sim0.9$ -- 1.5).
Only early and intermediate type galaxies 
inhabit the proximity of the colour selection for $z{\sim}5.7$ LBGs.
{\it Bottom:}
Colour-colour diagram of the EROs sample.
Green solid circles are objects classified as
Old Galaxies and red open triangles
correspond to Dusty Starburst \citep{miyazaki2003}.
The contamination in the colour window
is dominated by Old Galaxies, in agreement with 
the top plot.}
\label{f:colour-ellipticals}
\end{figure}

The 4000\AA\ Balmer break present in the spectrum 
of evolved galaxies at redshift $z{=}1.05$ would be observed 
at $\lambda{\sim}8200$\AA\ and could mimic the
\Lya\ break due to \Lya\ forest in the spectrum of
a $z{\sim}5.7$ LBG.
Figure \ref{f:colour-ellipticals} presents in the top
panel the evolution with redshift of the (R$_c$-z') and (i'-z')
colours of five different templates of red and evolved galaxies.
The long-dashed and dotted tracks correspond to 
the synthetic model template spectrum of a typical ES0 
galaxy and a more evolved elliptical galaxy, respectively,
from the \citet{bruzual2003} synthesis model. 
Open stars indicate redshift values $z = 0.9, 1.1, 1.3, 1.5, 1.7$ and 1.9,
for comparison with 
the composite spectra of observed galaxies.
The red, green and blue solid tracks correspond to the early, 
intermediate and late type galaxy composite spectra 
of the Gemini Deep Deep Survey (GDDS) \citep{abraham2004}. 
All tracks were calculated using the transmission
curve of SuprimeCam filters.
The red track samples the redshift range $z{=}0.92 - 1.70$, 
the green track samples  $z{=}0.96 - 1.48$ and the blue track 
samples $z{=}1.12 - 2.15$.
A comparison of the tracks shows that GDDS early 
and intermediate templates 
are fairly approximated by the ES0 synthetic template 
(long-dashed track) at the corresponding redshift
($0.9\leq z\leq1.5$, open stars), while a
more evolved stellar population 
(dotted track) will not reach the colour tracks of observed 
early type galaxies.
Finally, the late type galaxy template 
is very distant from the window 
of interest in the colour-colour diagram.
Thus, the more likely source of extragalactic contaminants in 
the colour-colour window of interest are galaxies 
of early to intermediate type.

\begin{figure}
\includegraphics[width=84mm]{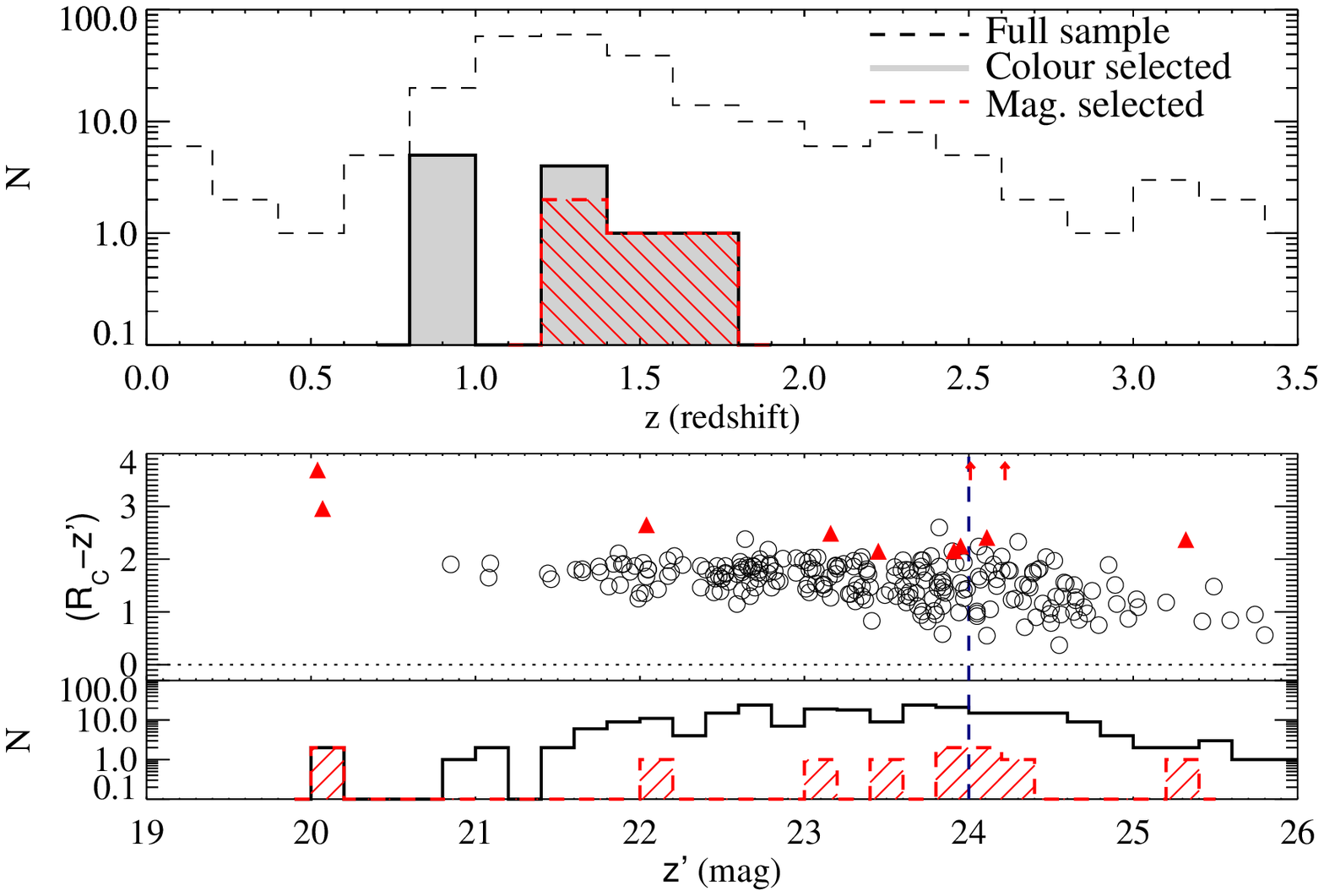}
\caption{{\it Top:}
Redshift distribution of the 
the EROs sample.
The black dashed histogram 
correspond to the
total sample of EROs from \citet{miyazaki2003},
the grey filled histogram indicates the redshift
of EROs in the colour-colour window of 
$z{\sim}5.7$ LBGs and the red filled 
histogram shows the sub-sample fainter than z'${=}24.0$.
The more likely source of contaminants seem to be in
the redshift range $1.2{<}z{<}1.8$. 
{\it Bottom:}
(R$_c$-z') -- z' colour-magnitude diagrams 
and z' band magnitude distribution of EROs. 
Contaminants of the colour selection window
are plotted with red solid triangles in the colour-magnitude
diagram and red filed histogram in the magnitude distribution
plot.
All contaminants are the reddest 
objects of the sample and only four of the eleven 
contaminant objects from the EROs sample are as 
faint as $z>5$ LBGs.}
\label{f:redshift-ellipticals}
\end{figure}

The bottom panel of Figure \ref{f:colour-ellipticals} 
presents the colour-colour diagram of 
247 Extremely Red Objects (EROs) of \citet{miyazaki2003}.
EROs classified as ``Old Galaxies" (OGs) are green filled
circles and ``Dusty Starburts" (DSs) are red open triangles.
Only eleven EROs are found inside the
colour window for $z{\sim}5.7$ LBGs and they represent
4.4\% of the total sample of \citet{miyazaki2003}.
Specifically, nine of them are classified as OGs and two of them as DSs, 
suggesting that the contamination is dominated by evolved 
galaxies rather than ``starburst-type" galaxies, in agreement 
with the result from GDDS data.

Figure \ref{f:redshift-ellipticals} shows the (R$_c$-z') vs. z' 
colour-magnitude diagram and the apparent z' magnitude 
distribution of EROs (bottom panel).
Open black circles represent the total population
and red triangles are the sources that contaminate
the colour window, which are among the reddest 
objects in the sample.
The limiting magnitude of the EROs sample is 
z'$_{lim}{=}25$ mag \citep[][]{miyazaki2003} 
and if we compare the magnitude distribution of EROs 
with the bright end of the luminosity function of  
LBGs at $z{\sim}6$ (z' ${\sim}24$ mag), we find that 
only 1.6\% EROs are red enough to contaminate the colour selection.
This is shown in the bottom panel of Figure \ref{f:redshift-ellipticals}
that presents the magnitude distribution of the EROs sample (black solid histogram) 
and the sub-sample that contaminates the colour-colour window (red filled histogram). 
Four objects are as faint as the expected high-redshift 
LBGs, all of them are classified as OGs.

The photometric redshifts of the EROs sample 
are plotted in the top panel of Figure \ref{f:redshift-ellipticals}. 
The dashed lines correspond to the
total sample and the solid line filled histogram correspond to
objects in the colour-colour window of $z{\sim}5.7$ LBGs.
The red dashed filled histogram shows the redshift of the four objects that also
meet the z' band magnitude criterion of $z{\sim}5.7$ LBGs
and suggests that the more likely source of contaminants is in
the redshift range $1.2{<}z{<}1.8$. This range corresponds
approximately to the redshift indicated by the four largest stars of the long dashed track 
in the top panel of Figure \ref{f:colour-ellipticals}.

In summary, the contamination from red galaxies 
seems to be dominated by relatively old galaxies with
spectroscopic features of early to intermediate type. 
Contamination maximises in the redshift range 
$1.2{<}z{<}1.8$ in which the observation
with broad-band photometry of the
4000\AA\ Balmer break can simulate the effect 
of the \Lya\ forest in LBGs at $z{>}5$.
Furthermore, only galaxies fainter than z' $\geq 24.0$ mag
are possible contaminants and the expected 
population of this type of galaxies at 
$1.2{<}z{<}1.8$ is not well known. 
Therefore, we can only attempt to remove
this source of contamination using a size-limited sample.

Considering the distribution of effective radius $R_e$ of
LBGs at $z>5$, which has been found smaller than 
1.5 kpc (or 0.45"), with a mean $\langle R_e\rangle{\sim}0.15$" 
\citep[e.g.][]{bouwens2006, huang2013}, we include a condition in the half-light radius
measured with \textsc{sextractor} $r_{hl} \leq 4.5''$. 
Additional conditions to remove lower redshift old galaxies are:
photometric area in the R$_c$ band $<$ 22 pixel$^2$ and 
S/G$_{\rm z\textrm'}\geq$ 0.01.
Spectroscopy follow-up 
of sources detected has been obtained and
data is being reduced. Thus, a more accurate estimate of the
contamination level of the selection criteria
will be presented in a forthcoming paper.

\begin{figure}
\includegraphics[width=84mm]{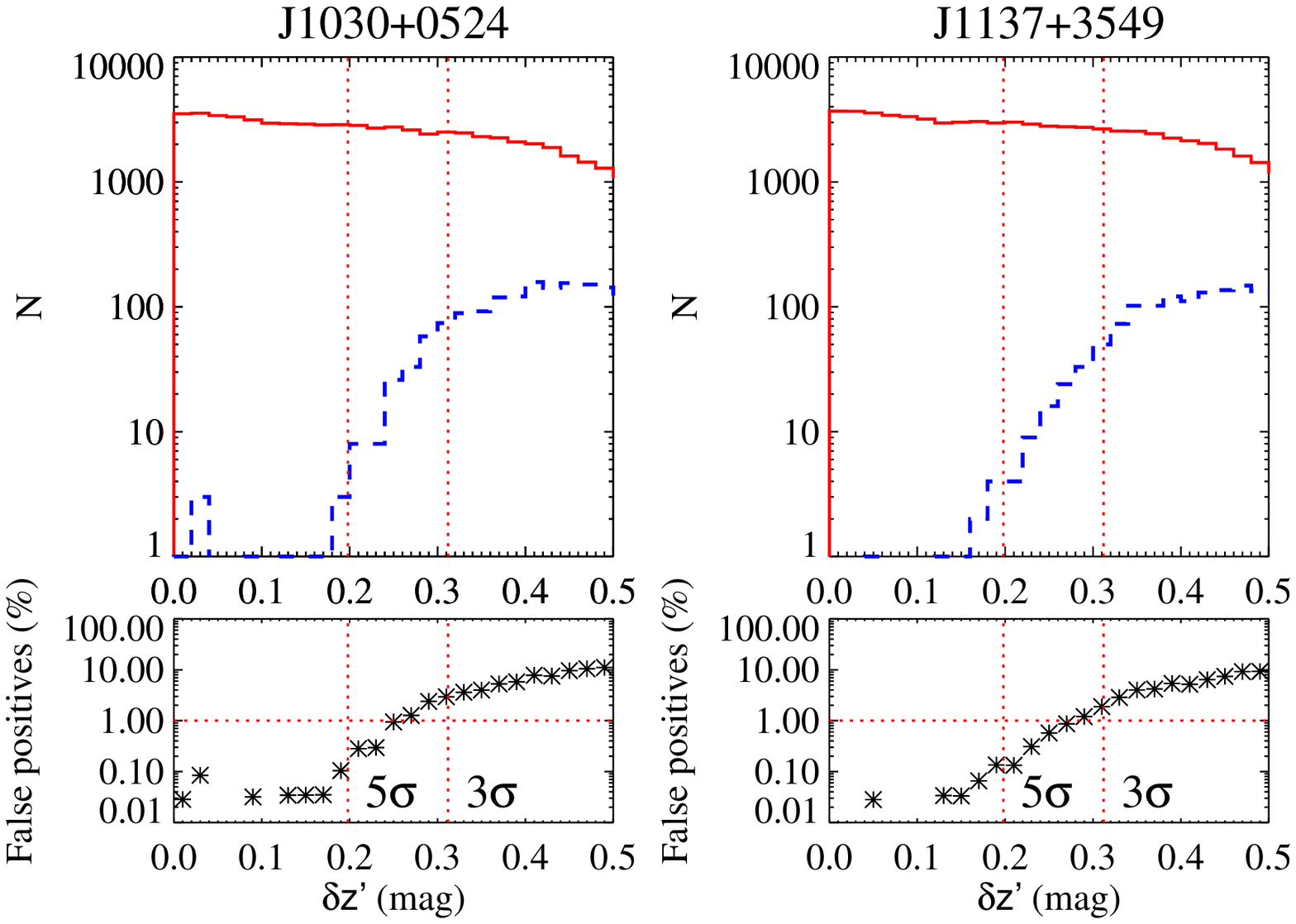}
 \caption{
 {\it Top:}  Distribution of photometric errors
 in the fields J1030+0524 ({\it left})
 and J1137+3549 ({\it right}).
The red solid histogram
corresponds to the positive image and the blue dashed histogram 
corresponds to the negative image. 
 {\it Bottom:} Percentage of false detection per error bin
obtained as the ratio of detections in the negative image 
over detections in the positive image. 
The horizontal dotted line indicates the
1\% contamination level. The vertical dotted line 
 indicates the error of a $5\sigma$ and $3\sigma$
detection, i.e. 0.198 and 0.312 mag, respectively. 
In both cases, a sample limited by S/N $\geq$ 5 ($\delta m_{z'} \leq 0.198$)
has less than 0.1$\%$ of contamination from false positive detections.
  } 
  \label{f:zhistb}
\end{figure}

\subsection{False detections from background fluctuations}

The histogram in Figure \ref{f:zhistb} 
shows the distribution of photometric errors 
in the science images (solid) and the 
negative images (dashed).
The contamination of false detections 
defined as the 
fraction of false-positives per error bin (bottom panel),
reaches $\sim$\,1\% for a detection magnitude error
of $\sim$\,0.27 magnitudes in both fields.
For sources with S/N $\geq 5$ ($5\sigma$)
the fraction is less than 0.1\%.
In other words, in a $5\sigma$ limited sample 
less than 1 detection in every 1000 sources is a noise fluctuation of the background.
Therefore, we limit our sample to sources with S/N $\geq 5$
and neglect this source of error.

\section{\\Selection Probability function }  \label{s:pmz}

The number of galaxies per apparent magnitude expected
from a given luminosity function $\Phi_{M}$ can be estimated as
\begin{equation}
N(m)=\int\limits_z\, \Phi[M(m,z)] P(m,z) \frac{dV}{dz}dz
\end{equation}
where $m$ is the apparent magnitude in the z' band, 
$P(m,z)$ is the probability of selecting an LBG at a redshift $z$
with a magnitude $m$ in our data set, $M$ is the 
absolute magnitude at 1350\AA\ and $dV/dz$
is the cosmological volume element.
In practice we calculate 
\begin{equation}
N_m=\Sigma_{k}\Phi_{k}V_{m,k}
\label{eq:counts}
\end{equation}
where $\Phi_{k}$ is the luminosity function binned in 0.1 magnitude
intervals and $V_{m,k}$ is the effective volume sampled by galaxies
with absolute magnitude $M_k$ and apparent magnitude $m$.
We adopted the luminosity function of i'-dropouts from \citet{bouwens2007}
and followed their procedure to calculate $V_{m,k}$ as
\begin{equation}
V_{m,k}{=}\int\limits_z\, \int_{m-0.05}^{m+0.05} W[M(m',z)-M_k] P(m',z)  \frac{dV}{dz}dz,
\label{eq:volumes}
\end{equation}
where
\[  W(x)= \left\{ \begin{array}{ll}
         0, & x{<}-0.05,\\
        1, & -0.05{<}x{<}0.05,\\
         0, & x{>}0.05. \end{array} \right.   
         \]      
         
\begin{figure}
\includegraphics[width=84mm]{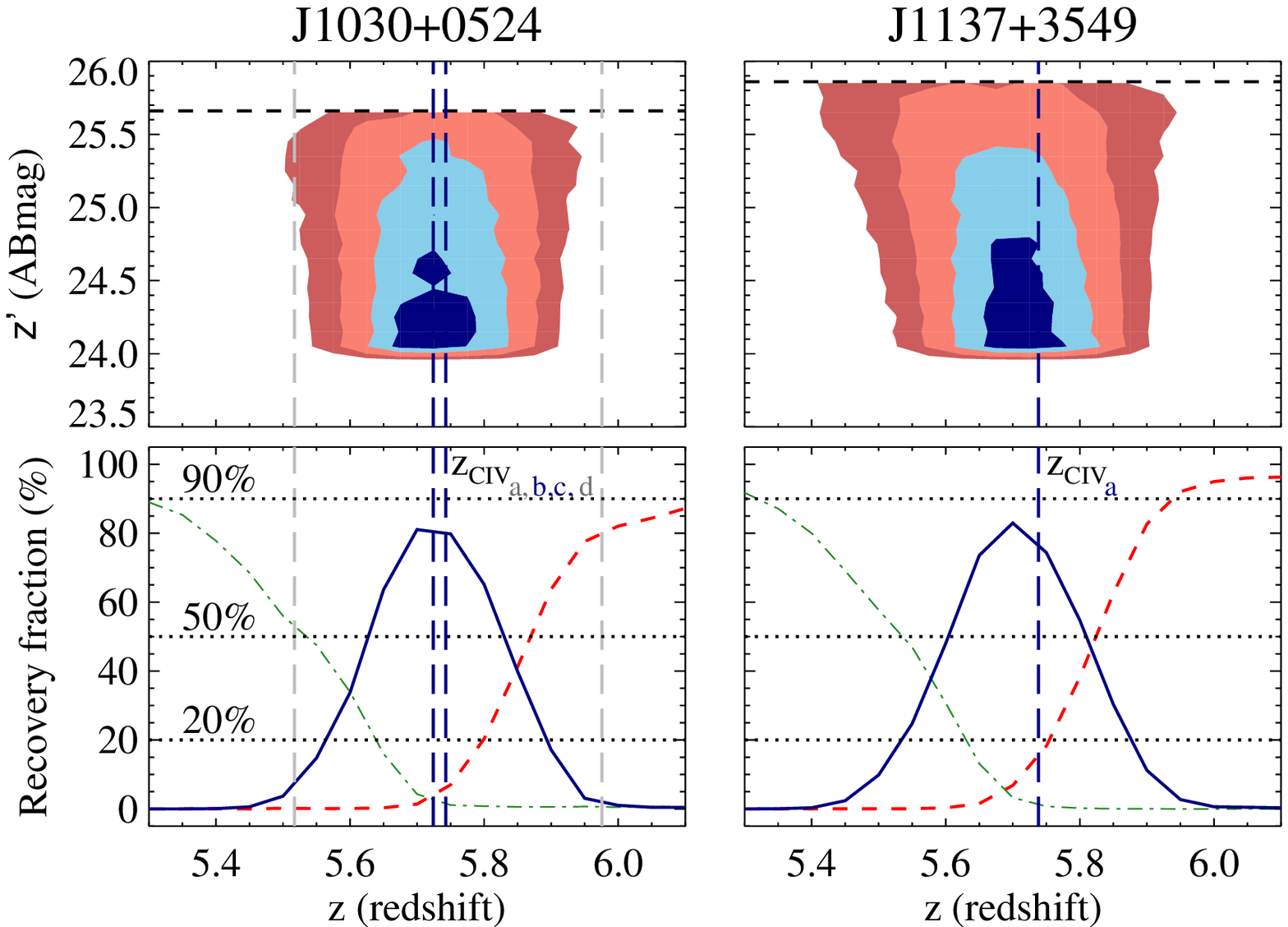} 
\caption{{\it Top:} Selection probability function 
$P(m,z)$ of the $z{\sim}5.7$ LBG selection introduced in this study.
The contours (outside-in) indicate probabilities of  
selection (including detection) of 
0.1, 0.25, 0.5 and 0.75.
{\it Bottom:}
Fraction of recovered objects as a function of redshift
obtained from the simulated observation of 10000 LBGs 
in each field and at each redshift step ($\Delta z=0.01$). 
The green dot-dashed line corresponds 
to objects recovered with the colour selection for 
R$_c$i'z'-LBGs of \citet{ouchi2004a}. 
The red dashed line corresponds to 
i'-dropout colour selected objects 
and the blue solid line corresponds to 
objects recovered with the $z{\sim}5.7$ LBG criteria.
In both fields, the blue line peaks at $z {\sim}5.7$ and the width at 
half maximum is ${\sim}0.2$. The vertical long-dashed lines 
show the redshift of the 
\civ\ systems known to date 
at $z{>}5$ in each field. In the left panel, 
the blue long-dashed lines indicate the 
redshift of two \civ\ systems at 
$z_{abs}{=}5.7242$ and 5.7428 \citep{ryan-weber2009,simcoe2011a}.
The grey long dashed lines indicates the systems at
$z_{abs}{=}5.5172 $ \citep{simcoe2011a,dodorico2013}
and $z_{abs}{=}5.9757$ \citep{dodorico2013}. 
In the right panel, the blue long-dashed vertical
line indicates the redshift of one \civ\
system at $z_{abs}{=}5.7383$ \citep{ryan-weber2009}. 
The present work focuses on objects in the 
environment of the $z_{abs} {\sim}5.7$ \civ\ 
systems only.}
\label{f:redshift-2fields}
\end{figure}

The selection  probability function $P(m,z)$ was obtained
from the simulated observations described in Appendix \ref{app:simulationobs},
using the expression
\begin{equation}
P(m,z)=\frac{N^{Recov}_{(m,z)}}{N^{Input}_{(m,z)}}
\end{equation}
where $N^{Recov}_{(m,z)}$ is the number of 
galaxies at a redshift $z$ recovered with magnitude 
$m$ and $N^{Input}_{(m,z)}$ is the number of 
input galaxies with magnitude $m$ and redshift $z$.
We recall that, since our selection aims for LBGs with EW(\Lya) $\leq$ 20\AA,
only these objects were used in the calculation.
Figure \ref{f:redshift-2fields} presents in the top panels
the $P(m,z)$ for \zlbg\ LBGs estimated in each field. 
The contours (outside-in)
correspond to probabilities of  
selection (including detection) of 0.1, 0.25, 0.5 and 0.75, respectively.
The bottom panels show the recovery fraction of
artificial LBGs as a function of redshift.
The dot-dashed line correspond to objects selected 
with the $z{\sim}5$ Ri'z'-LBG selection of \citet{ouchi2004a}, 
the solid line to the $z{\sim}5.7$ LBG selection from
this work and the dashed line to the $z{\sim}6$ standard i'-dropout selection.
Our selection criteria provides a ${>}$ 50\% recovery fraction in
a focused redshift window ($z\sim0.2$) around $z{\sim}5.7$
absorbers.
The conventional $z{\sim}5$ and $z{\sim}6$ LBG selection
criteria (green dot-dashed and red dashes lines in Figure \ref{f:redshift-2fields})
are clearly inadequate probes of the environment of the $z{\sim}5.7$
\civ\ systems.

\section{\\Magnitudes and colours of LBGs }\label{app:tables}
\clearpage
\begin{table*}
\begin{minipage}{170mm} 
\caption{$z{\sim}5.7$ LBGs - J1030+0524}
\label{t:J1030_LBGs}
\begin{tabular}{rrrrrrrrr}
\hline
  \multicolumn{1}{c}{R.A. (J2000)} &
  \multicolumn{1}{c}{DEC. (J2000)} &
  \multicolumn{1}{c}{R$_c$ (2")} &
  \multicolumn{1}{c}{i' (2")} &
  \multicolumn{1}{c}{z' (2")} &
  \multicolumn{1}{c}{NB$_{\civ}$ (2")} &
  \multicolumn{1}{c}{z' (MAG\_AUTO)} &
  \multicolumn{1}{c}{r$_{\rm hl}$ (pixels)} &
  \multicolumn{1}{c}{S/G$_{\text z'}$} \\
\hline
  10:29:34.05 & +05:36:09.6 & $\simgt$28.35 & 26.32$\pm$0.32 & 25.25$\pm$0.32 & 25.96$\pm$0.45 & 25.38$\pm$0.16 & 1.99 & 0.82\\
  10:29:36.17 & +05:25:44.1 & 27.37$\pm$0.50& 26.05$\pm$0.26 & 25.03$\pm$0.27 & 25.94$\pm$0.45 & 25.10$\pm$0.14 & 2.06 & 0.2\\
  10:29:39.44 & +05:27:38.8 & 27.26$\pm$0.46& 26.04$\pm$0.26 & 24.89$\pm$0.24 & 25.48$\pm$0.31 & 24.97$\pm$0.11 & 2.08 & 0.81\\
  10:29:44.06 & +05:14:27.8 & 27.56$\pm$0.58& 26.35$\pm$0.33 & 25.40$\pm$0.36 & 25.84$\pm$0.41 & 25.47$\pm$0.19 & 2.15 & 0.26\\
  10:29:47.74 & +05:37:16.6 & 27.74$\pm$0.66& 26.26$\pm$0.31 & 25.25$\pm$0.32 & 26.12$\pm$0.51 & 25.29$\pm$0.16 & 2.12 & 0.47\\
  10:29:51.01 & +05:09:07.8 & $\simgt$28.35 & 26.46$\pm$0.36 & 25.38$\pm$0.36 & 26.15$\pm$0.52 & 25.49$\pm$0.19 & 1.94 & 0.6\\
  10:29:51.67 & +05:26:50.0 & 27.53$\pm$0.56& 26.24$\pm$0.30 & 25.45$\pm$0.38 & 25.84$\pm$0.41 & 25.51$\pm$0.19 & 1.99 & 0.89\\
  10:29:53.43 & +05:27:18.3 & $\simgt$28.35 & 26.63$\pm$0.41 & 25.52$\pm$0.40 & 26.50$\pm$0.67 & 25.57$\pm$0.19 & 2.05 & 0.62\\
  10:29:56.18 & +05:18:48.9 & 27.36$\pm$0.50& 25.67$\pm$0.19 & 24.62$\pm$0.19 & 25.20$\pm$0.25 & 24.64$\pm$0.09 & 2.12 & 0.81\\
  10:29:59.19 & +05:34:46.4 & 27.76$\pm$0.66& 26.26$\pm$0.31 & 25.38$\pm$0.36 & 26.04$\pm$0.48 & 25.38$\pm$0.17 & 2.11 & 0.88\\
  10:30:06.72 & +05:16:48.8 & 27.26$\pm$0.46& 25.58$\pm$0.18 & 24.78$\pm$0.22 & 25.39$\pm$0.29 & 24.84$\pm$0.11 & 2.14 & 0.41\\
  10:30:06.88 & +05:19:22.3 & 27.13$\pm$0.42& 25.68$\pm$0.19 & 24.82$\pm$0.23 & 24.91$\pm$0.20 & 24.94$\pm$0.12 & 1.95 & 0.34\\
  10:30:06.91 & +05:39:05.6 & $\simgt$28.35 & 26.33$\pm$0.32 & 25.36$\pm$0.35 & 25.69$\pm$0.37 & 25.49$\pm$0.18 & 2.05 & 0.76\\
  10:30:07.75 & +05:20:44.0 & 27.77$\pm$0.67& 26.32$\pm$0.32 & 25.37$\pm$0.35 & $\simgt$27.49 & 25.49$\pm$0.18 & 1.96 & 0.61\\
  10:30:12.71 & +05:19:07.8 & 27.57$\pm$0.58& 26.14$\pm$0.28 & 25.15$\pm$0.30 & 26.24$\pm$0.56 & 25.15$\pm$0.15 & 2.22 & 0.39\\
  10:30:13.79 & +05:15:00.6 & $\simgt$28.35 & 26.48$\pm$0.37 & 25.47$\pm$0.38 & $\simgt$27.49 & 25.53$\pm$0.19 & 2.07 & 0.89\\
  10:30:16.30 & +05:32:36.6 & 27.85$\pm$0.70& 26.10$\pm$0.27 & 25.02$\pm$0.27 & 25.88$\pm$0.43 & 25.08$\pm$0.13 & 2.09 & 0.59\\
  10:30:18.93 & +05:18:33.9 & 27.73$\pm$0.65& 26.21$\pm$0.29 & 25.46$\pm$0.38 & 25.92$\pm$0.44 & 25.53$\pm$0.19 & 2.07 & 0.33\\
  10:30:19.80 & +05:17:53.0 & 27.44$\pm$0.53& 26.03$\pm$0.25 & 24.79$\pm$0.22 & 25.51$\pm$0.32 & 24.86$\pm$0.11 & 2.03 & 0.24\\
  10:30:21.45 & +05:36:49.9 & $\simgt$28.35 & 25.94$\pm$0.24 & 24.92$\pm$0.25 & 24.53$\pm$0.14 & 24.93$\pm$0.13 & 2.21 & 0.55\\
  10:30:35.45 & +05:24:20.5 & 27.60$\pm$0.59& 26.10$\pm$0.27 & 25.29$\pm$0.33 & 26.45$\pm$0.65 & 25.34$\pm$0.15 & 1.96 & 0.81\\
  10:30:41.26 & +05:17:53.6 & 27.86$\pm$0.71& 26.17$\pm$0.29 & 25.21$\pm$0.31 & 26.12$\pm$0.51 & 25.25$\pm$0.16 & 2.1 & 0.67\\
  10:30:46.12 & +05:11:23.0 & $\simgt$28.35 & 26.10$\pm$0.27 & 24.93$\pm$0.25 & 25.35$\pm$0.28 & 24.99$\pm$0.16 & 2.18 & 0.86\\
  10:30:47.84 & +05:11:58.3 & 27.89$\pm$0.72& 25.87$\pm$0.22 & 24.57$\pm$0.18 & 25.13$\pm$0.24 & 24.61$\pm$0.09 & 2.16 & 0.44\\
  10:30:48.69 & +05:26:37.3 & 27.39$\pm$0.51& 25.87$\pm$0.22 & 24.88$\pm$0.24 & 25.80$\pm$0.40 & 24.94$\pm$0.12 & 2.15 & 0.62\\
  10:30:50.36 & +05:35:06.0 & 27.40$\pm$0.51& 25.90$\pm$0.23 & 24.85$\pm$0.23 & 25.83$\pm$0.41 & 24.95$\pm$0.14 & 2.22 & 0.09\\
  10:30:52.61 & +05:08:05.9 & $\simgt$28.35 & 26.38$\pm$0.34 & 25.39$\pm$0.36 & 26.06$\pm$0.49 & 25.38$\pm$0.17 & 2.23 & 0.38\\
  10:30:53.95 & +05:32:32.9 & $\simgt$28.35 & 26.02$\pm$0.25 & 25.18$\pm$0.30 & 26.00$\pm$0.47 & 25.24$\pm$0.16 & 2.18 & 0.4\\
  10:31:02.36 & +05:33:08.1 & 27.72$\pm$0.65& 26.08$\pm$0.26 & 25.25$\pm$0.32 & 25.55$\pm$0.33 & 25.31$\pm$0.17 & 2.11 & 0.27\\
  10:31:02.97 & +05:14:21.0 & $\simgt$28.35 & 25.95$\pm$0.24 & 25.14$\pm$0.29 & 26.27$\pm$0.57 & 25.24$\pm$0.16 & 2.11 & 0.55\\
  10:31:10.51 & +05:25:52.9 & 27.18$\pm$0.44& 25.94$\pm$0.24 & 24.78$\pm$0.22 & 25.57$\pm$0.34 & 24.83$\pm$0.11 & 2.2 & 0.17\\
  10:31:11.22 & +05:14:43.7 & 26.99$\pm$0.38& 25.43$\pm$0.15 & 24.70$\pm$0.20 & 25.18$\pm$0.24 & 24.76$\pm$0.11 & 2.16 & 0.53\\
  10:31:17.39 & +05:17:32.2 & 27.61$\pm$0.60& 26.05$\pm$0.26 & 24.99$\pm$0.26 & 26.24$\pm$0.56 & 25.07$\pm$0.13 & 2.05 & 0.88\\
\hline\end{tabular}
\end{minipage}
\end{table*}

\begin{table*}
\begin{minipage}{170mm} 
\caption{$z{\sim}5.7$ LBGs - J1137+3549}
\label{t:J1137_LBGs}
\begin{tabular}{rrrrrrrrr}
\hline
  \multicolumn{1}{c}{R.A. (J2000)} &
  \multicolumn{1}{c}{DEC. (J2000)} &
  \multicolumn{1}{c}{R$_c$ (2.4")} &
  \multicolumn{1}{c}{i' (2.4")} &
  \multicolumn{1}{c}{z' (2.4")} &
  \multicolumn{1}{c}{NB$_{\civ}$ (2.4")} &
  \multicolumn{1}{c}{z' (MAG\_AUTO)} &
  \multicolumn{1}{c}{r$_{\rm hl}$ (pixels)} &
  \multicolumn{1}{c}{S/G$_{\rm z\textrm'}$} \\
\hline
  11:36:04.48 & +35:42:50.3 & $\simgt$28.00 & 26.54$\pm$0.45 & 25.62$\pm$0.45 & $\simgt$27.39 & 25.65$\pm$0.18 & 1.92 & 0.37\\
  11:36:05.05 & +35:42:51.1 & 27.20$\pm$0.53& 25.96$\pm$0.29 & 25.24$\pm$0.34 & 25.75$\pm$0.44& 25.26$\pm$0.15 & 2.2 & 0.18\\
  11:36:09.38 & +35:52:08.7 & $\simgt$28.00 & 25.68$\pm$0.23 & 24.75$\pm$0.23 & 25.16$\pm$0.28& 24.69$\pm$0.08 & 2.14 & 0.44\\
  11:36:10.19 & +35:55:25.0 & 26.83$\pm$0.41& 25.60$\pm$0.21 & 24.78$\pm$0.23 & 24.97$\pm$0.24& 24.82$\pm$0.09 & 2.12 & 0.88\\
  11:36:10.22 & +35:45:29.8 & 27.48$\pm$0.65& 26.11$\pm$0.33 & 24.90$\pm$0.25 & 25.86$\pm$0.47& 24.95$\pm$0.1 & 2.05 & 0.81\\
  11:36:10.91 & +35:46:48.1 & $\simgt$28.00 & 26.35$\pm$0.39 & 25.42$\pm$0.39 & $\simgt$27.39 & 25.38$\pm$0.17 & 2.2 & 0.04\\
  11:36:14.96 & +35:45:05.0 & 27.08$\pm$0.49& 25.55$\pm$0.21 & 24.45$\pm$0.18 & 25.15$\pm$0.27& 24.45$\pm$0.07 & 2.23 & 0.18\\
  11:36:15.05 & +35:44:56.0 & 27.53$\pm$0.67& 26.19$\pm$0.35 & 25.27$\pm$0.34 & $\simgt$27.39 & 25.35$\pm$0.13 & 1.9 & 0.64\\
  11:36:15.71 & +35:43:46.1 & $\simgt$28.00 & 26.28$\pm$0.37 & 25.21$\pm$0.33 & 25.87$\pm$0.48& 25.43$\pm$0.16 & 1.88 & 0.79\\
  11:36:16.16 & +35:37:44.0 & 27.45$\pm$0.64& 25.88$\pm$0.27 & 24.66$\pm$0.21 & 25.28$\pm$0.30& 24.65$\pm$0.08 & 2.19 & 0.18\\
  11:36:22.50 & +35:45:04.7 & $\simgt$28.00 & 26.79$\pm$0.55 & 25.53$\pm$0.42 & $\simgt$27.39 & 25.56$\pm$0.16 & 2.05 & 0.66\\
  11:36:23.01 & +35:57:54.3 & 26.97$\pm$0.45& 25.49$\pm$0.20 & 24.69$\pm$0.22 & 25.37$\pm$0.33& 24.70$\pm$0.09 & 2.16 & 0.24\\
  11:36:23.04 & +35:56:35.0 & $\simgt$28.00 & 26.34$\pm$0.39 & 25.43$\pm$0.39 & 25.80$\pm$0.46& 25.56$\pm$0.17 & 2.07 & 0.29\\
  11:36:23.77 & +35:38:28.1 & 27.14$\pm$0.51& 25.46$\pm$0.19 & 24.47$\pm$0.18 & 25.66$\pm$0.41& 24.46$\pm$0.06 & 2.11 & 0.41\\
  11:36:27.82 & +35:58:16.8 & $\simgt$28.00 & 26.71$\pm$0.51 & 25.50$\pm$0.41 & 26.40$\pm$0.70& 25.58$\pm$0.17 & 1.89 & 0.45\\
  11:36:33.27 & +35:40:39.9 & $\simgt$28.00 & 26.05$\pm$0.31 & 25.20$\pm$0.33 & 25.58$\pm$0.39& 25.23$\pm$0.12 & 2.0 & 0.19\\
  11:36:36.40 & +35:56:28.8 & $\simgt$28.00 & 26.16$\pm$0.34 & 24.92$\pm$0.26 & 26.09$\pm$0.56& 24.91$\pm$0.1 & 2.14 & 0.15\\
  11:36:37.51 & +35:57:07.7 & $\simgt$28.00 & 26.29$\pm$0.38 & 25.09$\pm$0.30 & 25.58$\pm$0.38& 25.15$\pm$0.14 & 2.08 & 0.22\\
  11:36:39.52 & +35:52:29.5 & 27.35$\pm$0.59& 26.05$\pm$0.31 & 24.95$\pm$0.27 & 26.02$\pm$0.53& 24.94$\pm$0.1 & 2.04 & 0.54\\
  11:36:41.69 & +35:44:40.3 & 27.56$\pm$0.69& 26.01$\pm$0.30 & 24.79$\pm$0.23 & 25.57$\pm$0.38& 24.73$\pm$0.08 & 2.19 & 0.84\\
  11:36:48.09 & +35:58:20.1 & $\simgt$28.00 & 26.27$\pm$0.37 & 25.31$\pm$0.36 & 26.29$\pm$0.65& 25.38$\pm$0.14 & 1.86 & 0.61\\
  11:36:49.63 & +35:58:55.0 & $\simgt$28.00 & 26.79$\pm$0.54 & 25.77$\pm$0.50 & $\simgt$27.39 & 25.74$\pm$0.2 & 2.19 & 0.3\\
  11:36:53.00 & +35:39:14.8 & 27.05$\pm$0.48& 25.54$\pm$0.20 & 24.54$\pm$0.19 & 25.10$\pm$0.26& 24.51$\pm$0.07 & 2.23 & 0.62\\
  11:36:53.54 & +35:44:16.2 & 26.81$\pm$0.40& 25.60$\pm$0.21 & 24.73$\pm$0.22 & 25.54$\pm$0.37& 24.74$\pm$0.08 & 2.03 & 0.86\\
  11:36:54.53 & +36:02:12.8 & $\simgt$28.00 & 26.41$\pm$0.41 & 25.70$\pm$0.48 & 25.89$\pm$0.49& 25.73$\pm$0.19 & 2.01 & 0.62\\
  11:36:56.06 & +35:57:43.6 & $\simgt$28.00 & 25.95$\pm$0.29 & 25.02$\pm$0.28 & 25.95$\pm$0.51& 25.25$\pm$0.13 & 2.01 & 0.85\\
  11:36:57.16 & +35:57:46.3 & 27.44$\pm$0.63& 26.17$\pm$0.34 & 25.41$\pm$0.39 & $\simgt$27.39 & 25.44$\pm$0.14 & 1.84 & 0.75\\
  11:36:58.77 & +35:46:06.3 & $\simgt$28.00 & 26.69$\pm$0.51 & 25.75$\pm$0.50 & $\simgt$27.39 & 25.74$\pm$0.19 & 1.8 & 0.22\\
  11:37:01.18 & +35:51:19.1 & $\simgt$28.00 & 25.84$\pm$0.26 & 25.02$\pm$0.28 & 25.84$\pm$0.47& 25.05$\pm$0.1 & 2.0 & 0.53\\
  11:37:01.87 & +35:48:46.8 & $\simgt$28.00 & 26.41$\pm$0.41 & 25.58$\pm$0.44 & $\simgt$27.39 & 25.61$\pm$0.19 & 2.15 & 0.31\\
  11:37:02.15 & +36:01:29.8 & 27.61$\pm$0.71& 25.80$\pm$0.25 & 24.77$\pm$0.23 & $\simgt$27.39 & 24.75$\pm$0.09 & 2.2 & 0.34\\
  11:37:02.90 & +35:42:19.1 & 27.39$\pm$0.61& 26.17$\pm$0.34 & 25.00$\pm$0.28 & 26.29$\pm$0.65& 25.01$\pm$0.1 & 2.0 & 0.88\\
  11:37:04.59 & +35:59:40.2 & $\simgt$28.00 & 26.62$\pm$0.48 & 25.53$\pm$0.42 & $\simgt$27.39 & 25.46$\pm$0.15 & 1.96 & 0.37\\
  11:37:14.81 & +36:01:31.8 & 27.09$\pm$0.49& 25.42$\pm$0.18 & 24.39$\pm$0.17 & 25.13$\pm$0.27& 24.37$\pm$0.06 & 2.2 & 0.2\\
  11:37:15.30 & +35:58:13.5 & $\simgt$28.00 & 26.51$\pm$0.44 & 25.70$\pm$0.48 & 26.09$\pm$0.56& 25.71$\pm$0.18 & 1.95 & 0.89\\
  11:37:26.84 & +36:00:09.5 & $\simgt$28.00 & 26.16$\pm$0.34 & 25.32$\pm$0.36 & 26.35$\pm$0.68& 25.45$\pm$0.17 & 2.15 & 0.18\\
  11:37:26.86 & +36:00:13.4 & $\simgt$28.00 & 26.63$\pm$0.49 & 25.71$\pm$0.48 & $\simgt$27.39 & 25.7 $\pm$0.18 & 1.78 & 0.79\\
  11:37:27.92 & +35:56:39.4 & $\simgt$28.00 & 26.48$\pm$0.43 & 25.36$\pm$0.37 & $\simgt$27.39 & 25.37$\pm$0.13 & 1.89 & 0.8\\
  11:37:30.52 & +36:01:40.7 & $\simgt$28.00 & 26.13$\pm$0.33 & 25.27$\pm$0.35 & 26.31$\pm$0.66& 25.22$\pm$0.15 & 2.21 & 0.13\\
  11:37:32.15 & +35:50:55.8 & $\simgt$28.00 & 26.60$\pm$0.48 & 25.60$\pm$0.44 & 26.44$\pm$0.72& 25.69$\pm$0.19 & 1.92 & 0.33\\
  11:37:32.97 & +35:48:00.8 & $\simgt$28.00 & 26.55$\pm$0.46 & 25.29$\pm$0.35 & $\simgt$27.39 & 25.27$\pm$0.16 & 2.16 & 0.03\\
  11:37:34.76 & +35:58:41.1 & 27.51$\pm$0.66& 26.28$\pm$0.37 & 25.50$\pm$0.41 & $\simgt$27.39 & 25.56$\pm$0.17 & 1.96 & 0.42\\
  11:37:37.94 & +35:54:18.3 & 27.22$\pm$0.54& 25.64$\pm$0.22 & 24.73$\pm$0.22 & 25.07$\pm$0.26& 24.72$\pm$0.08 & 2.07 & 0.66\\
  11:37:39.30 & +35:43:56.3 & 27.21$\pm$0.54& 25.85$\pm$0.26 & 25.13$\pm$0.31 & 25.73$\pm$0.43& 25.14$\pm$0.12 & 2.21 & 0.24\\
  11:37:46.20 & +35:49:29.1 & $\simgt$28.00 & 26.75$\pm$0.53 & 25.47$\pm$0.40 & 26.01$\pm$0.53& 25.52$\pm$0.18 & 2.08 & 0.16\\
  11:37:51.38 & +35:51:27.0 & $\simgt$28.00 & 26.20$\pm$0.35 & 25.44$\pm$0.39 & 26.14$\pm$0.58& 25.56$\pm$0.18 & 2.03 & 0.12\\
  11:37:51.51 & +35:55:29.3 & $\simgt$28.00 & 26.72$\pm$0.52 & 25.42$\pm$0.39 & $\simgt$27.39 & 25.46$\pm$0.15 & 2.01 & 0.16\\
  11:37:51.68 & +35:49:48.6 & 27.63$\pm$0.72& 26.29$\pm$0.37 & 25.58$\pm$0.44 & $\simgt$27.39 & 25.65$\pm$0.17 & 1.82 & 0.72\\
  11:37:55.35 & +35:49:50.8 & $\simgt$28.00 & 26.38$\pm$0.40 & 25.40$\pm$0.38 & 26.46$\pm$0.74& 25.47$\pm$0.15 & 1.96 & 0.82\\
  11:37:55.36 & +35:50:40.7 & 27.29$\pm$0.57& 25.46$\pm$0.19 & 24.63$\pm$0.21 & 25.30$\pm$0.31& 24.66$\pm$0.08 & 2.22 & 0.89\\
  11:37:56.47 & +35:51:07.7 & 27.59$\pm$0.70& 26.28$\pm$0.37 & 25.14$\pm$0.31 & 26.37$\pm$0.69& 25.14$\pm$0.12 & 2.22 & 0.14\\
  11:37:57.73 & +36:01:37.3 & 27.60$\pm$0.71& 26.32$\pm$0.38 & 25.47$\pm$0.40 & 25.98$\pm$0.52& 25.49$\pm$0.16 & 1.98 & 0.38\\
  11:38:01.07 & +35:48:49.6 & $\simgt$28.00 & 26.56$\pm$0.46 & 25.66$\pm$0.46 & $\simgt$27.39 & 25.75$\pm$0.2 & 1.83 & 0.42\\
  11:38:01.10 & +35:45:03.0 & $\simgt$28.00 & 26.42$\pm$0.41 & 25.49$\pm$0.41 & 26.38$\pm$0.69& 25.61$\pm$0.19 & 2.06 & 0.03\\
  11:38:10.96 & +35:57:24.4 & $\simgt$28.00 & 26.54$\pm$0.45 & 25.39$\pm$0.38 & 25.90$\pm$0.49& 25.50$\pm$0.14 & 1.78 & 0.85\\
  11:38:12.29 & +35:45:47.3 & $\simgt$28.00 & 26.21$\pm$0.35 & 25.18$\pm$0.32 & $\simgt$27.39 & 25.28$\pm$0.14 & 2.06 & 0.57\\
  11:38:13.10 & +36:01:44.5 & $\simgt$28.00 & 25.99$\pm$0.29 & 25.28$\pm$0.35 & 26.01$\pm$0.53& 25.33$\pm$0.17 & 2.13 & 0.52\\
  11:38:14.84 & +35:56:54.5 & $\simgt$28.00 & 25.76$\pm$0.25 & 24.78$\pm$0.23 & 25.35$\pm$0.32& 24.83$\pm$0.09 & 1.88 & 0.65\\
  11:38:20.67 & +35:45:47.6 & $\simgt$28.00 & 26.57$\pm$0.46 & 25.67$\pm$0.47 & $\simgt$27.39 & 25.72$\pm$0.19 & 2.09 & 0.36\\
  11:38:24.99 & +35:57:23.0 & 27.46$\pm$0.64& 25.78$\pm$0.25 & 24.74$\pm$0.22 & 25.63$\pm$0.40& 24.76$\pm$0.08 & 1.95 & 0.74\\
  11:38:25.35 & +36:00:55.2 & $\simgt$28.00 & 25.92$\pm$0.28 & 25.06$\pm$0.29 & 25.61$\pm$0.39& 25.12$\pm$0.12 & 2.06 & 0.47\\
\hline\end{tabular}
\end{minipage}
\end{table*}

\begin{table*}
\begin{minipage}{180mm} 
\caption{$z{\sim}6.0$ i'-dropouts - J1030+0524}
\label{t:J1030_idrops}
\begin{tabular}{rrrrrrrrr}
\hline
  \multicolumn{1}{c}{R.A. (J2000)} &
  \multicolumn{1}{c}{DEC. (J2000)} &
  \multicolumn{1}{c}{R$_c$ (2")} &
  \multicolumn{1}{c}{i' (2")} &
  \multicolumn{1}{c}{z' (2")} &
  \multicolumn{1}{c}{NB$_{\civ}$ (2")} &
  \multicolumn{1}{c}{z' (MAG\_AUTO)} &
  \multicolumn{1}{c}{r$_{\rm hl}$ (pixels)} &
  \multicolumn{1}{c}{S/G$_{\rm z\textrm'}$} \\
\hline
  10:29:35.15 & +05:13:02.1 & $\simgt$28.35 & $\simgt$28.04  & 25.44$\pm$0.37 & $\simgt$27.49 & 25.51$\pm$0.19 & 2.21 & 0.86\\
  10:29:36.19 & +05:27:40.7 & 27.14$\pm$0.42& 26.91$\pm$0.51 & 25.33$\pm$0.34 & $\simgt$27.49 & 25.37$\pm$0.18 & 2.2 & 0.43\\
  10:29:36.28 & +05:31:57.4 & $\simgt$28.35 & $\simgt$28.04  & 25.33$\pm$0.34 & $\simgt$27.49 & 25.35$\pm$0.16 & 2.05 & 0.71\\
  10:29:45.84 & +05:23:25.0 & $\simgt$28.35 & 27.11$\pm$0.58 & 25.23$\pm$0.32 & 26.16$\pm$0.52& 25.30$\pm$0.18 & 2.14 & 0.1\\
  10:29:51.74 & +05:23:22.4 & $\simgt$28.35 & 26.56$\pm$0.39 & 24.87$\pm$0.24 & $\simgt$27.49 & 24.90$\pm$0.11 & 2.14 & 0.8\\
  10:29:53.47 & +05:24:06.4 & 27.28$\pm$0.47& 26.19$\pm$0.29 & 24.70$\pm$0.20 & 25.59$\pm$0.34& 24.75$\pm$0.1 & 2.11 & 0.46\\
  10:29:56.83 & +05:39:17.2 & $\simgt$28.35 & 26.64$\pm$0.41 & 25.30$\pm$0.33 & 26.46$\pm$0.65& 25.31$\pm$0.17 & 2.16 & 0.07\\
  10:30:08.49 & +05:32:45.7 & $\simgt$28.35 & $\simgt$28.04  & 25.29$\pm$0.33 & $\simgt$27.49 & 25.40$\pm$0.16 & 2.08 & 0.87\\
  10:30:09.23 & +05:09:59.4 & $\simgt$28.35 & 27.34$\pm$0.69 & 25.33$\pm$0.34 & 26.45$\pm$0.65& 25.36$\pm$0.17 & 2.2 & 0.89\\
  10:30:17.21 & +05:26:08.6 & 27.11$\pm$0.41& 26.56$\pm$0.39 & 25.13$\pm$0.29 & $\simgt$27.49 & 25.18$\pm$0.15 & 2.08 & 0.54\\
  10:30:17.75 & +05:33:51.8 & 27.26$\pm$0.46& 26.36$\pm$0.33 & 25.02$\pm$0.26 & 26.13$\pm$0.51& 25.10$\pm$0.14 & 2.21 & 0.66\\
  10:30:22.26 & +05:26:22.8 & $\simgt$28.35 & 26.61$\pm$0.41 & 25.20$\pm$0.31 & 25.94$\pm$0.45& 25.22$\pm$0.15 & 2.16 & 0.42\\
  10:30:23.64 & +05:18:44.3 & $\simgt$28.35 & 26.96$\pm$0.53 & 25.50$\pm$0.39 & 25.96$\pm$0.45& 25.61$\pm$0.19 & 1.88 & 0.8\\
  10:30:24.09 & +05:24:20.6 & 27.18$\pm$0.44& 28.04$\pm$0.92 & 25.16$\pm$0.30 & $\simgt$27.49 & 25.23$\pm$0.18 & 2.2 & 0.04\\
  10:30:24.89 & +05:08:13.0 & 27.24$\pm$0.45& 26.95$\pm$0.52 & 25.41$\pm$0.37 & $\simgt$27.49 & 25.50$\pm$0.2 & 2.14 & 0.63\\
  10:30:26.95 & +05:33:57.1 & $\simgt$28.35 & 26.75$\pm$0.45 & 25.43$\pm$0.37 & $\simgt$27.49 & 25.53$\pm$0.19 & 2.02 & 0.48\\
  10:30:29.22 & +05:07:21.9 & $\simgt$28.35 & 26.89$\pm$0.50 & 24.79$\pm$0.22 & 25.81$\pm$0.41& 24.90$\pm$0.11 & 1.95 & 0.65\\
  10:30:35.46 & +05:15:08.8 & 27.69$\pm$0.63& 26.67$\pm$0.42 & 25.30$\pm$0.33 & 25.79$\pm$0.40& 25.42$\pm$0.17 & 2.08 & 0.87\\
  10:30:37.37 & +05:11:21.6 & 27.36$\pm$0.50& 26.90$\pm$0.50 & 25.52$\pm$0.40 & 26.59$\pm$0.71& 25.59$\pm$0.2 & 2.09 & 0.59\\
  10:30:38.61 & +05:10:11.8 & 27.24$\pm$0.45& 26.78$\pm$0.46 & 25.47$\pm$0.38 & $\simgt$27.49 & 25.47$\pm$0.19 & 2.14 & 0.18\\
  10:30:54.79 & +05:34:00.0 & $\simgt$28.35 & 26.01$\pm$0.25 & 24.51$\pm$0.17 & 25.06$\pm$0.22& 24.57$\pm$0.07 & 2.02 & 0.89\\
  10:31:06.75 & +05:34:34.5 & 27.10$\pm$0.41& 26.55$\pm$0.38 & 25.14$\pm$0.29 & 25.91$\pm$0.44& 25.23$\pm$0.15 & 2.1 & 0.86\\
  10:31:14.28 & +05:16:39.4 & $\simgt$28.35 & 26.39$\pm$0.34 & 25.07$\pm$0.28 & 25.37$\pm$0.29& 25.15$\pm$0.14 & 2.03 & 0.48\\
\hline\end{tabular}
\end{minipage}
\end{table*}
\begin{table*}
\begin{minipage}{180mm} 
\caption{$z{\sim}6.0$ i'-dropouts - J1137+3549}
\label{t:J1137_idrops}
\begin{tabular}{rrrrrrrrr}
\hline
  \multicolumn{1}{c}{R.A. (J2000)} &
  \multicolumn{1}{c}{DEC. (J2000)} &
  \multicolumn{1}{c}{R$_c$ (2.4")} &
  \multicolumn{1}{c}{i' (2.4")} &
  \multicolumn{1}{c}{z' (2.4")} &
  \multicolumn{1}{c}{NB$_{\civ}$ (2.4")} &
  \multicolumn{1}{c}{z' (MAG\_AUTO)} &
  \multicolumn{1}{c}{r$_{\rm hl}$ (pixels)} &
  \multicolumn{1}{c}{S/G$_{\rm z\textrm'}$} \\
\hline
  11:35:56.83 & +36:01:36.2 & 27.02$\pm$0.47& 26.89$\pm$0.58& 25.29$\pm$0.35 & $\simgt$27.39 & 25.26$\pm$0.14 & 2.13 & 0.07\\
  11:35:57.21 & +35:52:17.9 & $\simgt$28.00 & 26.10$\pm$0.32& 24.78$\pm$0.23 & 25.27$\pm$0.30& 24.80$\pm$0.08 & 1.92 & 0.88\\
  11:35:59.87 & +35:53:38.0 & $\simgt$28.00 & 27.13$\pm$0.69& 25.21$\pm$0.33 & $\simgt$27.39 & 25.39$\pm$0.17 & 2.22 & 0.29\\
  11:36:00.11 & +36:02:56.3 & 27.29$\pm$0.57& 26.25$\pm$0.36& 24.89$\pm$0.25 & 25.87$\pm$0.48& 24.82$\pm$0.09 & 2.18 & 0.77\\
  11:36:09.70 & +36:00:09.9 & 27.11$\pm$0.5 & 26.92$\pm$0.6 & 25.60$\pm$0.44 & $\simgt$27.39 & 25.59$\pm$0.17 & 2.02 & 0.44\\
  11:36:14.72 & +35:49:10.3 & 26.98$\pm$0.45& 27.02$\pm$0.64& 25.70$\pm$0.48 & $\simgt$27.39 & 25.67$\pm$0.19 & 2.08 & 0.22\\
  11:36:15.92 & +36:02:10.0 & $\simgt$28.00 & $\simgt$27.61 & 25.82$\pm$0.52 & $\simgt$27.39 & 25.79$\pm$0.2 & 1.86 & 0.72\\
  11:36:18.30 & +35:53:42.3 & 27.20$\pm$0.54& 26.79$\pm$0.55& 25.45$\pm$0.40 & 26.41$\pm$0.70& 25.49$\pm$0.16 & 2.18 & 0.73\\
  11:36:19.77 & +35:41:44.6 & $\simgt$28.00 & 27.10$\pm$0.68& 25.46$\pm$0.40 & $\simgt$27.39 & 25.54$\pm$0.16 & 1.88 & 0.86\\
  11:36:21.47 & +35:59:09.1 & $\simgt$28.00 & 26.97$\pm$0.62& 25.29$\pm$0.35 & 26.42$\pm$0.71& 25.34$\pm$0.15 & 2.13 & 0.18\\
  11:36:23.02 & +35:56:16.7 & 27.38$\pm$0.61& 27.00$\pm$0.63& 25.61$\pm$0.45 & 26.33$\pm$0.67& 25.73$\pm$0.19 & 1.76 & 0.53\\
  11:36:23.18 & +35:45:11.7 & $\simgt$28.00 & $\simgt$27.61 & 25.53$\pm$0.42 & 26.41$\pm$0.70& 25.56$\pm$0.17 & 1.77 & 0.63\\
  11:36:24.63 & +35:38:30.1 & 26.93$\pm$0.44& $\simgt$27.61 & 25.42$\pm$0.39 & 25.90$\pm$0.49& 25.54$\pm$0.16 & 1.96 & 0.25\\
  11:36:26.71 & +35:45:07.7 & 27.40$\pm$0.62& 26.60$\pm$0.48& 25.11$\pm$0.30 & $\simgt$27.39 & 25.10$\pm$0.12 & 2.16 & 0.4\\
  11:36:27.91 & +36:02:43.2 & $\simgt$28.00 & $\simgt$27.61 & 25.22$\pm$0.33 & 25.91$\pm$0.49& 25.36$\pm$0.14 & 1.96 & 0.88\\
  11:36:28.84 & +35:55:43.7 & $\simgt$28.00 & 26.87$\pm$0.58& 25.47$\pm$0.40 & 26.15$\pm$0.59& 25.79$\pm$0.19 & 1.81 & 0.78\\
  11:36:32.68 & +35:44:52.8 & 27.60$\pm$0.71& 26.81$\pm$0.55& 25.48$\pm$0.41 & 26.05$\pm$0.55& 25.51$\pm$0.16 & 2.02 & 0.63\\
  11:36:33.03 & +35:44:08.0 & 27.55$\pm$0.68& 26.77$\pm$0.54& 25.28$\pm$0.35 & 25.76$\pm$0.44& 25.31$\pm$0.15 & 2.11 & 0.17\\
  11:36:36.98 & +36:03:20.4 & 27.43$\pm$0.63& 26.99$\pm$0.63& 25.33$\pm$0.36 & 25.93$\pm$0.50& 25.33$\pm$0.17 & 2.22 & 0.08\\
  11:36:37.51 & +35:40:00.6 & 27.49$\pm$0.66& 26.90$\pm$0.59& 25.32$\pm$0.36 & $\simgt$27.39 & 25.37$\pm$0.16 & 2.21 & 0.03\\
  11:36:39.18 & +36:01:56.0 & 27.40$\pm$0.62& 26.87$\pm$0.58& 25.36$\pm$0.37 & $\simgt$27.39 & 25.35$\pm$0.16 & 2.19 & 0.1\\
  11:36:41.70 & +35:58:03.4 & 26.98$\pm$0.45& 26.05$\pm$0.31& 24.74$\pm$0.22 & 26.17$\pm$0.60& 24.76$\pm$0.09 & 2.04 & 0.34\\
  11:36:46.01 & +35:46:56.3 & 27.01$\pm$0.46& 26.88$\pm$0.58& 25.37$\pm$0.37 & 26.22$\pm$0.62& 25.48$\pm$0.17 & 1.81 & 0.03\\
  11:36:46.39 & +35:47:33.1 & 27.65$\pm$0.73& 26.93$\pm$0.61& 25.53$\pm$0.42 & $\simgt$27.39 & 25.68$\pm$0.18 & 2.17 & 0.84\\
  11:36:46.56 & +35:46:38.2 & 27.45$\pm$0.64& 26.43$\pm$0.42& 24.91$\pm$0.26 & 26.07$\pm$0.56& 24.94$\pm$0.1 & 2.14 & 0.05\\
  11:37:04.48 & +35:56:41.4 & 27.10$\pm$0.5 & 26.04$\pm$0.31& 24.68$\pm$0.21 & 25.85$\pm$0.47& 24.69$\pm$0.08 & 2.12 & 0.32\\
  11:37:07.49 & +35:47:51.6 & 26.82$\pm$0.4 & 26.68$\pm$0.5 & 25.32$\pm$0.36 & $\simgt$27.39 & 25.55$\pm$0.16 & 2.14 & 0.79\\
  11:37:08.51 & +35:46:17.3 & $\simgt$28.00 & 26.84$\pm$0.57& 25.30$\pm$0.35 & $\simgt$27.39 & 25.33$\pm$0.14 & 2.0 & 0.26\\
  11:37:10.26 & +35:52:35.1 & $\simgt$28.00 & 27.00$\pm$0.63& 25.24$\pm$0.34 & $\simgt$27.39 & 25.34$\pm$0.14 & 1.99 & 0.23\\
  11:37:11.07 & +35:41:07.0 & $\simgt$28.00 & 27.02$\pm$0.64& 25.72$\pm$0.49 & 25.59$\pm$0.39& 25.75$\pm$0.19 & 1.89 & 0.88\\
  11:37:13.89 & +35:48:11.3 & $\simgt$28.00 & 27.20$\pm$0.73& 25.67$\pm$0.47 & $\simgt$27.39 & 25.70$\pm$0.19 & 1.93 & 0.47\\
  11:37:14.83 & +35:48:00.9 & 27.69$\pm$0.75& 27.13$\pm$0.69& 25.40$\pm$0.38 & $\simgt$27.39 & 25.42$\pm$0.18 & 2.17 & 0.05\\
  11:37:14.97 & +35:52:57.3 & $\simgt$28.00 & $\simgt$27.61 & 25.54$\pm$0.42 & $\simgt$27.39 & 25.50$\pm$0.17 & 2.19 & 0.14\\
  11:37:15.23 & +35:53:01.3 & 27.13$\pm$0.51& 26.82$\pm$0.56& 25.48$\pm$0.41 & $\simgt$27.39 & 25.46$\pm$0.16 & 2.13 & 0.51\\
  11:37:28.25 & +35:42:09.3 & $\simgt$28.00 & $\simgt$27.61 & 25.66$\pm$0.46 & $\simgt$27.39 & 25.64$\pm$0.19 & 2.06 & 0.13\\
  11:37:34.46 & +35:48:41.1 & $\simgt$28.00 & 26.38$\pm$0.4 & 24.80$\pm$0.24 & 26.44$\pm$0.72& 24.80$\pm$0.09 & 2.1 & 0.74\\
  11:37:36.27 & +36:02:52.7 & 27.05$\pm$0.48& 26.81$\pm$0.56& 25.48$\pm$0.41 & $\simgt$27.39 & 25.47$\pm$0.16 & 2.17 & 0.27\\
  11:37:43.39 & +35:45:24.5 & 27.59$\pm$0.7 & $\simgt$27.61 & 25.61$\pm$0.45 & $\simgt$27.39 & 25.64$\pm$0.18 & 2.02 & 0.47\\
  11:37:43.95 & +36:02:20.1 & $\simgt$28.00 & $\simgt$27.61 & 25.44$\pm$0.39 & 26.44$\pm$0.72& 25.40$\pm$0.14 & 2.17 & 0.6\\
  11:37:49.57 & +35:48:16.2 & 27.03$\pm$0.47& 26.65$\pm$0.49& 25.32$\pm$0.36 & $\simgt$27.39 & 25.32$\pm$0.13 & 2.04 & 0.81\\
  11:37:50.60 & +35:56:35.9 & $\simgt$28.00 & 27.14$\pm$0.7 & 25.64$\pm$0.46 & $\simgt$27.39 & 25.59$\pm$0.17 & 2.07 & 0.76\\
  11:37:54.64 & +36:00:49.1 & $\simgt$28.00 & 26.70$\pm$0.51& 25.35$\pm$0.37 & 25.78$\pm$0.45& 25.42$\pm$0.15 & 1.93 & 0.76\\
  11:37:58.65 & +35:57:07.7 & $\simgt$28.00 & $\simgt$27.61 & 25.55$\pm$0.43 & 25.81$\pm$0.46& 25.56$\pm$0.17 & 2.06 & 0.32\\
  11:38:00.02 & +35:57:43.6 & 26.90$\pm$0.43& 26.77$\pm$0.54& 25.38$\pm$0.37 & $\simgt$27.39 & 25.35$\pm$0.16 & 2.18 & 0.15\\
  11:38:00.77 & +35:42:06.5 & 26.88$\pm$0.42& 27.13$\pm$0.69& 25.71$\pm$0.48 & $\simgt$27.39 & 25.66$\pm$0.19 & 1.88 & 0.81\\
  11:38:09.18 & +35:57:12.8 & $\simgt$28.00 & $\simgt$27.61 & 25.62$\pm$0.45 & $\simgt$27.39 & 25.62$\pm$0.17 & 1.88 & 0.67\\
  11:38:09.39 & +35:43:27.5 & $\simgt$28.00 & $\simgt$27.61 & 25.29$\pm$0.35 & $\simgt$27.39 & 25.42$\pm$0.16 & 1.97 & 0.17\\
  11:38:11.87 & +35:53:46.4 & 27.39$\pm$0.61& 25.84$\pm$0.26& 24.33$\pm$0.16 & 25.48$\pm$0.36& 24.33$\pm$0.06 & 2.05 & 0.67\\
  11:38:15.11 & +35:58:17.4 & $\simgt$28.00 & 26.70$\pm$0.51& 25.39$\pm$0.38 & $\simgt$27.39 & 25.46$\pm$0.17 & 1.92 & 0.01\\
  11:38:17.38 & +35:56:31.7 & $\simgt$28.00 & $\simgt$27.61 & 25.69$\pm$0.47 & $\simgt$27.39 & 25.71$\pm$0.19 & 2.11 & 0.26\\
  11:38:17.56 & +35:56:43.6 & 26.92$\pm$0.43& 26.72$\pm$0.52& 25.42$\pm$0.39 & $\simgt$27.39 & 25.49$\pm$0.16 & 2.19 & 0.59\\
  11:38:18.53 & +35:37:37.7 & 26.83$\pm$0.4 & 26.93$\pm$0.6 & 25.28$\pm$0.35 & 25.87$\pm$0.48& 25.39$\pm$0.15 & 2.01 & 0.62\\
  11:38:24.73 & +35:48:53.0 & 27.42$\pm$0.62& $\simgt$27.61 & 25.73$\pm$0.49 & 26.29$\pm$0.65& 25.70$\pm$0.18 & 2.01 & 0.78\\
  11:38:26.22 & +36:00:51.7 & 26.86$\pm$0.41& 26.63$\pm$0.48& 25.31$\pm$0.36 & 26.10$\pm$0.57& 25.42$\pm$0.16 & 2.18 & 0.15\\
  11:38:27.31 & +35:55:11.9 & $\simgt$28.00 & $\simgt$27.61 & 25.56$\pm$0.43 & 26.05$\pm$0.55& 25.53$\pm$0.16 & 1.98 & 0.8\\
  11:38:37.42 & +35:37:57.4 & $\simgt$28.00 & 26.69$\pm$0.51& 25.17$\pm$0.32 & 26.28$\pm$0.65& 25.20$\pm$0.13 & 2.13 & 0.16\\
\hline\end{tabular}
\end{minipage}
\end{table*}

\begin{table*}
\begin{minipage}{180mm} 
\caption{$z{\sim}5.7$ LAEs - J1030+0524}
\label{t:J1030_LAEs}
\begin{tabular}{rrrrrrrrr}
\hline
  \multicolumn{1}{c}{R.A. (J2000)} &
  \multicolumn{1}{c}{DEC. (J2000)} &
  \multicolumn{1}{c}{R$_c$ (2")} &
  \multicolumn{1}{c}{i' (2")} &
  \multicolumn{1}{c}{z' (2")} &
  \multicolumn{1}{c}{NB$_{\civ}$ (2")} &
  \multicolumn{1}{c}{NB$_{\civ}$ (MAG\_AUTO)} &
  \multicolumn{1}{c}{r$_{\rm hl}$ (pixels)} &
  \multicolumn{1}{c}{S/G$_{\rm NB}$} \\
\hline
  10:29:44.84 & +05:37:48.1 & $\simgt$28.35 & $\simgt$28.04 & $\simgt$27.49 & 25.40$\pm$0.29 & 25.39$\pm$0.19 & 2.64 & 0.02\\
  10:29:44.86 & +05:29:55.7 & 27.95$\pm$0.75& $\simgt$28.04 & 25.94$\pm$0.54& 25.47$\pm$0.31 & 25.21$\pm$0.17 & 4.24 & 0.5\\
  10:29:47.11 & +05:33:19.4 & $\simgt$28.35 & $\simgt$28.04 & $\simgt$27.49 & 25.57$\pm$0.34 & 25.34$\pm$0.15 & 3.33 & 0.71\\
  10:29:49.86 & +05:24:38.0 & $\simgt$28.35 & 26.88$\pm$0.5 & $\simgt$27.49 & 24.82$\pm$0.18 & 24.58$\pm$0.09 & 2.8 & 0.51\\
  10:29:50.00 & +05:24:17.5 & $\simgt$28.35 & 27.09$\pm$0.58& $\simgt$27.49 & 24.49$\pm$0.14 & 24.42$\pm$0.07 & 2.57 & 0.16\\
  10:29:51.23 & +05:20:57.3 & $\simgt$28.35 & 27.01$\pm$0.54& $\simgt$27.49 & 25.23$\pm$0.26 & 25.29$\pm$0.16 & 2.44 & 0.07\\
  10:29:54.70 & +05:10:50.7 & $\simgt$28.35 & 27.00$\pm$0.54& 26.05$\pm$0.59& 25.06$\pm$0.22 & 24.82$\pm$0.15 & 3.56 & 0.0\\
  10:29:56.44 & +05:39:35.3 & 27.93$\pm$0.74& 26.49$\pm$0.37& 25.76$\pm$0.48& 24.59$\pm$0.15 & 23.88$\pm$0.10 & 5.3 & 0.0\\
  10:29:56.65 & +05:11:37.1 & $\simgt$28.35 & $\simgt$28.04 & $\simgt$27.49 & 26.01$\pm$0.47 & 25.29$\pm$0.19 & 5.11 & 0.11\\
  10:29:56.85 & +05:21:36.7 & $\simgt$28.35 & 26.09$\pm$0.5 & 26.19$\pm$0.65& 25.25$\pm$0.26 & 24.80$\pm$0.12 & 4.17 & 0.3\\
  10:29:57.91 & +05:36:37.9 & $\simgt$28.35 & $\simgt$28.04 & $\simgt$27.49 & 25.63$\pm$0.35 & 25.71$\pm$0.19 & 2.08 & 0.7\\
  10:29:58.40 & +05:15:08.1 & $\simgt$28.35 & 26.61$\pm$0.4 & $\simgt$27.49 & 24.50$\pm$0.14 & 24.55$\pm$0.09 & 2.42 & 0.28\\
  10:29:59.21 & +05:08:13.4 & $\simgt$28.35 & 27.35$\pm$0.69& $\simgt$27.49 & 25.06$\pm$0.22 & 24.87$\pm$0.13 & 3.18 & 0.15\\
  10:29:59.36 & +05:21:55.9 & $\simgt$28.35 & $\simgt$28.04 & $\simgt$27.49 & 25.52$\pm$0.32 & 25.49$\pm$0.16 & 2.22 & 0.72\\
  10:30:02.54 & +05:32:48.9 & $\simgt$28.35 & $\simgt$28.04 & $\simgt$27.49 & 24.92$\pm$0.20 & 25.01$\pm$0.11 & 2.11 & 0.91\\
  10:30:04.19 & +05:23:43.3 & $\simgt$28.35 & $\simgt$28.04 & 26.20$\pm$0.65& 25.01$\pm$0.21 & 24.51$\pm$0.12 & 4.26 & 0.0\\
  10:30:05.80 & +05:15:23.4 & $\simgt$28.35 & $\simgt$28.04 & 25.89$\pm$0.52& 25.76$\pm$0.39 & 25.17$\pm$0.17 & 4.89 & 0.26\\
  10:30:06.36 & +05:17:42.1 & $\simgt$28.35 & $\simgt$28.04 & $\simgt$27.49 & 24.99$\pm$0.21 & 25.07$\pm$0.13 & 2.14 & 0.41\\
  10:30:12.11 & +05:13:04.1 & $\simgt$28.35 & 27.02$\pm$0.55& $\simgt$27.49 & 25.59$\pm$0.34 & 25.62$\pm$0.18 & 2.14 & 0.74\\
  10:30:13.08 & +05:24:39.5 & $\simgt$28.35 & $\simgt$28.04 & $\simgt$27.49 & 25.30$\pm$0.27 & 25.13$\pm$0.16 & 2.99 & 0.0\\
  10:30:14.88 & +05:11:00.7 & $\simgt$28.35 & 27.23$\pm$0.64& $\simgt$27.49 & 25.21$\pm$0.25 & 25.17$\pm$0.14 & 2.43 & 0.22\\
  10:30:15.69 & +05:15:50.9 & $\simgt$28.35 & 25.90$\pm$0.23& 25.69$\pm$0.45& 24.34$\pm$0.12 & 24.36$\pm$0.07 & 2.42 & 0.88\\
  10:30:15.72 & +05:34:59.6 & $\simgt$28.35 & 27.27$\pm$0.65& 26.03$\pm$0.58& 25.15$\pm$0.24 & 25.03$\pm$0.13 & 2.59 & 0.71\\
  10:30:15.73 & +05:27:37.6 & $\simgt$28.35 & $\simgt$28.04 & $\simgt$27.49 & 25.40$\pm$0.29 & 25.24$\pm$0.14 & 2.67 & 0.73\\
  10:30:18.26 & +05:09:43.2 & $\simgt$28.35 & 26.53$\pm$0.38& $\simgt$27.49 & 25.12$\pm$0.23 & 25.17$\pm$0.13 & 2.34 & 0.88\\
  10:30:18.58 & +05:08:22.0 & $\simgt$28.35 & 27.34$\pm$0.69& $\simgt$27.49 & 25.34$\pm$0.28 & 25.48$\pm$0.13 & 1.73 & 0.91\\
  10:30:19.05 & +05:30:03.9 & 27.54$\pm$0.57& 26.70$\pm$0.43& 25.66$\pm$0.44& 24.66$\pm$0.16 & 24.60$\pm$0.09 & 2.67 & 0.31\\
  10:30:20.52 & +05:14:22.3 & $\simgt$28.35 & 26.68$\pm$0.43& 26.09$\pm$0.61& 25.15$\pm$0.24 & 25.19$\pm$0.11 & 1.99 & 0.94\\
  10:30:21.44 & +05:36:50.0 & $\simgt$28.35 & 25.95$\pm$0.24& 24.93$\pm$0.25& 24.53$\pm$0.14 & 24.44$\pm$0.08 & 2.78 & 0.17\\
  10:30:21.54 & +05:32:56.3 & $\simgt$28.35 & 26.24$\pm$0.3 & 25.92$\pm$0.54& 24.23$\pm$0.11 & 24.19$\pm$0.05 & 2.38 & 0.97\\
  10:30:23.11 & +05:33:41.5 & $\simgt$28.35 & $\simgt$28.04 & $\simgt$27.49 & 25.50$\pm$0.32 & 25.29$\pm$0.19 & 3.13 & 0.19\\
  10:30:25.15 & +05:30:36.8 & $\simgt$28.35 & $\simgt$28.04 & 26.40$\pm$0.75& 25.08$\pm$0.23 & 25.00$\pm$0.19 & 2.98 & 0.04\\
  10:30:27.68 & +05:24:19.9 & $\simgt$28.35 & $\simgt$28.04 & 25.89$\pm$0.52& 25.04$\pm$0.22 & 25.12$\pm$0.15 & 2.51 & 0.17\\
  10:30:28.03 & +05:32:32.9 & $\simgt$28.35 & $\simgt$28.04 & $\simgt$27.49 & 25.45$\pm$0.31 & 25.55$\pm$0.13 & 1.87 & 0.65\\
  10:30:33.41 & +05:23:41.8 & 27.67$\pm$0.62& 26.36$\pm$0.33& 25.44$\pm$0.37& 24.66$\pm$0.16 & 24.13$\pm$0.08 & 4.68 & 0.11\\
  10:30:35.72 & +05:30:07.0 & $\simgt$28.35 & 26.54$\pm$0.38& 25.61$\pm$0.42& 24.90$\pm$0.19 & 24.41$\pm$0.11 & 4.52 & 0.1\\
  10:30:36.90 & +05:17:08.3 & $\simgt$28.35 & 26.85$\pm$0.49& 26.40$\pm$0.75& 24.27$\pm$0.11 & 24.35$\pm$0.06 & 2.02 & 0.97\\
  10:30:37.94 & +05:23:04.6 & $\simgt$28.35 & $\simgt$28.04 & $\simgt$27.49 & 25.69$\pm$0.37 & 25.61$\pm$0.18 & 2.73 & 0.59\\
  10:30:40.39 & +05:16:18.1 & $\simgt$28.35 & $\simgt$28.04 & $\simgt$27.49 & 25.35$\pm$0.28 & 25.42$\pm$0.17 & 2.31 & 0.71\\
  10:30:40.80 & +05:27:17.4 & $\simgt$28.35 & 26.49$\pm$0.37& 25.38$\pm$0.36& 24.31$\pm$0.12 & 24.02$\pm$0.08 & 3.74 & 0.01\\
  10:30:45.50 & +05:37:39.0 & $\simgt$28.35 & 27.37$\pm$0.70& $\simgt$27.49 & 25.20$\pm$0.25 & 25.02$\pm$0.19 & 3.32 & 0.0\\
  10:30:58.73 & +05:32:44.2 & $\simgt$28.35 & 26.90$\pm$0.50& 25.93$\pm$0.54& 24.52$\pm$0.14 & 24.28$\pm$0.08 & 3.16 & 0.47\\
  10:30:59.06 & +05:13:29.4 & $\simgt$28.35 & $\simgt$28.040& $\simgt$27.49 & 25.63$\pm$0.35 & 25.64$\pm$0.19 & 2.51 & 0.58\\
  10:31:00.18 & +05:10:39.3 & $\simgt$28.35 & 26.89$\pm$0.50& $\simgt$27.49 & 25.50$\pm$0.32 & 25.64$\pm$0.15 & 1.78 & 0.68\\
  10:31:07.19 & +05:11:37.6 & 27.58$\pm$0.58& 25.93$\pm$0.23& 25.25$\pm$0.32& 24.10$\pm$0.10 & 23.85$\pm$0.07 & 3.42 & 0.03\\
\hline\end{tabular}
\end{minipage}
\end{table*}

\begin{table*}
\begin{minipage}{170mm} 
\caption{$z{\sim}5.7$ LAEs - J1137+3549}
\label{t:J1137_LAEs}
\begin{tabular}{rrrrrrrrr}
\hline
  \multicolumn{1}{c}{R.A. (J2000)} &
  \multicolumn{1}{c}{DEC. (J2000)} &
  \multicolumn{1}{c}{R$_c$ (2.4")} &
  \multicolumn{1}{c}{i' (2.4")} &
  \multicolumn{1}{c}{z' (2.4")} &
  \multicolumn{1}{c}{NB$_{\civ}$ (2.4")} &
  \multicolumn{1}{c}{NB$_{\civ}$ (MAG\_AUTO)} &
  \multicolumn{1}{c}{r$_{\rm hl}$ (pixels)} &
  \multicolumn{1}{c}{S/G$_{\rm NB}$} \\
\hline
  11:35:55.80 & +35:48:58.2 & $\simgt$28.0  & 25.95$\pm$0.29& 25.44$\pm$0.39& 24.23$\pm$0.13 & 24.16$\pm$0.12 & 4.32 & 0.1\\
  11:36:37.38 & +35:58:31.6 & 27.26$\pm$0.56& 26.02$\pm$0.3 & 25.1 $\pm$0.3 & 24.55$\pm$0.17 & 24.04$\pm$0.17 & 5.94 & 0.0\\
  11:36:37.61 & +35:41:03.1 & $\simgt$28.0  & 26.77$\pm$0.54& 25.5 $\pm$0.41& 24.39$\pm$0.14 & 23.89$\pm$0.12 & 5.86 & 0.01\\
  11:36:39.09 & +35:43:46.4 & $\simgt$28.0  & 25.95$\pm$0.29& 25.3 $\pm$0.35& 23.89$\pm$0.09 & 23.88$\pm$0.06 & 3.72 & 0.91\\
  11:36:45.27 & +35:40:40.4 & $\simgt$28.0  & $\simgt$27.61 & $\simgt$27.39 & 24.70$\pm$0.19 & 24.71$\pm$0.17 & 3.9 & 0.67\\
  11:36:47.09 & +35:44:02.2 & $\simgt$28.0  & 26.16$\pm$0.34& $\simgt$27.39 & 23.82$\pm$0.09 & 23.87$\pm$0.06 & 3.47 & 0.94\\
  11:36:49.74 & +35:57:59.2 & 27.64$\pm$0.73& 26.37$\pm$0.4 & 25.96$\pm$0.58& 24.71$\pm$0.19 & 24.59$\pm$0.13 & 3.9 & 0.5\\
  11:37:01.31 & +35:49:17.4 & $\simgt$28.0  & $\simgt$27.61 & $\simgt$27.39 & 24.77$\pm$0.20 & 24.88$\pm$0.23 & 3.944 & 0.2\\
  11:37:18.60 & +35:53:00.4 & $\simgt$28.0  & $\simgt$27.61 & $\simgt$27.39 & 24.69$\pm$0.19 & 24.49$\pm$0.14 & 4.15 & 0.68\\
  11:37:39.82 & +35:52:38.2 & $\simgt$28.0  & $\simgt$27.61 & $\simgt$27.39 & 24.98$\pm$0.24 & 24.72$\pm$0.14 & 5.23 & 0.48\\
  11:37:48.12 & +35:45:07.7 & 27.62$\pm$0.72& 26.28$\pm$0.37& 25.59$\pm$0.44& 24.51$\pm$0.16 & 24.42$\pm$0.10 & 3.86 & 0.62\\
  11:37:58.70 & +35:56:44.3 & $\simgt$28.0  & 26.46$\pm$0.43& 26.12$\pm$0.65& 24.37$\pm$0.14 & 24.33$\pm$0.12 & 4.2 & 0.08\\
  11:38:05.27 & +35:58:11.5 & 26.43$\pm$0.29& 25.45$\pm$0.19& 24.26$\pm$0.15& 24.11$\pm$0.11 & 23.47$\pm$0.09 & 6.47 & 0.0\\
  11:38:07.69 & +35:54:42.6 & $\simgt$28.0  & 26.81$\pm$0.56& $\simgt$27.39 & 24.98$\pm$0.24 & 24.95$\pm$0.19 & 3.53 & 0.66\\
  11:38:29.86 & +35:54:22.7 & $\simgt$28.0  & $\simgt$27.61 & $\simgt$27.39 & 24.74$\pm$0.19 & 24.49$\pm$0.18 & 5.09 & 0.0\\
\hline\end{tabular}
\end{minipage}
\end{table*}


\begin{thebibliography}{99}

\bibitem[\protect\citeauthoryear{Abraham et al.}{2004}]{abraham2004} Abraham, R., G., Glazebrook, K., McCarthy, P., Crampton, D., 
Murowinski, R.,  J¿rgensen, I., Roth, K. et al. 2004, AJ, 127, 2455
\bibitem[\protect\citeauthoryear{Adelberger et al.}{2003}]{adelberger2003} Adelberger, K., Steidel, C., 
Shapley, A., \& Pettini, M. 2003, ApJ, 584, 45
\bibitem[\protect\citeauthoryear{Adelberger et al.}{2005a}]{adelberger2005a} Adelberger, K. , Steidel, C. , Pettini, M., 
Shapley, A., Reddy, N., \& Erb, D. 2005, ApJ, 619, 697
\bibitem[\protect\citeauthoryear{Adelberger et al.}{2005b}]{adelberger2005b} Adelberger, K.
L., Shapley, A. E., Steidel, C. C., Pettini, M., Erb, D. K. \& Reddy, N. A. 2005, ApJ, 629, 636 
\bibitem[\protect\citeauthoryear{Ajiki et al.}{2003}]{ajiki2003} Ajiki, M. et al. 2003, ApJ, 126, 2091
\bibitem[\protect\citeauthoryear{Bielby et al.}{2013}]{bielby2013} Bielby, R., Hill, M. D., Shanks, T., 
Crighton, N. H. M., Infante, L., Bornancini, C. G., Francke, H., HŽraudeau, P., Lambas, D. G. et al. 2013, MNRAS, 430, 425
\bibitem[\protect\citeauthoryear{Bertin \& Arnouts }{1996}]{bertin1996} Bertin, E. \& Arnouts, S. 1996, A\&AS, 117, 393 
\bibitem[\protect\citeauthoryear{Becker et al.}{2001}]{becker2001} Becker, R. et al. 2001, AJ, 122, 2850
\bibitem[\protect\citeauthoryear{Becker et al.}{2006}]{becker2006} Becker, G. D.,
Sargent, W. L. W.,  Rauch, M. \& Simcoe, R. 2006, ApJ, 640, 69
\bibitem[\protect\citeauthoryear{Becker, Rauch \& Sargent}{2007}]{becker2007} Becker,
G. D., Rauch, M. \& Sargent, W. L. W. 2007, ApJ, 662, 72
\bibitem[\protect\citeauthoryear{Becker, Rauch \& Sargent}{2009}]{becker2009} Becker,
G. D., Rauch, M. \& Sargent, W. L. W. 2009, ApJ, 698, 1010
\bibitem[\protect\citeauthoryear{Bochanski et al.}{2010}] {bochanski2010} Bochanski, J., Hawley, S.,
Covey, K., West, A., Reid, N., Golimowski, D. \& Ivezi\'c, Z., 2010, AJ, 139, 2679
\bibitem[\protect\citeauthoryear{Bochanski et al.}{2012}] {bochanski2012} Bochanski, J., Hawley, S., 
Covey, K., West, A., Reid, N., Golimowski, D., \& Ivezi\'c, Z. 2012, AJ, 143, 152
\bibitem[\protect\citeauthoryear{Bolton \& Haehnelt}{2007}]{bolton2007} Bolton, J. \& Haehnelt, M. 2007, MNRAS, 382, 325
\bibitem[\protect\citeauthoryear{Bolton \& Haehnelt}{2013}]{bolton2013} Bolton, J. \& Haehnelt, M. 2013, MNRAS, 429, 1695
\bibitem[\protect\citeauthoryear{Bouwens et al.}{2004}]{bouwens2004} Bouwens, R., Illingworth, G.,
Blakeslee, J., Broadhurst, T., \& Franx, M. 2004, ApJL, 611, L1
\bibitem[\protect\citeauthoryear{Bouwens et al.}{2006}]{bouwens2006} Bouwens, R. J.,
Illingworth, G. D., Blakeslee, J. P., \& Franx, M. 2006, ApJ, 653, 53
\bibitem[\protect\citeauthoryear{Bouwens et al.}{2007}]{bouwens2007} Bouwens, R. J.,
Illingworth, G. D., Franx, M. \& Ford, H. 2007, ApJ, 670, 928
\bibitem[\protect\citeauthoryear{Bouwens et al.}{2012}]{bouwens2012} Bouwens, R. J.,
et al. 2012, ApJ, 754, 83
\bibitem[\protect\citeauthoryear{Bradshaw et al.}{2013}]{bradshaw2013} Bradshaw, E. J., Almaini, O., 
Hartley, W. G., Smith, K. T., Conselice, C. J., Dunlop, J. S., Simpson, C., Chuter, R. W., et al. 2013, MNRAS, 433, 194
\bibitem[\protect\citeauthoryear{Brown et al.} {2004}]{brown2004} Brown, W., Geller, M.,
Kenyon, S., Beers, T., Kurtz, M. \& Roll, J. 2004, AJ, 127, 1555
\bibitem[\protect\citeauthoryear{Bruzual \& Charlot}{2003}]{bruzual2003} Bruzual, G. \& Charlot, S. 2003, MNRAS, 344, 1000
\bibitem[\protect\citeauthoryear{Bunker et al.}{2004}]{bunker2004} Bunker, A. J., Stanway, 
E. R., Ellis, R. S. \& McMahon, R. G. 2004, MNRAS, 355, 374
\bibitem[\protect\citeauthoryear{Bunker et al.}{2013}]{bunker2013} Bunker, A. J., Caruana, J., 
Wilkins, S. M., Stanway, E. R., Lorenzoni, S., Lacy, M., Jarvis, M.J., \& Hickey, S. 2013, MNRAS, 430, 3314
\bibitem[\protect\citeauthoryear{Cai et al.}{2014}]{cai2014} Cai, Z.-Y., Lapi, A., 
Bressan, A., De Zotti, G., Negrello, M., and Danese, L. 2014, ApJ, 785, 65
\bibitem[\protect\citeauthoryear{Carollo et al.}{2007}] {carollo2007} Carollo, D., Beers, T.,
Lee, Y., Chiba, M., Norris, J., Wilhelm, R., Sivarani, T. et al. 2007, Nature, 450, 1020
\bibitem[\protect\citeauthoryear{Caruana et al.}{2012}]{caruana2012} Caruana, J., et al. 2012, MNRAS, 427, 3055
\bibitem[\protect\citeauthoryear{Cassata et al.}{2011}] {cassata2011} Cassata, P., et al. 2011, A\&A, 525, 143
\bibitem[\protect\citeauthoryear{Cen \& Chisari}{2011}]{cen2011} Cen, R., Chisari,
N. E. 2011, ApJ, 731, 11
\bibitem[\protect\citeauthoryear{Choudhury, Haehnelt \& Regan}{2009}]{choudhury2009} Choudhury, T., 
Haehnelt, M. \& Regan, J. 2009, MNRAS, 394, 960
\bibitem[\protect\citeauthoryear{Cl\'ement et al.}{2011}]{clement2012} Cl\'ement, B.,
et al. 2012, A\&A, 538, 66
\bibitem[\protect\citeauthoryear{Cooke et al.}{2006}]{cooke2006} Cooke, J., Wolfe, A. M., 
Gawiser, E., \& Prochaska, J. X. 2006, ApJ, 652, 994
\bibitem[\protect\citeauthoryear{Cooke}{2009}]{cooke2009} Cooke, J. 2009, ApJ, 704, L62
\bibitem[\protect\citeauthoryear{Cooke et al.}{2010}]{cooke2010} Cooke, J., Berrier, J.,
Barton, E., Bullock, J., \& Wolfe, A. 2010, MNRAS, 403, 1020
\bibitem[\protect\citeauthoryear{Cooke, Omori \& Ryan-Weber}{2013}]{cooke2013} Cooke, J., Omori, Y., \& Ryan-Weber, E. 2013, MNRAS, 433, 2122
\bibitem[\protect\citeauthoryear{Cooksey et al.}{2010}]{cooksey2010} Cooksey, K., L., 
Thom, C., Prochaska, J., X., and Chen, H.-W. 2010, ApJ, 708, 868
\bibitem[\protect\citeauthoryear{Curtis-Lake et al.}{2012}]{curtis-lake2012} Curtis-Lake, E., et al. 2012, MNRAS, 422, 1425
\bibitem[\protect\citeauthoryear{Dessauges-Zavadsky et al.}{2010}]{dessauges-zavadsky2010}
Dessauges-Zavadsky, M., D'Odorico, S., Schaerer, D., Modigliani, A., Tapken,
C., \& Vernet, J.\ 2010, A\&A, 510, 26 
\bibitem[\protect\citeauthoryear{Diaz et al.}{2011}]{diaz2011} Diaz, C. G.,
Ryan-Weber, E., Cooke, J., Pettini, M. \& Madau, P. 2011, MNRAS, 418, 820
\bibitem[\protect\citeauthoryear{D'Odorico et al.}{2013}]{dodorico2013} D'Odorico, V., Cupani, G.,
Cristiani, S., Maiolino, R., Molaro, P., Nonino, M., Centuri\'on, M., et al. 2013, MNRAS, 435, 1198
\bibitem[\protect\citeauthoryear{Dressler et al.}{2011}]{dressler2011} Dressler, A., 
Martin, C.L., Henry, A., Sawicki, M., \& McCarthy, P. 2011, ApJ, 740, 71
\bibitem[\protect\citeauthoryear{Fan et al.}{2006}]{fan2006a} Fan, X. et al. 2006,
AJ, 132, 117
\bibitem[\protect\citeauthoryear{Faucher-Gigu\`ere et al.}{2008}] {faucher-giguere2008} Faucher-Gigu\`ere, C., 
Lidz, A., Hernquist, L., \& Zaldarriaga, M. 2008, ApJ, 688, 85
\bibitem[\protect\citeauthoryear{Ferrara \& Loeb}{2013}]{ferrara2013} Ferrara, A.,\& Loeb, A. 2013,
MNRAS, 431, 2826
\bibitem[\protect\citeauthoryear{Finkelstein et al.}{2012a}] {finkelstein2012a} Finkelstein, S. L., et al. 2012, ApJ, 756, 164
\bibitem[\protect\citeauthoryear{Finkelstein et al.}{2012b}] {finkelstein2012b} Finkelstein, S. L., et al. 2012, ApJ, 758, 93.
\bibitem[\protect\citeauthoryear{Finlator et al.}{2009}]{finlator2009b} Finlator, K.,
O\"zel, F., Dav\'e, R. \& Oppenheimer, B. 2009, MNRAS, 400, 1049
\bibitem[\protect\citeauthoryear{Finlator et al.}{2013}]{finlator2013} Finlator, K.,
Mu\~noz, J., Oppenheimer, B., Peng Oh, S., O\"zel, F., \& Dav\'e., R. 2013, MNRAS, 436, 1818
\bibitem[\protect\citeauthoryear{Fontana et al.}{2010}]{fontana2010} Fontana, A., et al. 2010, ApJL, 725, L205 
\bibitem[\protect\citeauthoryear{Fontanot et al.}{2014}]{fontanot2014} Fontanot, F., Cristiani, S., Pfrommer, C., 
Cupani, G., \& Vanzella, E. 2014, MNRAS, 438, 2097
\bibitem[\protect\citeauthoryear{Gawiser et al.}{2006}]{gawiser2006} Gawiser, E., van Dokkum, P., Gronwall, C., Ciardullo, R.,
Blanc, G., Castander, F., Feldmeier, J., et al. 2006, ApJL, 642, L13
\bibitem[\protect\citeauthoryear{Giallongo \& Cristiani}{1990}]{giallongo1990} Giallongo, E. \& Cristiani, S., 1990, MNRAS, 247, 696
\bibitem[\protect\citeauthoryear{Giavalisco et al.}{2004}]{giavalisco2004} Giavalisco, M.,
et al. 2004, ApJ, 600, L93 
\bibitem[\protect\citeauthoryear{Gonz\'alez et al.}{2011}]{gonzalez2011} Gonz\'alez, V., Labb\'e, I., 
Bouwens, R.J., Illingworth, G., Franx, M., \& Kriek, M. 2011, ApJL, 735, L34
\bibitem[\protect\citeauthoryear{Goto et al.}{2011}] {goto2011} Goto, T., Utsumi, Y., Hattori, T., 
Miyazaki, S. \& Yamauchi, C. 2011, MNRAS, 415, L1
\bibitem[\protect\citeauthoryear{Griffen et al.}{2012}] {griffen2013} Griffen, B., 
Drinkwater, M., Iliev, I., Thomas, P. \& Mellema, G. 2013, MNRAS, 431, 3087
\bibitem[\protect\citeauthoryear{Gunn \& Peterson}{1965}]{gunnpeterson1965} Gunn, J. E. \& Peterson, B. A. 1965, ApJ, 142, 1633
\bibitem[\protect\citeauthoryear{Heckman et al.}{2000}]{heckman2000} Heckman, T. M., 
Lehnert, M. D., Strickland, D. K., \& Armus, L.  2000, ApJS, 129, 493 
\bibitem[\protect\citeauthoryear{Hildebrandt et al.}{2009}]{hildebrandt2009} Hildebrandt, H., 
Pielorz, J., Erben, T., van Waerbeke, L., Simon, P., \& Capak., P. 2009, A\&A, 498, 725
\bibitem[\protect\citeauthoryear{Hu et al.}{2010}]{hu2010} Hu, E., Cowie, L., Barger, 
A., Capak, P., Kakazu, Y., \& Trouille, L. 2010, ApJ, 725, 394
\bibitem[\protect\citeauthoryear{Huang et al.}{2013}]{huang2013} Huang, K., Ferguson, H., 
Ravindranath, S., \& Su, J. 2013, ApJ, 765, 68.
\bibitem[\protect\citeauthoryear{Jaacks et al.}{2012}] {jaacks2012} Jaacks, J., Choi, J.-H., 
Nagamine, K., Thompson, R., \& Varghese, S. 2012, MNRAS, 420, 1606
\bibitem[\protect\citeauthoryear{Jensen et al.}{2013}] {jensen2013} Jensen, H., Laursen, P.,
Mellema, G., Iliev, I., Sommer-Larsen, J. \& Shapiro, P. 2013, MNRAS, 428, 1366
\bibitem[\protect\citeauthoryear{Jones, Stark \& Ellis}{2012}]{jones2012} Jones, T., 
Stark, D., \& Ellis., R. 2012, ApJ, 751, 51
\bibitem[\protect\citeauthoryear{Juri\'c et al.}{2008}] {juric2008} Juri\'c, M., Ivezi\'c, \v{Z}, 
Brooks, A., Lupton, R., Schlegel, D., Finkbeiner, D., Padmanabhan, N., et al. 2008, ApJ, 673, 864
\bibitem[\protect\citeauthoryear{Kashikawa et al.}{2006}]{kashikawa2006a} Kashikawa, N.,
et al. 2006, ApJ, 637, 631
\bibitem[\protect\citeauthoryear{Kashikawa et al.}{2011}]{kashikawa2011} Kashikawa, N.,
et al. 2011, ApJ, 734, 119
\bibitem[\protect\citeauthoryear{Keating et al.}{2014}] {keating2014} Keating, L., Haehnelt, M., 
Becker, G. \& Bolton, J., 2014, MNRAS, 438, 1820
\bibitem[\protect\citeauthoryear{Kim et al.}{2009}]{kim2009} Kim, S., et al. 2009,
ApJ, 695, 809 
\bibitem[\protect\citeauthoryear{Komatsu et al.}{2011}]{komatsu2011} Komatsu, E., et al. 2011 ApJS, 192, 18
\bibitem[\protect\citeauthoryear{Kuhlen \& Faucher-Gigu\`ere}{2012}]{kuhlen2012} Kuhlen, M., \&
Faucher-Gigu\`ere, C.-A. 2012, MNRAS 423, 862
\bibitem[\protect\citeauthoryear{Larson et al.}{2011}]{larson2011} Larson, D., et al.
2011, ApJS, 192, 16
\bibitem[\protect\citeauthoryear{Lee et al.}{2006}]{lee2006} Lee, K., Giavalisco, M.,
Gnedin, O., Somerville, R., Ferguson, H., Dickinson, \& M., Ouchi., M., 2006, ApJ, 642, 63
\bibitem[\protect\citeauthoryear{Lidman et al.}{2012}]{lidman2012} Lidman, C., Hayes, M.,
Jones, D., Schaerer, D., Westra, E., Tapken, C., Meisenheimer, K., \& Verhamme., 
A. 2012, MNRAS, 420, 1946
\bibitem[\protect\citeauthoryear{Lu \& Zuo}{1994}]{lu1994} Lu, L. \& Zuo, L. 1994, ApJ, 426, 502 
\bibitem[\protect\citeauthoryear{Malhotra et al.}{2012}]{malhotra2012} Malhotra, S., 
Rhoads, J., Finkelstein, S., Hathi, N., Nilsson, K., McLinden, E., \& Pirzkal, N. 2012, ApJ, 750, L36
\bibitem[\protect\citeauthoryear{Martin}{2005}]{martin2005} Martin, C. L. 2005, ApJ, 621, 227
\bibitem[\protect\citeauthoryear{Martin}{2006}]{martin2006} Martin, C. L. 2006, ApJ, 647, 222
\bibitem[\protect\citeauthoryear{Martin et al.}{2010}]{martin2010} Martin, C. L., Scannapieco, E., Ellison, S. L., 
Hennawi, J. F., Djorgovski, S. G., \& Fournier, A. P. 2010, ApJ, 721, 174
\bibitem[\protect\citeauthoryear{McLure et al.}{2011}]{mclure2011} McLure, R. J., et al. 2011, MNRAS, 418, 2074
\bibitem[\protect\citeauthoryear{Mesinger \& Furlanetto}{2009}] {mesinger2009}
Mesinger, A. \& Furlanetto, S. 2009, MNRAS, 400, 1461 
\bibitem[\protect\citeauthoryear{Metchev et al.}{2008}] {metchev2008} Metchev, S., Kirkpatrick, J., Berriman, G., \& Looper, D. 2008, ApJ, 676, 1281
\bibitem[\protect\citeauthoryear{Miralda-Escud\'e}{2003}] {miralda-escude2003} Miralda-Escud\'e, J. 2003, ApJ, 597, 66
\bibitem[\protect\citeauthoryear{Miyazaki et al.} {2002}] {miyazaki2002} Miyazaki, S., Komiyama, Y.,
Sekiguchi, M., Okamura, S., Doi, M., Furusawa, H., Hamabe, M., et al. 2002, PASJ, 54, 833.
\bibitem[\protect\citeauthoryear{Miyazaki et al.} {2003}] {miyazaki2003} Miyazaki, M., Shimasaku, K., Kodama, T., 
Okamura, S., Furusawa, H., Ouchi, M., Nakata, F., et al. 2003, PASJ, 55, 1079
\bibitem[\protect\citeauthoryear{Mortlock et al.}{2011}] {mortlock2011} Mortlock, D., J., et al. 2011, Nature, 474, 616
\bibitem[\protect\citeauthoryear{Oesch et al.}{2010}]{oesch2010} Oesch, P., Bouwens, R., Carollo, C., 
Illingworth, G., Trenti, M., Stiavelli, M., Magee, D., LabbŽ, I., \& Franx, M. 2010, ApJL, 709, L21
\bibitem[\protect\citeauthoryear{Ono et al.}{2012}]{ono2012a} Ono, Y., et al. 2012, ApJ, 744, 83
\bibitem[\protect\citeauthoryear{Oppenheimer \& Dav\'e}{2006}]{oppenheimer2006}
Oppenheimer, B. D. \& Dav\'e, R. 2006, MNRAS, 373, 1265
\bibitem[\protect\citeauthoryear{Oppenheimer \& Dav\'e}{2008}]{oppenheimer2008}
Oppenheimer, B. D. \& Dav\'e, R. 2008, MNRAS, 387, 577
\bibitem[\protect\citeauthoryear{Oppenheimer, Dav\'e \& Finlator}{2009}]{oppenheimer2009}
Oppenheimer, B. D., Dav\'e, R. \& Finlator, K. 2009, MNRAS, 396, 729
\bibitem[\protect\citeauthoryear{Ouchi et al.}{2004a}]{ouchi2004a} Ouchi, M. et al.
2004a, ApJ, 611, 660 
\bibitem[\protect\citeauthoryear{Ouchi et al.}{2004b}]{ouchi2004b} Ouchi, M. et al.
2004b, ApJ, 611, 685 
\bibitem[\protect\citeauthoryear{Ouchi et al.}{2005a}]{ouchi2005a} Ouchi, M. et al.
2005, ApJL, 620, L1
\bibitem[\protect\citeauthoryear{Ouchi et al.}{2005b}]{ouchi2005b} Ouchi, M. et al.
2005, ApJL, 635, L117 
\bibitem[\protect\citeauthoryear{Ouchi et al.}{2008}]{ouchi2008} Ouchi, M. et al.
2008, ApJS, 176, 301 
\bibitem[\protect\citeauthoryear{Ouchi et al.}{2010}]{ouchi2010} Ouchi, M. et al.
2010, ApJ, 723, 869 
\bibitem[\protect\citeauthoryear{Pentericci et al.}{2011}]{pentericci2011} Pentericci,
L., et al. 2011, ApJ, 743, 132
\bibitem[\protect\citeauthoryear{Pettini et al.}{2002}]{pettini2002} Pettini, M., Rix, 
S. A., Steidel, C. C., Adelberger, K. L., Hunt, M. P., \& Shapley, A. E.\ 2002,
ApJ, 569, 742 
\bibitem[\protect\citeauthoryear{Porciani \& Madau}{2005}]{porciani2005}
Porciani, C. \& Madau, P. 2005, ApJ, 625, L43
\bibitem[\protect\citeauthoryear{Reichart}{2001}]{reichart2001} Reichart, D. 2001, ApJ, 553, 235
\bibitem[\protect\citeauthoryear{Robertson et al.}{2013}]{robertson2013} Robertson, B. E., et al. 2013, ApJ, 768, 71
\bibitem[\protect\citeauthoryear{Rupke, Veilleux \& Sanders}{2005}]{rupke2005}
Rupke, D. S., Veilleux, S. \& Sanders, D. B. 2005, ApJS, 160, 115
\bibitem[\protect\citeauthoryear{Ryan-Weber et al.}{2009}]{ryan-weber2009} Ryan-Weber, E.
V., Pettini, M., Madau, P. \& Zych, B. J. 2009, MNRAS, 395, 1476
\bibitem[\protect\citeauthoryear{Schechter}{1976}]{schechter1976} Schechter, P. 1976, ApJ, 203, 297
\bibitem[\protect\citeauthoryear{Schenker et al.}{2012}]{schenker2012a} Schenker, M.,
et al. 2011, ApJ, 744, 179
\bibitem[\protect\citeauthoryear{Schlegel, Finkbeiner \& Davis}{1998}]{schlegel1998}
Schlegel, D.J., Finkbeiner, D. P. \& Davis, M. 1998, ApJ, 500, 525
\bibitem[\protect\citeauthoryear{Shapley et al.}{2003}]{shapley2003} Shapley, A. E.,
Steidel, C. C., Pettini, M., \& Adelberger, K. L.\ 2003, ApJ, 588, 65
\bibitem[\protect\citeauthoryear{Shimasaku et al.}{2006}]{shimasaku2006} Shimasaku, K., 
et al. 2006, PASJ, 58, 313 
\bibitem[\protect\citeauthoryear{Simcoe}{2011}]{simcoe2011a} Simcoe R. A. 2011, ApJ, 738, 159
\bibitem[\protect\citeauthoryear{Simcoe et al.}{2011}]{simcoe2011b} Simcoe R. A., et
al.  2011, ApJ, 743, 21
\bibitem[\protect\citeauthoryear{Schroeder, Mesinger \& Haiman}{2013}] {schroeder2013} Schroeder, J., 
Mesinger, A. \& Haiman, Z. 2013, MNRAS, 428, 3058
\bibitem[\protect\citeauthoryear{Stanway et al.}{2004}]{stanway2004} Stanway, E., et al. 2004, ApJL, 604, L13
\bibitem[\protect\citeauthoryear{Stanway et al.}{2007}]{stanway2007} Stanway, E., et al. 2007, MNRAS, 376, 727
\bibitem[\protect\citeauthoryear{Stanway, Bremer \& Lehnert}{2008}]{stanway2008} Stanway, 
E., Bremer, M. \& Lehnert, M. 2008, MNRAS, 385, 493
\bibitem[\protect\citeauthoryear{Stark et al.}{2009}]{stark2009} Stark, D., Ellis, R., 
Bunker, A., Bundy, K., Targett, T., Benson, A. \& Lacy, M. 2009, ApJ, 697, 1493
\bibitem[\protect\citeauthoryear{Stark et al.}{2010}]{stark2010} Stark, D. P., Ellis, 
R. S., Chiu, K., Ouchi, M. \& Bunker, A. 2010, MNRAS, 408, 1628
\bibitem[\protect\citeauthoryear{Stark, Ellis \& Ouchi}{2011}]{stark2011} Stark, D. P., 
Ellis, R. S. \& Ouchi, M. 2011, ApJL, 728, L2
\bibitem[\protect\citeauthoryear{Stark et al.}{2013}]{stark2013} Stark, D., Schenker, M., 
Ellis, R., Robertson, B., McLure, R., \& Dunlop, J. 2013, ApJ, 763, 129
\bibitem[\protect\citeauthoryear{Steidel, Pettini \& Hamilton}{1995}]{steidel1995} Steidel, C. C.,
Pettini, M., \& Hamilton, D. 1995, AJ, 110, 2519
\bibitem[\protect\citeauthoryear{Steidel et al.}{1996}]{steidel1996}Steidel, C.C., Giavalisco, M., 
Pettini, M., Dickinson, M., \& Adelberger, K.L. 1996, ApJL, 462, L17.
\bibitem[\protect\citeauthoryear{Steidel et al.}{1998}]{steidel1998} Steidel, C., Adelberger, K., Dickinson, M.,
Giavalisco, M., Pettini, M., \& Kellogg, M. 1998, ApJ, 492, 428
\bibitem[\protect\citeauthoryear{Steidel et al.}{2000}]{steidel2000} Steidel, C. C., 
Adelberger, K., Shapley, A. E., Pettini, M., Dickinson, M. \& Giavalisco, M. 2000, ApJ, 532, 170
\bibitem[\protect\citeauthoryear{Steidel et al.}{2010}]{steidel2010} Steidel, C. C.,
Erb, D. K., Shapley, A. E., Pettini, M., Reddy, N., Bogosavljevi\'c, M., Rudie,
G. C. \& Rakic, O. 2010, ApJ, 717, 289
\bibitem[\protect\citeauthoryear{Stiavelli et al.}{2005}]{stiavelli2005} Stiavelli, M.,
et al. 2005, ApJL, 622, L1 
\bibitem[\protect\citeauthoryear{Tescari et al.}{2011}]{tescari2011} Tescari, E.,
Viel, M., D'Odorico, V., Cristiani, S., Calura, F., Borgani, S., \& Tornatore, L.
2011, MNRAS, 411, 826 
\bibitem[\protect\citeauthoryear{Trac, Cen \& Loeb}{2008}]{trac2008} Trac, H., Cen,
R. \& Loeb, A. 2008, ApJL, 689, L81
\bibitem[\protect\citeauthoryear{Treu et al.}{2013}]{treu2013} Treu, T., Schmidt, K. B., 
Trenti, M., Bradley, L. D., \& Stiavelli, M. 2013, ApJ, 775, L29
\bibitem[\protect\citeauthoryear{Vanzella et al.}{2009}]{vanzella2009} Vanzella, E., et al. 2009, ApJ, 695, 1163
\bibitem[\protect\citeauthoryear{West et al.}{2008}] {west2008} West, A., Hawley, S., 
Bochanski, J., Covey, K., Reid, I., Dhital, S., Hilton, E., \& Masuda, M. 2008, AJ, 135, 785
\bibitem[\protect\citeauthoryear{Weiner et al.}{2009}]{weiner2009} Weiner, B. J., et
al. 2009, ApJ, 692, 187 
\bibitem[\protect\citeauthoryear{Yagi et al.}{2002}]{yagi2002} Yagi, M., Kashikawa, N., 
Sekiguchi, M., Doi, M., Yasuda, N., Shimasaku, K., \& Okamura, S. 2002, AJ, 123, 66
\bibitem[\protect\citeauthoryear{Zahn et al.}{2012}]{zahn2012} Zahn, O., et al. 
2012, ApJ, 756, 65
\bibitem[\protect\citeauthoryear{Zheng et al.}{2014}]{zheng2014} Zheng, Z.-Y., 
Wang, J.-X., Malhotra, S., Rhoads, J. E., Finkelstein, S. L., \& Finkelstein, K. 2014, MNRAS, 439, 1101

\end{thebibliography}
\end{document}